\documentclass[onecolumn]{pasj00}
\draft
\usepackage{dcolumn}
\newcolumntype{d}[1]{D{.}{.}{#1}}
\begin{document}
\author{Kohji \textsc{Tomisaka} and Kengo \textsc{Tomida}}
\email{tomisaka@th.nao.ac.jp, tomida@th.nao.ac.jp}
\affil{National Astronomical Observatory Japan\\ and\\
 Department of Astronomy, School of Physical Sciences,\\
 Graduate University for Advanced Studies (SOKENDAI),\\
 Osawa 2-21-1, Mitaka, Tokyo 181-8588, Japan}
\title{Observational Identification of First Cores:\\
Non-LTE Radiative Transfer Simulation}
\Received{2011/4/5}
\Accepted{2011/6/9}
\SetRunningHead{K.~Tomisaka and K.~Tomida}{Observational Identification of First Cores}
\maketitle
\begin{abstract}
A first core is a first hydrostatic object formed in the course of dynamical
 contraction of a molecular cloud core.
Since the inflow pattern changes drastically before and after
 the first core formation, it is regarded as a milestone in the star 
 formation process. 
In order to identify the first core from a mapping observation,
 the features expected for the first core are studied 
 for CS rotation transitions at radio wavelengths.
The non-LTE radiation transfer is calculated for the results of 
 radiation magnetohydrodynamical  simulations 
 of the contraction of the magnetized molecular cloud
 core in rotation \citep{tomida2010a}. 
We use the Monte-Carlo method to solve the non-LTE radiation transfer
 in a nested grid hierarchy.
In the first core phase, an outflow arises from the vicinity of
 the first core due to the twisted magnetic field amplified by
 the rotation motion of the contracting gas disk.
The disk and outflow system has several characteristic observational features:
 (i) relatively opaque lines indicate
 asymmetry in the emission lines in which the blue side is stronger
 than the red side (an infall signature of the envelope);
 (ii) in the edge-on view, the disk has a signature of
 simultaneous rotation and infall, 
 i.e., the integrated intensity of the approaching side is brighter than that
 of the receding side
 and the gradient in the intensity-weighted velocity is larger
 in the approaching side;
 (iii) the observed outflow indicates rotation around the rotation
 axis.
The size of the outflow gives the approximate age
 after the first core is formed, since the outflow is not expected for
 the earlier runaway isothermal collapse phase.
\end{abstract}

\section{Introduction}

The star formation process continues to attract much attention.
In the process,  ``the first core'' is regarded as a key to understanding
 the star formation process, since it is a first hydrostatic object created in 
 the course of forming a new star. 
The first core was theoretically predicted by \citet{larson1969}
 as an object supported by gas pressure after the thermal
 radiation from the dust, which is a major coolant in interstellar 
 molecular gas with density $n\sim 10^4-10^{10}{\rm H_2\,cm^{-3}}$, becomes
 inefficient at a high density of $n\gtrsim 10^{10-11}{\rm H_2\,cm^{-3}}$.    
The high density gas increases in temperature due to gravitational compression
 and a hydrostatic balance is finally achieved between the self-gravity
 and the pressure gradient.
This forms a hydrostatic object, the first core.

The dynamical evolution is completely different before and after the first
 core formation \citep{whitworth1985,tomisaka1998}.
Before the first core formation, the contraction is well expressed
 by the so-called \citet{larson1969}--\citet{penston1969}
 (LP) similarity solution.
After the core formation, the solution resembles
 \citet{shu1977}'s inside-out collapse solution but is well expressed
 by another self-similar solution regarded as an analytic continuation of
 the LP solution, as given in \citet{whitworth1985}.
Magnetohydrodynamical
 simulations confirm the transition at the first core formation
 even in a model with a magnetic field and rotation motion
 \citep{tomisaka1998,machida2004}.
After the core formation, the magnetic field lines, which are essentially
 poloidal before the first core formation, begin to be distorted
 by the rotation motion and the toroidal component is amplified.
This transfers its angular momentum from the rotating disk
 to the gas just outside the first core
 and accelerates the gas, which explains the molecular outflows
 observed accompanying protostars.
Therefore, the formation of the first core is a milestone in the process
 of star formation.
   
Although more than 40 years have passed since Larson's
 (\yearcite{larson1969}) prediction,
 the first core has not been observed.
Recently however, a possible detection has been reported  
  in a dense core IRS2E in L1448 \citep{chen2010}.
An object has been discovered which is not visible with Spitzer ($3.6-70\mu {\rm m}$)
 but is a weak mm or sub-mm continuum source.
\citet{saigo2011} calculated the expected spectral energy distribution (SED)
 of dust thermal emissions for their hydrodynamic model of the first core
 \citep{saigo2008}.
They obtained a luminosity of $\sim 0.02\, L_{\odot}$ and
 effective temperature of $\sim$ a few 10\,K.
These expected features of the first core, i.e., low luminosity and cold SED,
 are consistent with the observation by \citet{chen2010}.
However, this object seems more evolved than the first core
 since it is associated with a high-velocity outflow 
 ($\sim 25 {\rm km\,s^{-1}}$).
Another candidate is Per-Bolo 58, a dense core in Perseus, found in Spitzer 70
 $\mu$m map \citep{enoch2010}.
Although this has also a similar SED to that expected from the first core,
 detection of 24 $\mu$m seems to indicate the object is older than the first core 
 and the envelope is being to be cleared.  
Therefore, further observations are required to confirm whether these objects are
 first cores.
\citet{saigo2008} predicted that a first core forms a flat disk due to
 rotation and that a dynamical instability develops in such a fast rotating
 disk \citep{durisen1986}.
This brings us an idea that
 a fine structure (such as spriral arms) should be observed in first
 core disks in high resolution observations with such as ALMA 
 \citep{tomida2010b,saigo2011}. 

Another signature of a first core is an accompanying molecular outflow
 (\authorcite{tomisaka1998} \yearcite{tomisaka1998}; \yearcite{tomisaka2002}). 
The molecular outflow is driven by the magnetic Lorentz force
 working just outside the first core \citep{tomisaka2002,banerjee2006,machida2007,commercon2010,tomida2010a}.
In this paper, we investigate the appearance of the molecular outflow
 in its early phase, which should allow identification of a 
 first core.
This is done by post-processing the numerical results of radiation
 MHD simulations.
We call this procedure ``observational visualization.'' It
 is a method of visualizing numerical results in order to
 understand the undergoing physics but also emphasizes the observational 
 expectation from the simulation.
We have already shown the expected polarization observation of
 the thermal emission from dust grains in the molecular outflow \citep{tomisaka2011}.
Comparison with observation, this enables us to conclude
 whether the molecular outflow is directly driven magnetically or
 is indirectly made by the entrainment mechanism from the interstellar jets.

The plan of the paper is as follows:
 In section 2 we describe the model and numerical method.
A non-LTE line transfer calculation is made for 
 previously obtained data from radiative MHD simulations
 of the contraction of a rotating magnetized cloud.
The model of the initial molecular cloud is described.
The method for the non-LTE radiative transfer calculations
 for the interstellar molecular lines is also described in this section. 
In section 3, we show the results of the observational visualization.
Section 4 is devoted to a discussion of the asymmetrical distribution
 with respect to the vertical axis and a comparison with the isothermal
 model.

\section{Model and Method\label{sec:2}}

\citet{tomida2010a} (hereafter Paper I) 
 have calculated the evolution of a rotating magnetized 
 cloud using radiation magnetohydrodynamical (RMHD) simulations.
In Paper I, we assumed a cloud in a hydrostatic balance  
 with a central density of $\rho_c=1.0\times 10^{-19}\,{\rm g\,cm^{-3}}$ or
 $n_0=1.6 \times 10^4\, {\rm H_2\,cm^{-3}}$,
 a uniform rotation rate of
 $\omega=0.1/t_{\rm ff}=1.5\times 10^{-14}\,{\rm rad~s^{-1}}$,
 and relatively weak  uniform magnetic field of 
 $B_z=1.1\,\mu\, {\rm G}$.
The angular momentum vector and the magnetic field direction are
 assumed to both be in the $z$-direction.

In Paper I, we employed the nested grid technique,
 in which the global structure is covered with a coarse grid
 while the central spatially fine structure 
 with high density is calculated with a fine grid.  
We numbered the grid level from $L=0$ (the coarsest)
 to $L=17$ (the finest).
The structures of the respective grid levels are self-similar.
All the levels of the grids are co-centered, i.e.,
 the centers of all the grids are placed at the same point
 (see Fig.~\ref{fig:nest-nonLTE}).
The number of grid cells of each level was chosen to be $64^3$
 and the size of cell $\Delta_L$ of the $L$-th level
 is given by that of the $L=0$ level $\Delta_0$ as $\Delta_L=2^{-L}\Delta_0$.

\begin{figure}[h]
\centering
(a)\hspace*{70mm}(b)\\
\includegraphics[width=80mm]{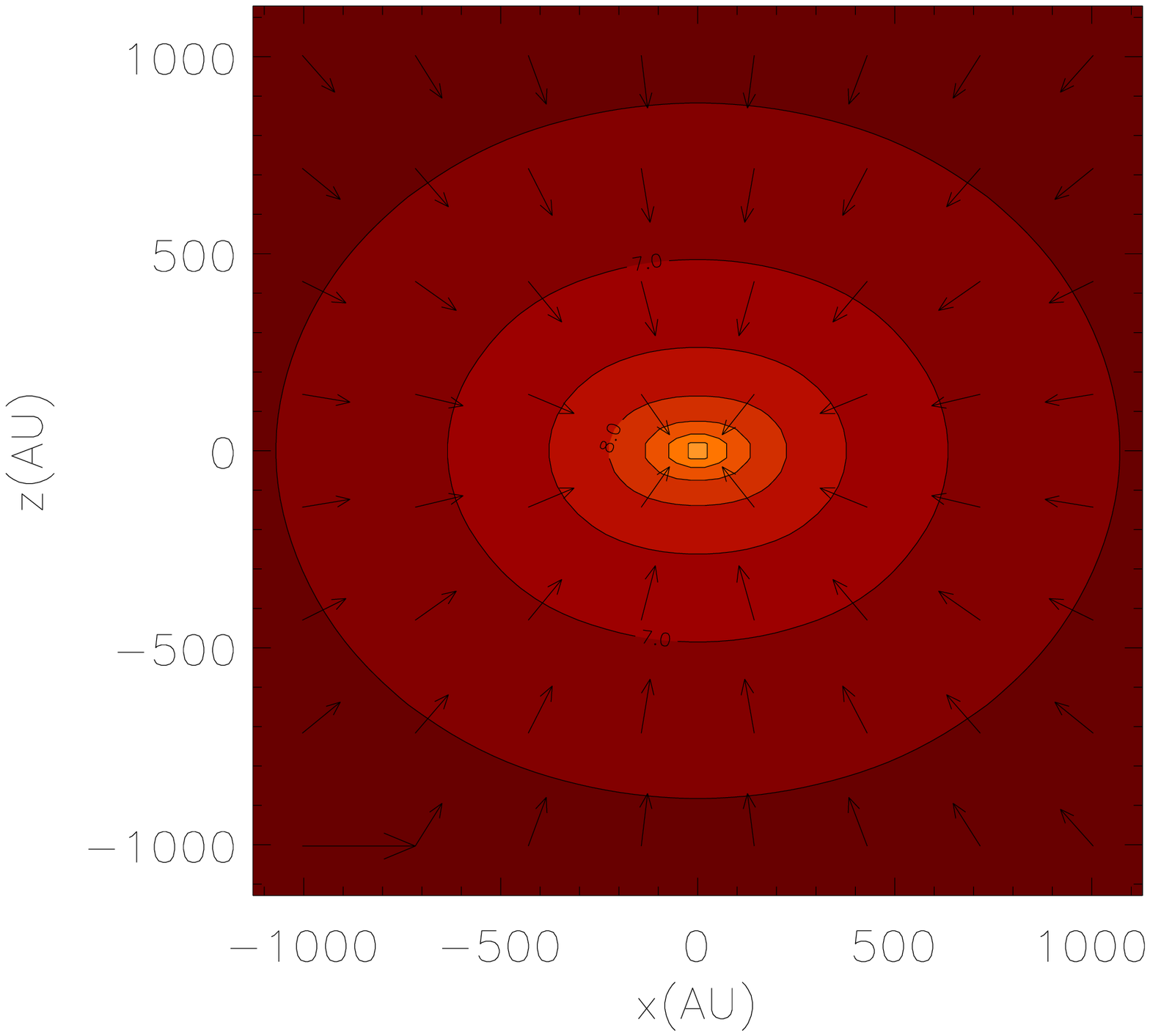}
\includegraphics[width=80mm]{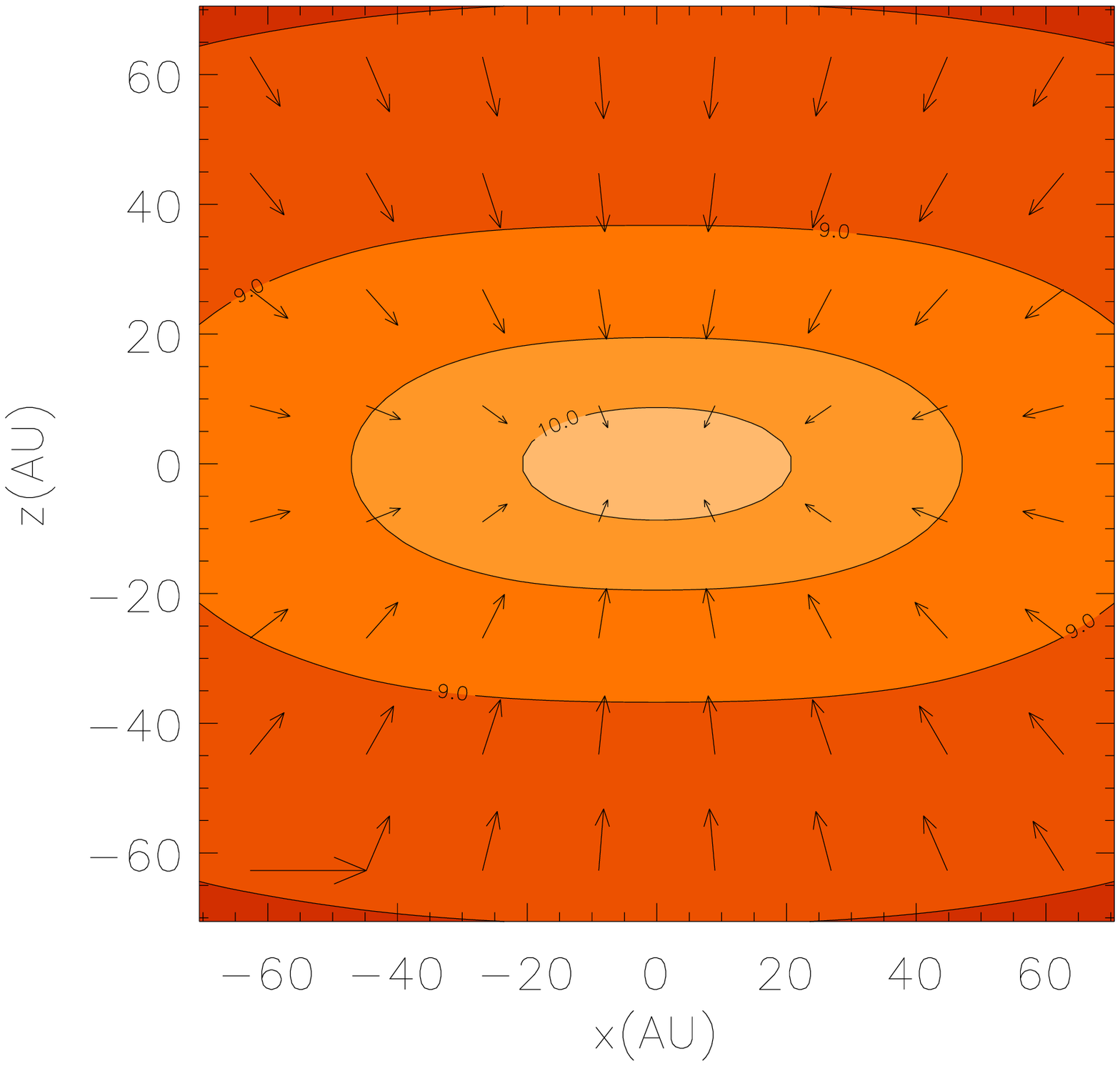}\\
(c)\hspace*{70mm}(d)\\
\includegraphics[width=80mm]{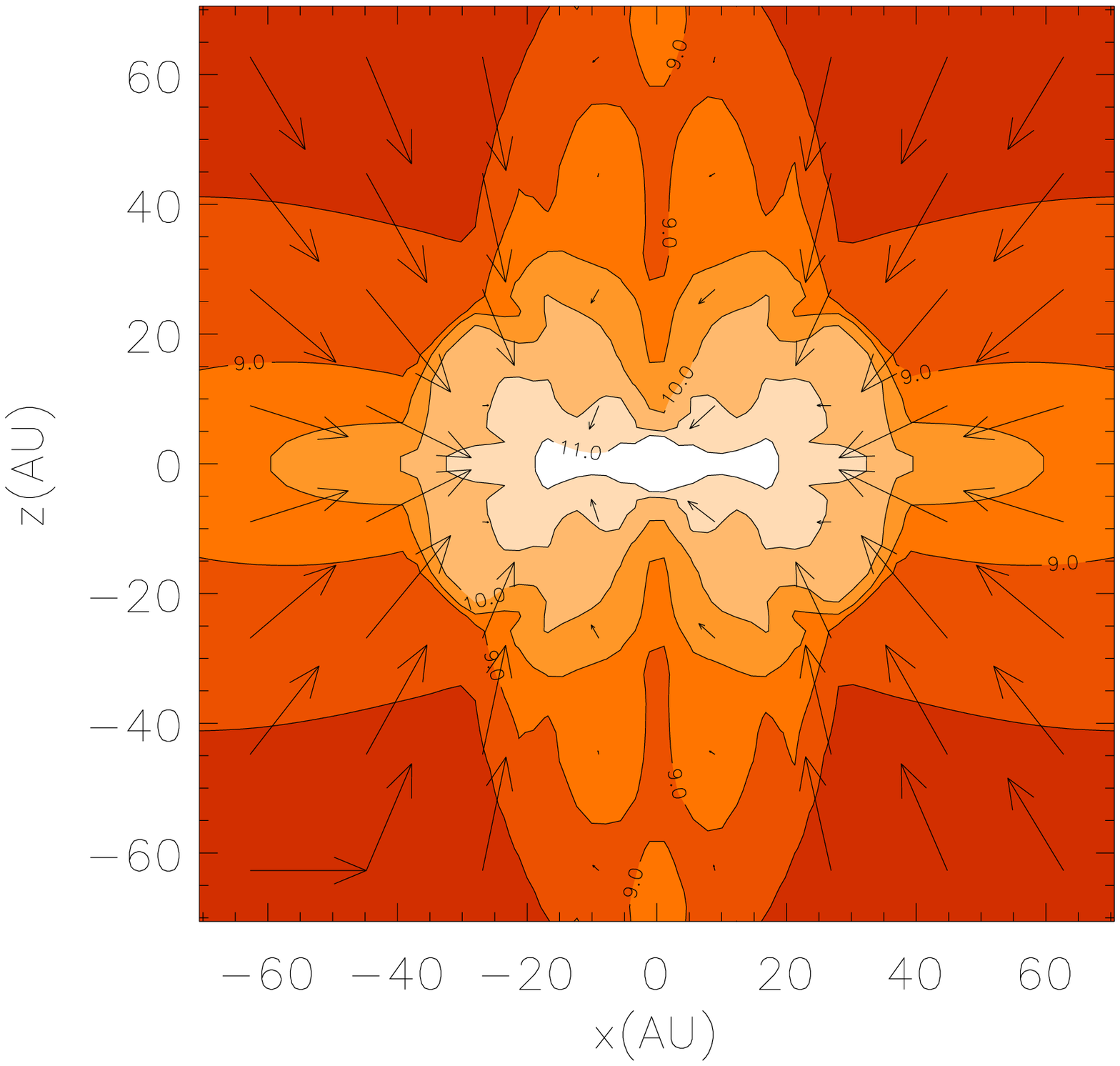}
\includegraphics[width=80mm]{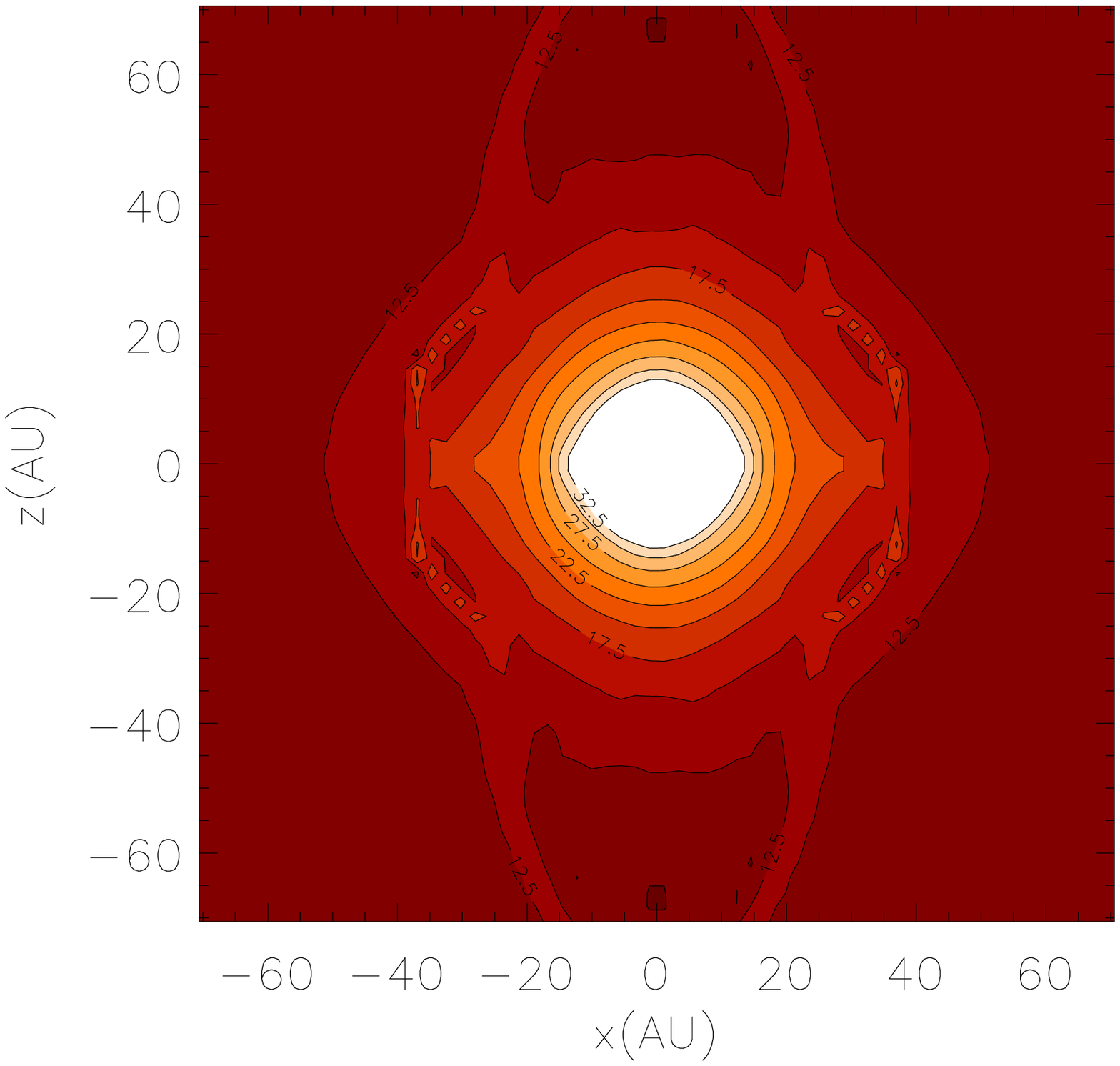}
\caption{\small\label{fig:physical}
Density structure of prestellar isothermal collapse phase
 at 2200-AU ({\it a}: Level 5) and 140-AU ({\it b}: Level 9) scales.
The contour levels are H$_2$ number density $n=10^{6.5}{\rm cm}^{-3}$, $10^{7}{\rm cm}^{-3}$, $10^{7.5}{\rm cm}^{-3},\ \ldots $. 
The central density reaches $n_c\sim 10^{10}{\rm H_2\,cm}^{-3}$ at this time.
In this phase, gas is essentially isothermal with $\simeq 10{\rm K}$,
 which is not shown in this figure.
In ({\it c}) we show the density distribution after the first core formation.
The kinetic temperature is plotted in ({\it d}) with contour levels
 $T=12.5 {\rm K}$, $15 {\rm K}$, $17.5 {\rm K},\ \ldots$.
The size of panels ({\it c}) and ({\it d}) is 140 AU $\times$ 140 AU 
 (Level 9).
The velocity field is shown by arrows.
We plot a velocity vector of $(v_x,v_z)=(1{\rm km\,s^{-1}},0)$ in the lower-left corner for comparison.}
\end{figure}

Before the second collapse phase $n \lesssim 10^{15}\,{\rm H_2\,cm^{-3}}$,
 the main coolant of the molecular gas
 is the dust thermal emission, rather than the gaseous line cooling emissions.
RMHD simulations of Paper I included the radiation transfer of 
 the thermal emissions using the gray flux-limited diffusion (FLD)
 approximation. 
We apply here the radiation transfer calculation of gaseous line emissions
 to the snapshot data of density, kinetic temperature, and velocity distributions
 obtained in Paper I.
Thus, we employ a so-called post-process to calculate the molecular
 line emissions.     
Figure \ref{fig:physical} shows the density ({\it a})--({\it c})
 and temperature ({\it d}) distributions obtained in Paper I.
The ages of these two stages are   
 $t=3.852\times 10^5 {\rm yr}$ ({\it a} and {\it b}:
 just before the first core formation)
 and $t=3.858\times 10^5 {\rm yr}$ ({\it c} and {\it d}: 
$\tau=6.45\times 10^2 {\rm yr}$ after the first core formation).
That is, Figures \ref{fig:physical}({\it a}) and ({\it b})
 represent the prestellar isothermal collapse phase, 
 while  Figures \ref{fig:physical}({\it c}) and ({\it d})
 correspond to the protostellar first core phase.
At the stage of Figure \ref{fig:physical}({\it c}) and ({\it d}),
 the first core has grown to the mass of $\sim 0.04 M_\odot$.
From the first core, an outflow with $1-2\ {\rm km\,s^{-1}}$ is ejected,
 of which the velocity is similar to the Kepler speed of the launching point
 \citep{kudoh1997}. 

\subsection{Non-LTE Radiation Transfer}
We calculate the level population of the rotation transitions of 
 a number of abundant molecules in molecular clouds
and solve the non-LTE radiative transfer problem for the rotation 
 transitions of the molecules.
Denoting the number density of the $J$ level (energy level $E(J)$) as $n_J$, 
 we can write the balance equation as
\begin{equation}
n_J\sum_{J'\ne J}R_{JJ'}=\sum_{J'\ne J} n_{J'}R_{J'J}
\hspace*{5mm} (J=0,1,\cdots,J_{\rm max}),
\label{eq:balance}
\end{equation}
where $R_{JJ'}$ represents the transition probability from $J$ to $J'$
as
\begin{equation}
 R_{JJ'}
\left\{
\begin{array}{ll}
=A_{JJ'}+B_{JJ'}{\cal J}_{\nu{JJ'}}+nC_{JJ'}&{\rm for}\ J>J',\\
=B_{JJ'}{\cal J}_{\nu JJ'}+nC_{JJ'}&{\rm for}\  J<J',
\end{array}
\right.
\label{eq:rate}
\end{equation}
where $A_{JJ'}$ and $B_{JJ'}$ represent Einstein's coefficients,
the former being the coefficient for spontaneous emission
 and the latter the coefficient for absorption ($J<J'$) and
 induced emission ($J>J'$).
$C_{JJ'}$ is the collisional transition rate from $J$ to $J'$
for collisions with H$_2$ molecules whose density is denoted by $n$. 
The average intensity of radiation with a frequency of
 $\nu=[E(J')-E(J)]/h$ is written as ${\cal J}_{\nu JJ'}$, where $h$ is the
 Plank constant.  
We take into account the energy levels from $J=0$ to $J=J_{\rm max}$.
Although $J_{\rm max}$ should be taken as  large as possible
 for completeness, in our simulations of cloud collapse and outflow
 we take $J_{\rm max}=10$.

To obtain the average intensity ${\cal J}_\nu$, we have to solve
 the radiation transfer equation for the specific intensity $I_\nu$: 
\begin{equation}
\frac{dI_\nu}{ds}=-\kappa_\nu I_\nu +\epsilon_\nu,
\label{eq:radiation transfer}
\end{equation}
where $\kappa_\nu$ and $\epsilon_\nu$ represent
 the monochromatic volume absorption coefficient and monochromatic volume emissivity.
Total absorption coefficient and emissivity for the transition between $J$ and $J'$
 are related to the level populations $n_J$ as
\begin{equation}
\kappa^{\rm (tot)}_{JJ'}=\frac{h\nu}{4\pi\Delta\nu}
 \left(n_{J'}B_{J'J}-n_JB_{JJ'}\right) \hspace*{1cm} (J > J'),
\end{equation}
\begin{equation}
\epsilon^{\rm (tot)}_{JJ'}= \frac{h\nu}{4\pi\Delta\nu}n_JA_{JJ'}
 \hspace*{1cm} (J > J').
\end{equation}
Here the Doppler width of the line due to the thermal and turbulent motion
 of molecules is expressed as
\begin{equation}
\Delta \nu=\frac{\nu_{JJ'}}{c}\sigma
\label{eqn:Delta_nu}
\end{equation}
using the microturbulence parameter $\sigma$.
Assuming a line profile function $\phi$ of 
\begin{equation}
\phi_{JJ'}\left(\frac{\nu-\nu_{0JJ'}}{\Delta \nu}\right)
=\frac{1}{\pi^{1/2}}
\exp\left[-\left(\frac{\nu-\nu_{0JJ'}}{\Delta \nu}\right)^2\right],
\end{equation}
where $\nu_{0JJ'}=[E(J)-E(J')]/h$,
 the absorption coefficient and emissivity for $\nu$ are written 
 using the total absorption coefficient and emissivity for specific
 transition $J\rightarrow J'$ as  
\begin{eqnarray}
\kappa_{\nu JJ'}&=&\phi_{JJ'}(\nu)\kappa^{\rm (tot)}_{JJ'}\\
\epsilon_{\nu JJ'}&=&\phi_{JJ'}(\nu)\epsilon^{\rm (tot)}_{JJ'}.
\end{eqnarray}
Thus, solving  equations (\ref{eq:balance}) and (\ref{eq:radiation transfer})
 is a nonlocal problem, or in other words, 
 physical states of the molecules in different places are coupled with
 each other by the radiation.

We solve this non-LTE radiative transfer problem
 by a Monte Carlo method similar to
 \citet{vandertak2000,hoge2000}.
The formalism is as follows:
The region is divided into uniform Cartesian cells of
 $N_{\rm grid}^3$.
In randomly chosen directions, rays are ejected from the center
 of each cell.
Along the ray, the transfer equation (\ref{eq:radiation transfer})
 is solved from the outer boundary to the cell center.
Before we finish integrating equation (\ref{eq:radiation transfer})
 for all rays from each cell,
 the level population, $n_J^{({\rm old})}$,
 and thus the emissivity and the absorption coefficients are both fixed.     
The number of rays per cell is chosen to be $N_{\rm Ray}=100$ in this
 calculation.
After completing the integration,
 we can obtain the average intensity ${\cal J}$ for each cell by
\begin{equation}
{\cal J}_{\nu JJ'}=\oint I_{\nu JJ'} d\Omega = \frac{1}{N_{\rm Ray}}\sum_{N=1}^{N_{\rm Ray}} I_{\nu JJ'}. 
\end{equation}
This gives new level populations, $n_J^{{\rm (new)}}$,
 consistent with the average intensity ${\cal J}$
 for use in equation (\ref{eq:balance}).
This cycle is repeated until the assumed level populations $n_J^{\rm (old)}$
 and the new populations $n_J^{\rm (new)}$ converge.
We assume the convergence criterion that a relative deviation of the populations 
 is smaller than $\epsilon_n=10^{-8}$,
$|n_J^{\rm (old)}-n_J^{\rm (new)}|/n_J^{\rm (old)}< \epsilon_n$.
The procedure is tested by a comparison with a test problem done by
 \citet{juvela1997} which has been used for the visualization of 
 interstellar turbulence driven by supernovae \citep{wada2005,yamada2007}.

Molecular data, such as the Einstein's $A$ and $B$ coefficients and the $C$
 coefficients in equation (\ref{eq:rate}), are 
 taken from the Leiden Atomic and Molecular
 Database (LAMDA)\footnote{
http://www.strw.leidenuniv.nl/$\sim$moldata/}
\citep{shoeier2005}.
The abundance of CS relative to H$_2$ is chosen to be
 $X_{\rm CS}=4\times 10^{-9}$.
Frequencies in laboratory frame of the respective transitions
 are shown in Table \ref{tbl:freq}.
The one-dimensional random velocity composed of the thermal plus turbulent 
 contributions is chosen to be $\sigma=100\,{\rm m\,s^{-1}}$
 of equation (\ref{eqn:Delta_nu})\footnote{
We changed $\sigma$ from $100\,{\rm m\,s^{-1}}$ to $300\,{\rm m\,s^{-1}}$.
Although obtained line-width and line-shape depend on $\sigma$,
 results are qualitatively the same,
 because the global velocity gradient has more significant effects
 on the optical depth than the local line width.
}
. 

\subsection{Radiative Transfer on Nested Grid}
In this subsection we present the method to solve the non-LTE radiative transfer
 on the nested grid hierarchy. 
When the cloud is in a vacuum, 
 the outer boundary condition should be 
 that the inwardly directed intensity is equal to that of
 the Cosmic Microwave Background (CMB; $T_{\rm B}=2.7\rm K$).   
In reality, the interstellar molecular cloud is immersed in
 the interstellar radiation field.
However, since we do not have sufficient information about
 the interstellar radiation field,
 we adopt the vacuum boundary condition in this paper.
In order to reduce the artificial effect of this boundary condition,
 we place the boundary as far as we can from the inner part
 which we are interested in by using the nested grid technique.
As shown in $\S$\ref{sec:2},
 the nested grid technique uses a special grid hierarchy
 as the grid spacing increases with departing from the center.
This enables us to place the outer boundary far from the center
 compared with the uniform grid spacing. 
%
In the nested grid hierarchy, 
 we assume the outermost grid $L=0$
 is in a vacuum
 filled with the CMB.
The outer boundary of the inner nested grids ($L \ge 1$) should be taken 
 from the intensity obtained from the coarser outer grids
 (see Fig.\ref{fig:nest-nonLTE}).
In this case, we calculate the level populations of the molecules
 in each nested grid step-by-step from the coarser level to the inner finer
 level as shown in Appendix.

\begin{figure}[h]
\centering
\includegraphics[width=120mm]{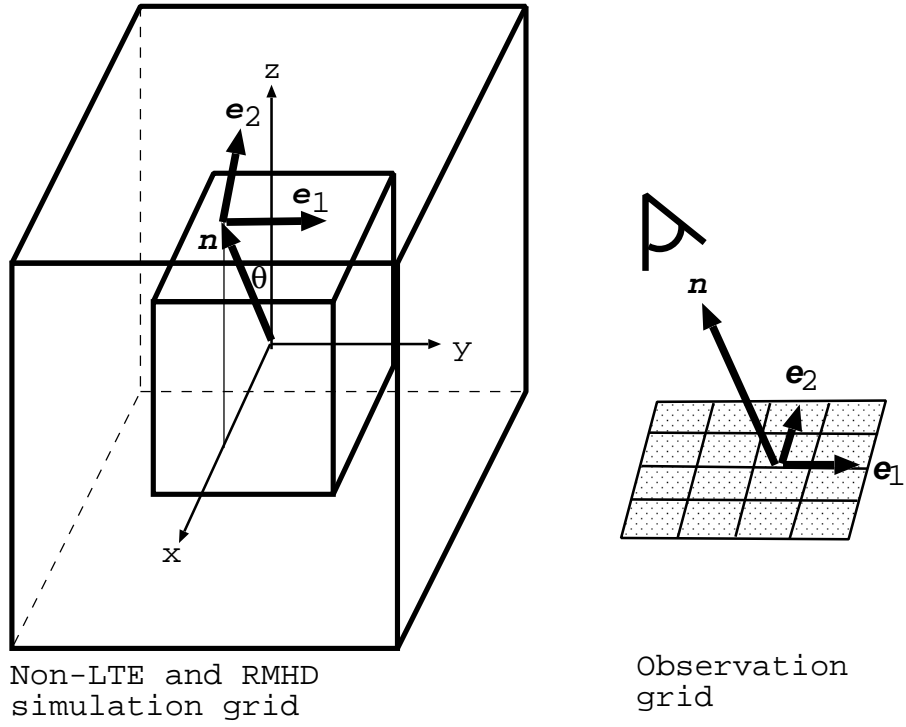}
\caption{\label{fig:grid}
Relationship between the grid
 used in the non-LTE and RMHD simulations (left: nested grid)
 and the observation grid (right).
Observations were made by integrating along the normal vector
 $\boldsymbol{n}$.
That is, the direction of the observation is specified by 
 $\boldsymbol{n}=(\sin\theta, 0, \cos \theta)$. 
The unit vectors which specifies the observation grid are given as 
 $\boldsymbol{e}_1=(0,1,0)$ and
 $\boldsymbol{e}_2=[\boldsymbol{e}_z
  -(\boldsymbol{e}_z\cdot \boldsymbol{n})\boldsymbol{n}]/
|\boldsymbol{e}_z-(\boldsymbol{e}_z\cdot \boldsymbol{n})\boldsymbol{n}|$. 
}
\end{figure}

Also, in the 3D non-LTE simulations, 
 the number of cells in each level of the nested grid is chosen
 to be equal to that of the RMHD simulation, $N_{\rm 1D}^3=64^3$.
The grid sizes of these simulations are chosen to be the same, $\Delta_L$.
The cell size is taken as $\Delta_L=1150\,{\rm AU}/2^L$.
Thus, in the model shown in Figure \ref{fig:physical} ($L=9$)
 the size is equal to $\Delta_9=2.25\,{\rm AU}$.
We use the Cartesian coordinate system $(x,y,z)$
 in calculating non-LTE radiative transfer, which is also used in
 RMHD simulation (Fig. \ref{fig:grid} left) .

In this paper, we consider axisymmetric objects.  
To reproduce the observations, we choose a special geometry
 in which  
 (1) the direction of the line of sight (LOS) is on the $x$--$z$ plane
 and (2) in the figures
 the horizontal direction coincides with the $y$-axis (see Fig. \ref{fig:grid}).
Under this specification, 
 the LOS is expressed as $\boldsymbol{n}=(\sin\theta,0,\cos\theta)$,
 where the direction of the LOS is specified only with the angle measured 
 from the $z$-axis, $\theta$, since the structure is axisymmetric.
If we express the directions of the observational grid as $\boldsymbol{e}_1$
 (horizontal) and $\boldsymbol{e}_2$ (vertical) (that is, any observational
 point is expressed with two integers $i$ and $j$ as
$\boldsymbol{x}=\Delta_L(i \boldsymbol{e}_1
+j \boldsymbol{e}_2)$), the unit vectors are given as 
$\boldsymbol{e}_1=(0,1,0)$ and $\boldsymbol{e}_2=[\boldsymbol{e}_z
  -(\boldsymbol{e}_z\cdot \boldsymbol{n})\boldsymbol{n}]/
|\boldsymbol{e}_z-(\boldsymbol{e}_z\cdot \boldsymbol{n})\boldsymbol{n}|$
 (see Fig. \ref{fig:grid}).
The case for $\theta=0^\circ$ is a pole-on view of the disk
 and that  for $\theta=90^\circ$ is edge-on.

\subsection{Efficiency of Nested Grid}

\begin{figure}
\centering
(a)\hspace*{70mm}(b)\\
\includegraphics[width=80mm]{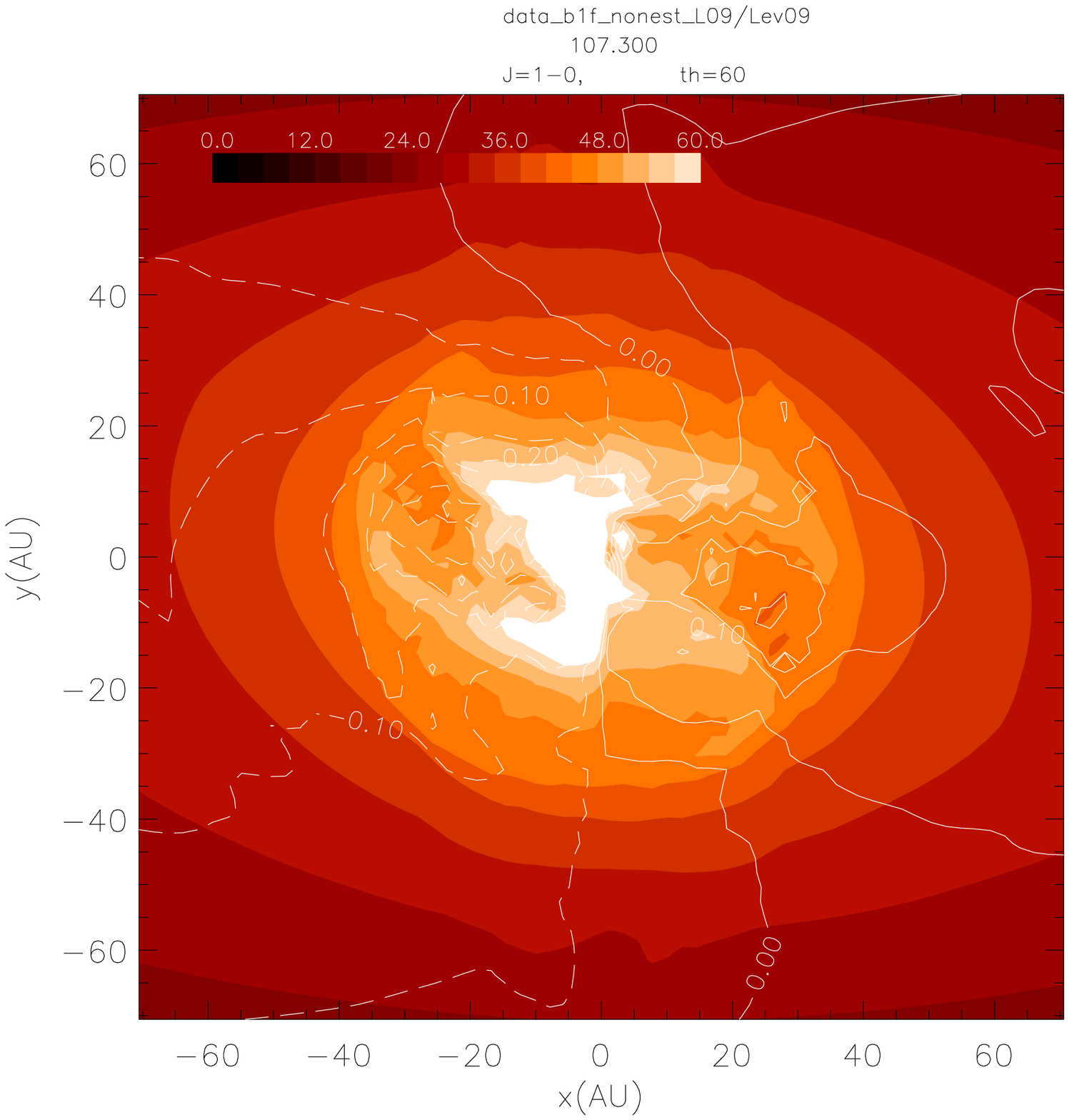}
\includegraphics[width=80mm]{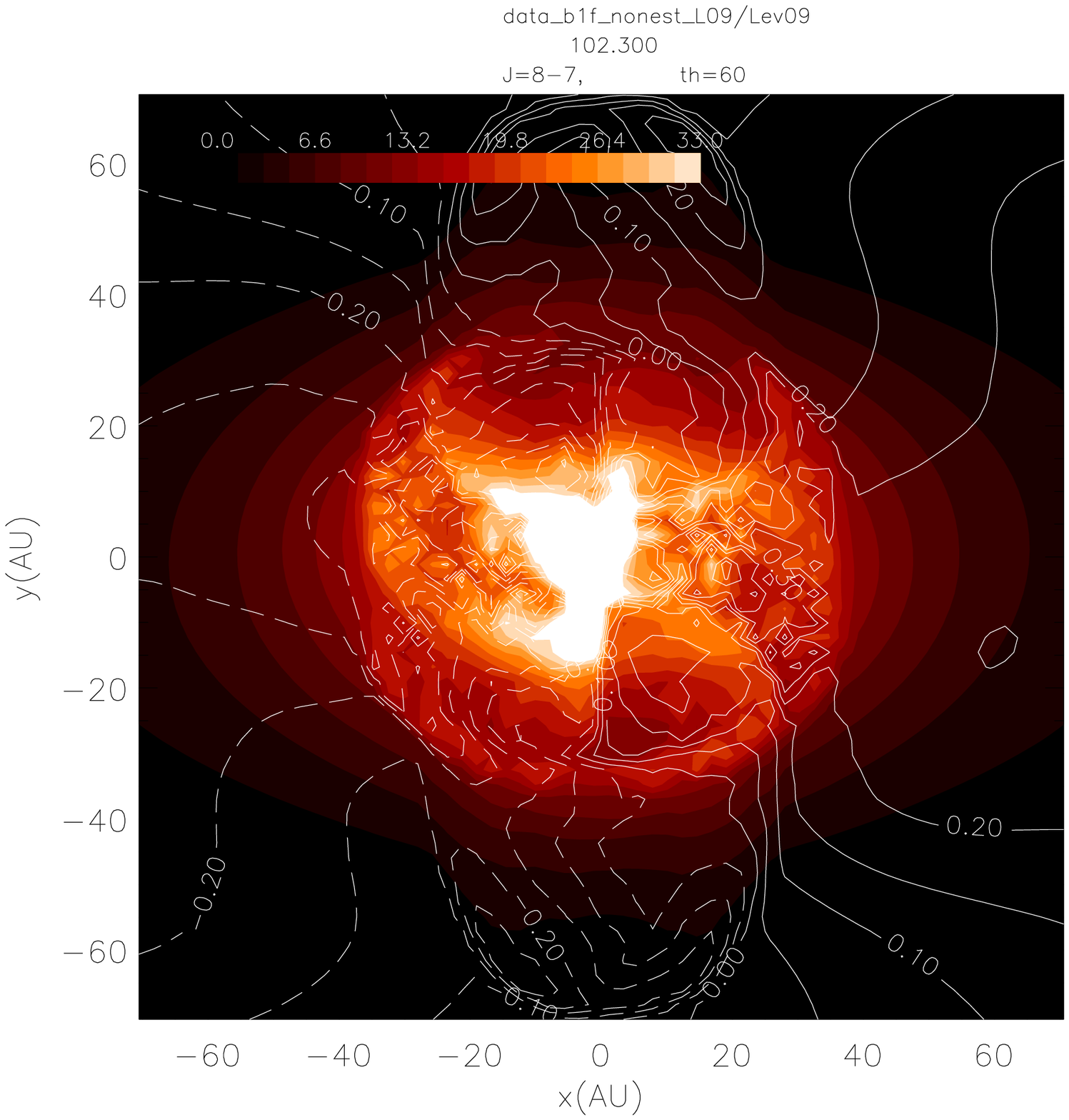}\\
(c)\hspace*{70mm}(d)\\
\includegraphics[width=80mm]{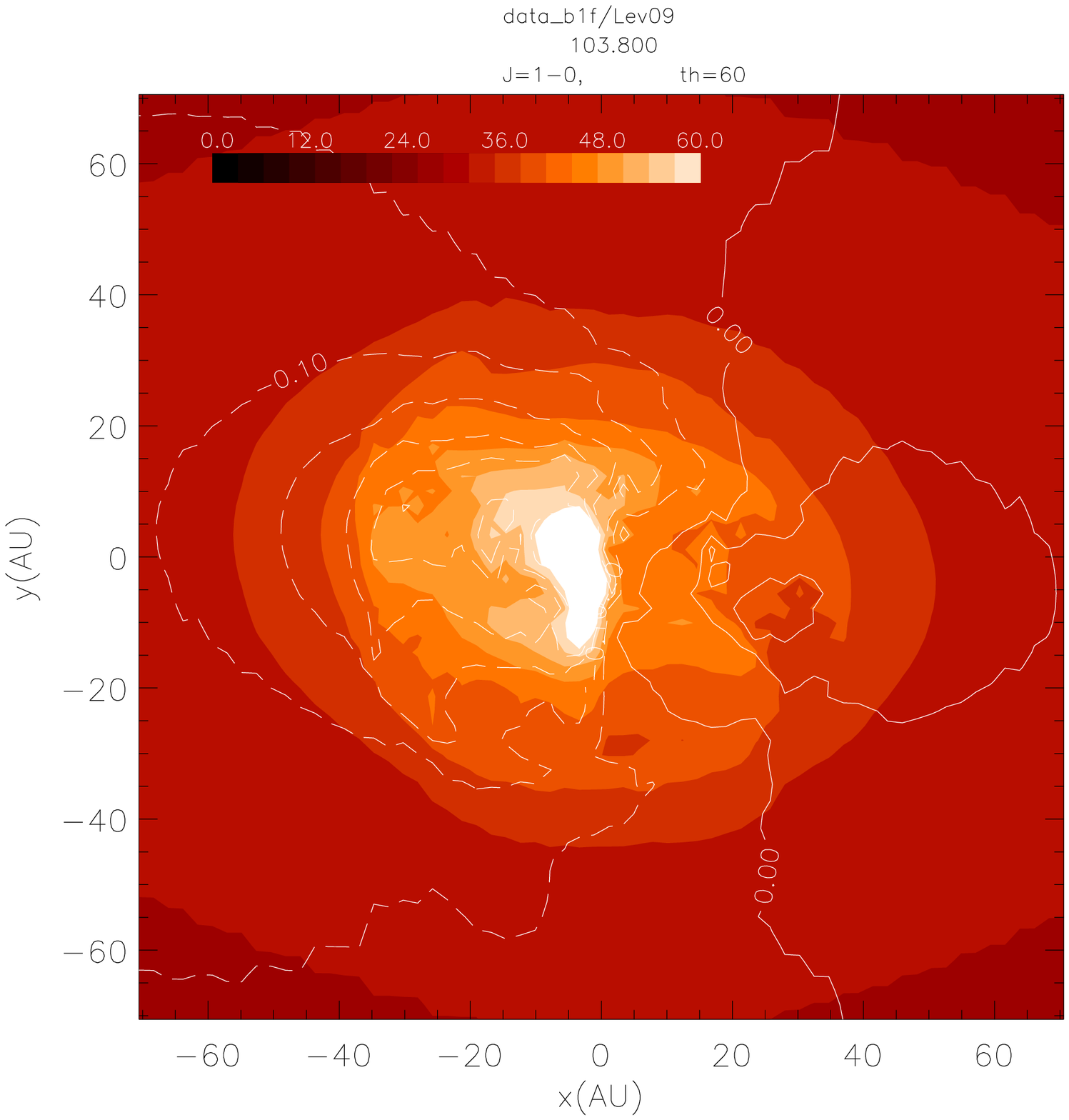}
\includegraphics[width=80mm]{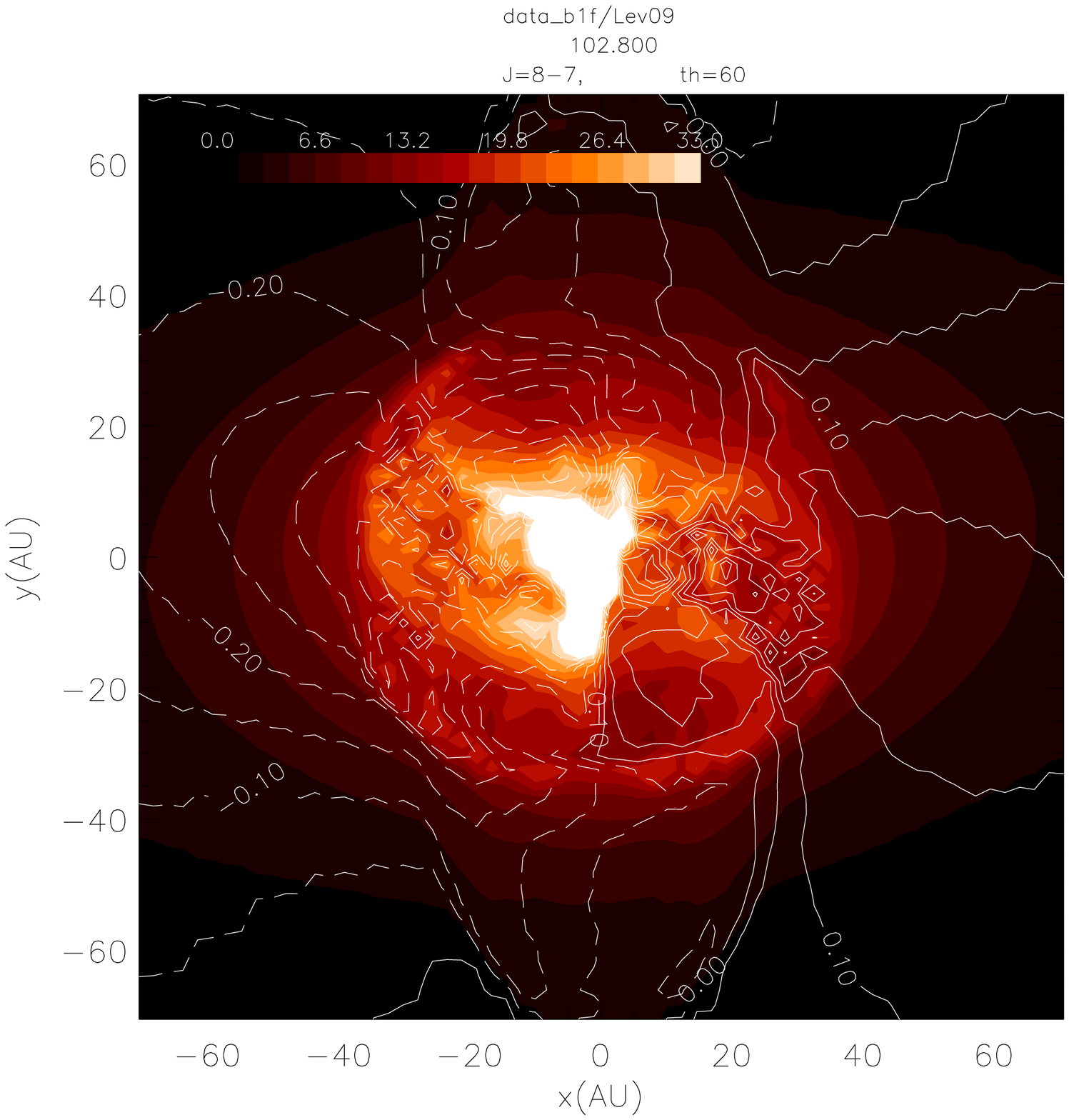}
\caption{\label{fig:meaning_of_nest}
Comparison of expected 
 integrated intensities with and without applying the nested grid method.
The integrated intensities based
 on the non-LTE radiation transfer for the model of Fig.\ref{fig:physical}
 are shown by the false color.
These are the CS $J=1$--$0$ ([a] and [c]) and  $J=8$--$7$ ([b] and [d]) transitions
 viewed from $\theta=60^\circ$.
The contour lines represent the intensity-weighted mean velocity $\langle V\rangle$.
The upper panels ([a] and [b]) indicate the case
 where the extracted $L=9$ level data are integrated
 without applying the nested grid method.
In the lower panels ([c] and [d]) the radiation transfer is calculated
 using all the data for levels $0\le L\le 9$.
Color bars near the upper-left corner
 represent the levels of integrated intensities in 
 ${\rm K\,km\,s^{-1}}$.
The $x$- and $y$-axes represent $\boldsymbol{e}_1$ and
 $\boldsymbol{e}_2$ axes in Fig. 3, respectively.
}
\end{figure}

To clarify the efficiency of the nested grid
 we compare results obtained with and without 
 the nested grid radiation transfer for the model of 
 Figure \ref{fig:physical}.
The integrated intensity distribution (color)
\begin{equation}
\int_{V_{\rm min}}^{V_{\rm max}}T_{\rm B} dV,
\label{eq:integI}
\end{equation}
 and the intensity weighted velocity (contours)
\begin{equation}
\langle V\rangle\equiv \int_{V_{\rm min}}^{V_{\rm max}}VT_{\rm B} dV/\int_{V_{\rm min}}^{V_{\rm max}}T_{\rm B} dV,
\label{eq:Vave}
\end{equation}
where $V$ and $T_{\rm B}$ represent the velocity along the LOS
 and the corresponding brightness temperature,
are shown for CS $J=1$--$0$ (left panels: {\it a} and {\it c})
 and $J=8$--$7$ (right panels: {\it b} and {\it d}) 
 lines in Figure \ref{fig:meaning_of_nest}.
The upper panels ({\it a} and {\it b}) are obtained without the nested
 grid method for $L=9$ density, temperature, and velocity distributions,
 to which we apply the CMB boundary condition
 on the outer boundary of $L=9$.
In contrast, 
 the lower panels ({\it c} and {\it d}) show the result
 with the nested grid technique.
That is, the outer boundary is properly placed on the $L=0$ grid and 
 the intensity at the outer boundary of $L=9$ is obtained consistently.
Comparing ({\it a}) and ({\it c}), we can see that
 the integrated intensity of the $J=1$--$0$ line seems to be very centrally peaked
 in the calculation without the nested grid method.
Both the gradient in the integrated intensity and the
 velocity gradient $\nabla \langle V\rangle$ are overestimated
 in the calculation without the nested grid method.
This shows that not accounting for the envelope outside $L=9$ has a
 strong impact on the intensity distribution for a relatively
 optical thick case of the $J=1$--$0$ line. 
On the other hand, the difference in the integrated intensities 
 between ({\it b}) and ({\it d}) is
 relatively small compared with that between ({\it a}) and ({\it c}).
This is due to the low optical thickness in the envelope for this high
 excitation line $J=8$--$7$,
 since such emissions occur only in the central part of the disk
 and the $J=7$ level is not excited in the outer low-temperature envelope.
This seems to explain the similarity in ({\it b}) and ({\it d}).
The comparison indicates that it is important to use proper boundary 
 conditions in the non-LTE calculations.  
That is, even if
 the excitation temperature $T_{\rm ex}$ is obtained accurately
 (for example, LTE is established) for a grid level $L$ without the
 nested grid technique,
 the optical thickness $\tau$ outside the level $L$ cannot be 
 ignored in estimating the line intensity.
In other words, the intensity of the optically thick lines is 
 qualitatively inaccurate if we ignore the outer grids.  
That is to say, even in such a case, without applying the nested grid method,
 a proper results for the non-LTE radiation transfer is not obtained.

Since we assume $\theta=60^\circ$ here, the obtained structures
 seen in ({\it a}) and ({\it b}) must be affected by the fact that the
 LOSs passing the central part and those passing uppermost and lowermost
 parts have different path lengths in the $L=9$ level. 
However, we place the outer boundary far from the target level
 of the grid, $L=9$, by the nested grid technique.
Thus, this effect of the artificial boundary is removed in
 ({\it c}) and ({\it d}). 
 
\section{Results}

\subsection{Spectral Change between Prestellar and Protostellar Phases}

\begin{figure}
\centering
(a)\hspace*{70mm}(b)\\
\includegraphics[width=80mm]{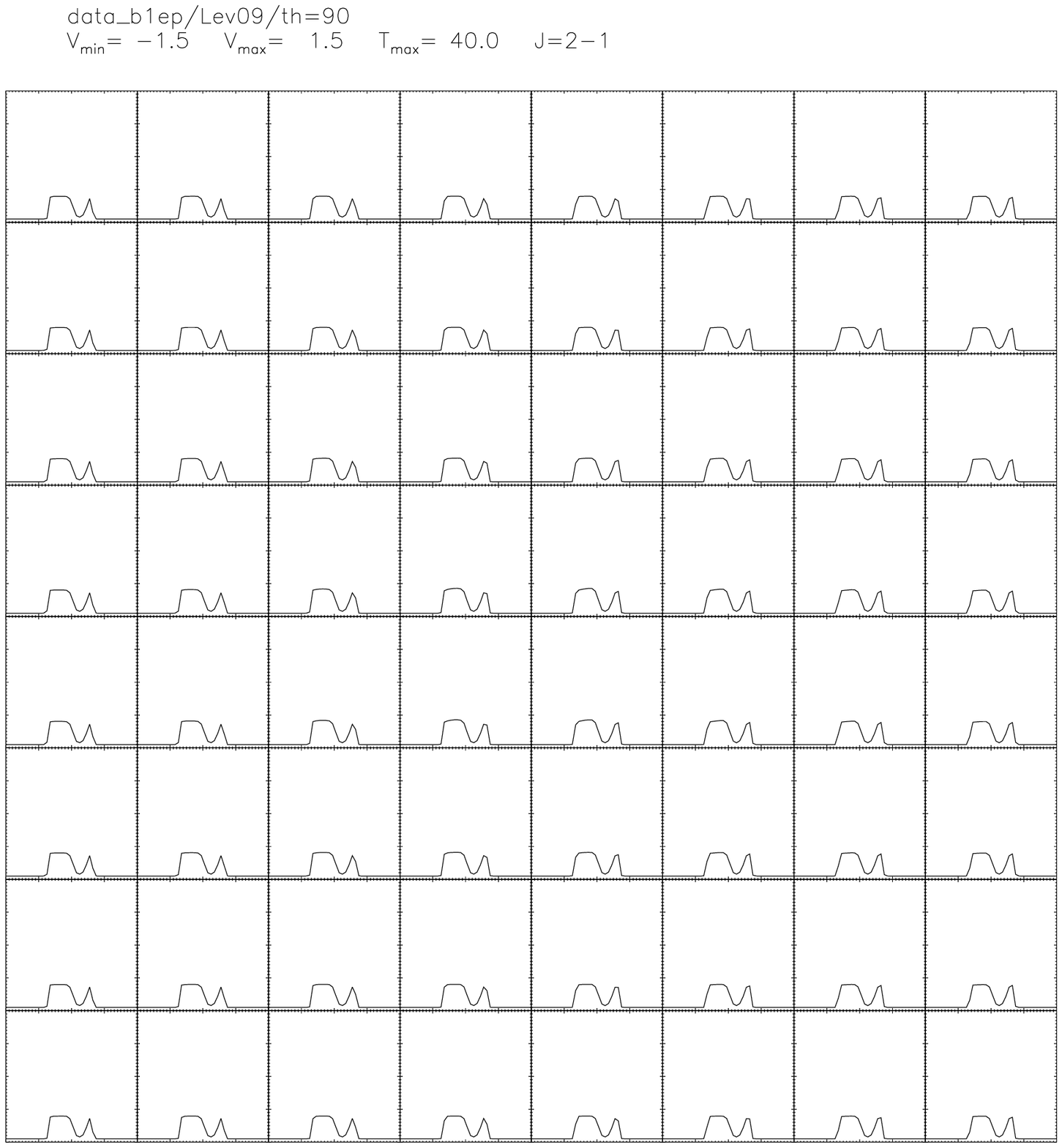}
\includegraphics[width=80mm]{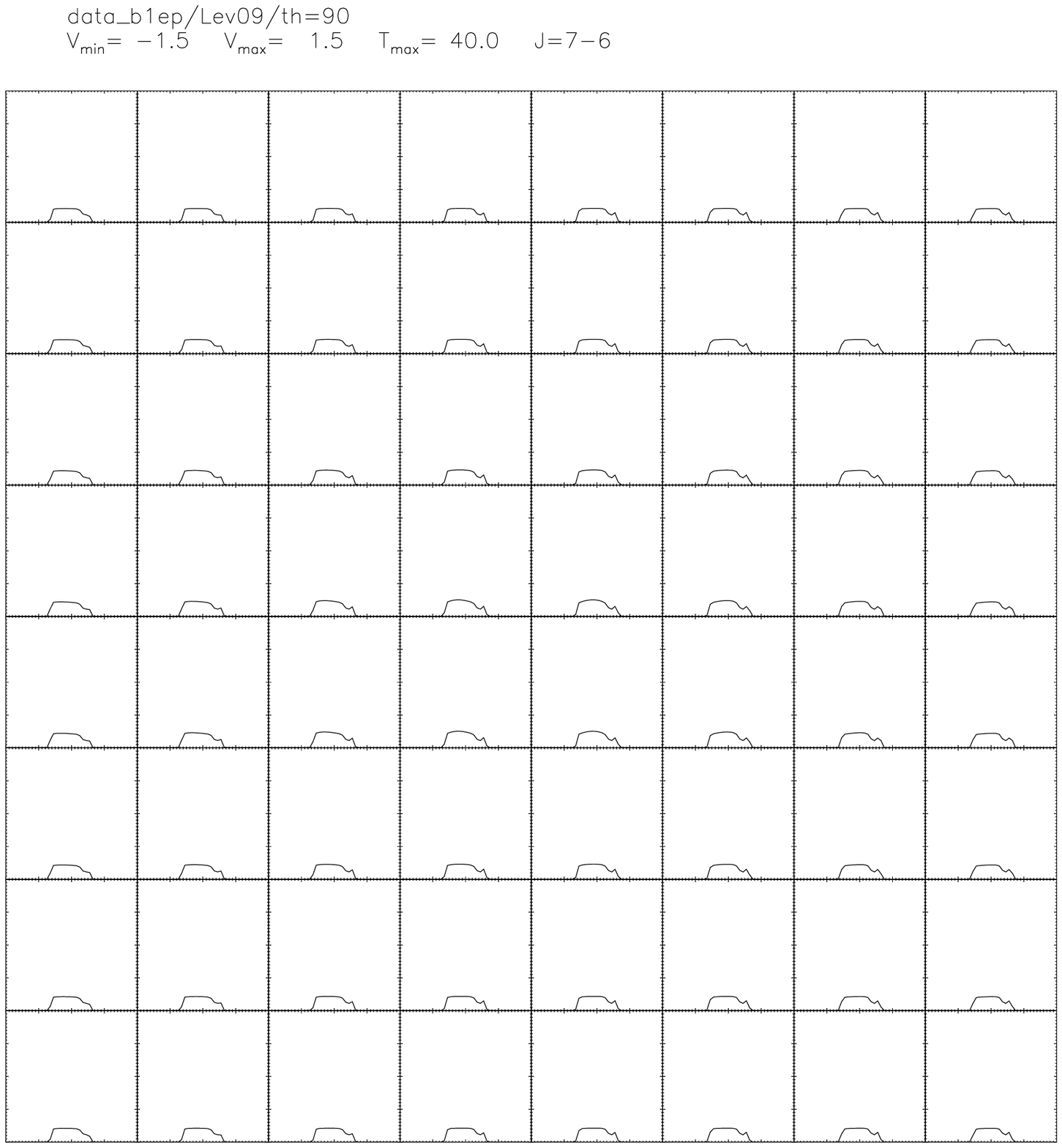}\\
(c)\hspace*{70mm}(d)\\
\includegraphics[width=80mm]{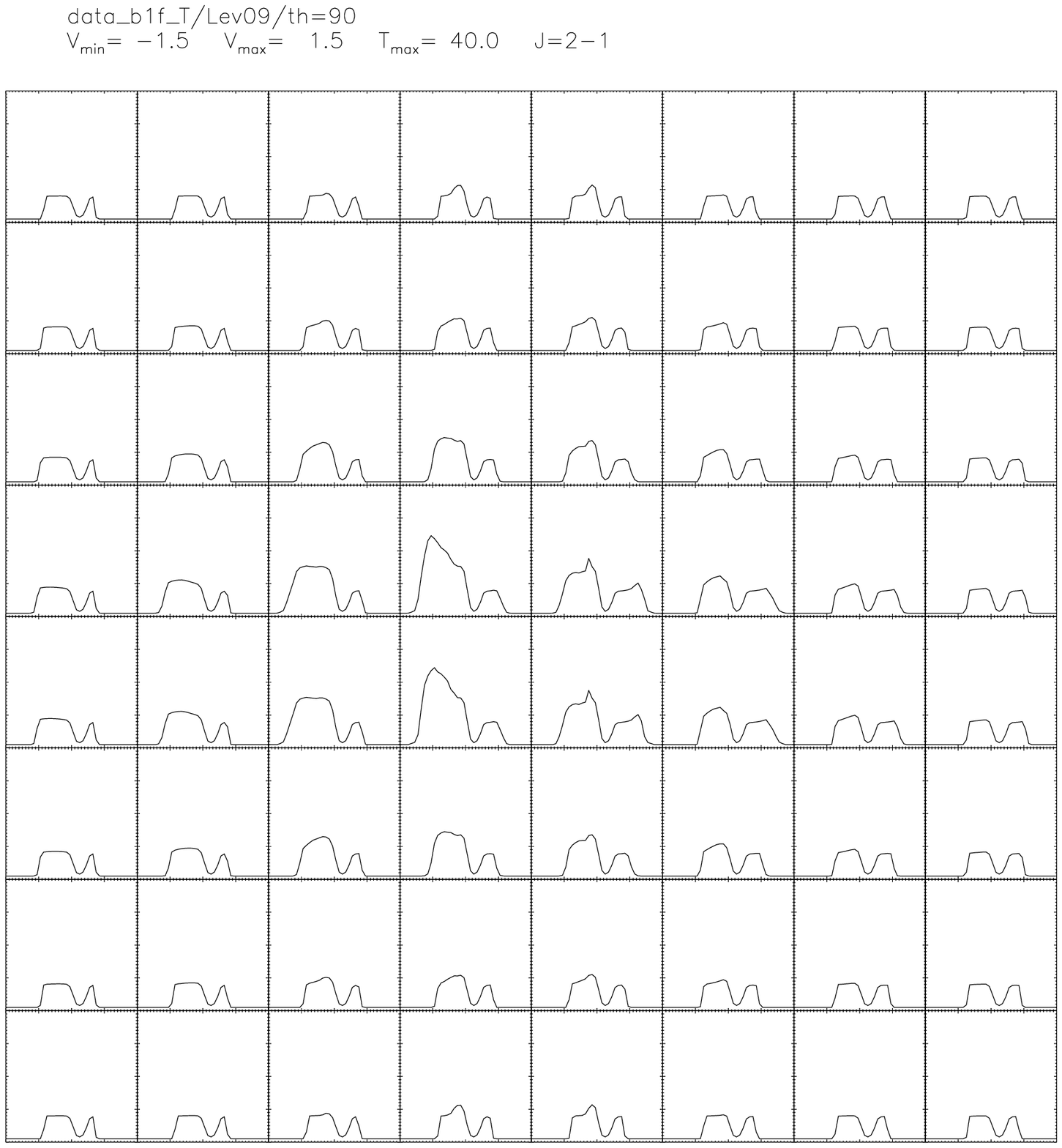}
\includegraphics[width=80mm]{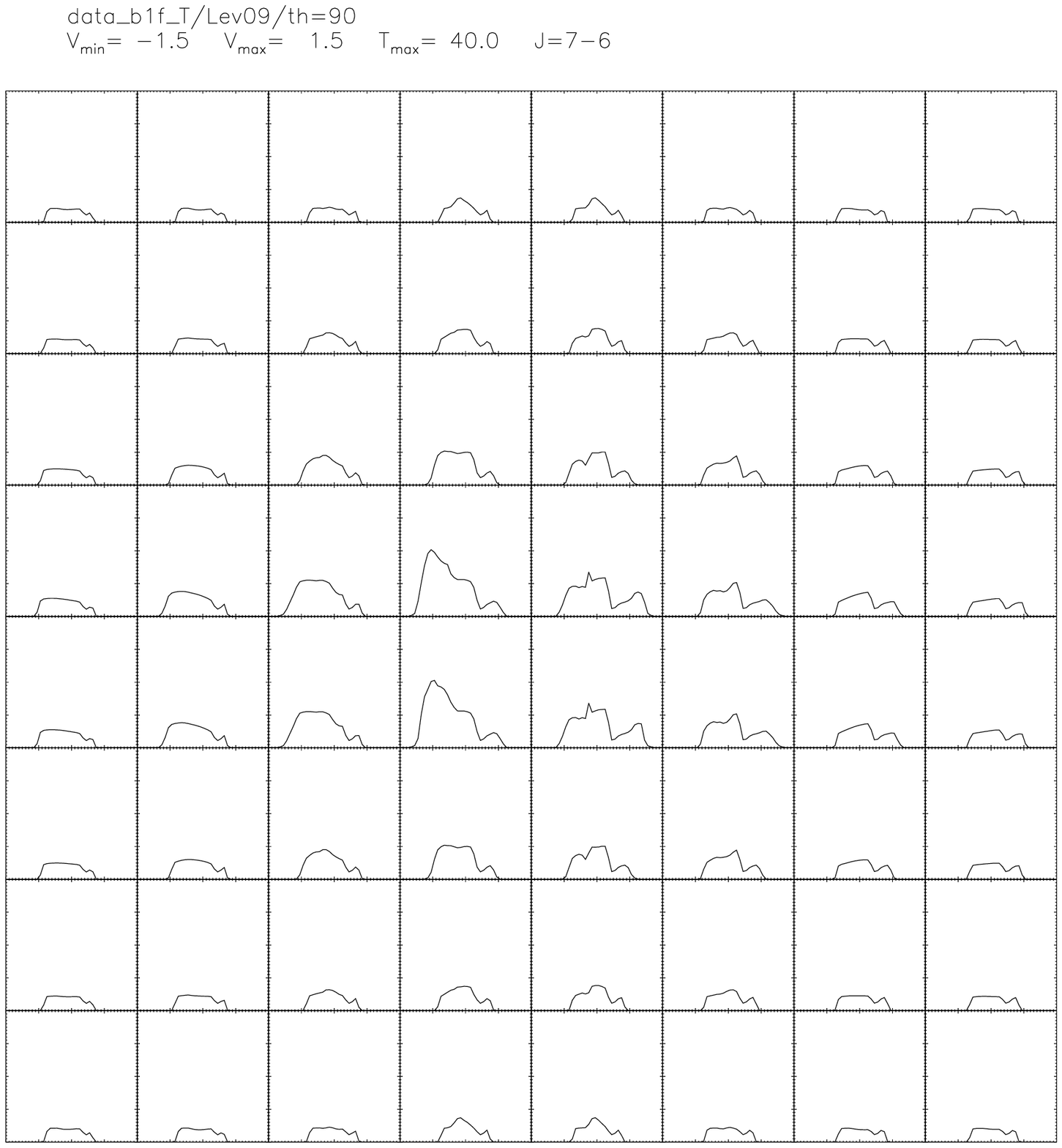}
\caption{\label{fig:spectrum}
Spectra observed from edge-on $\theta=90^\circ$ in the $L=9$ level.
Each spectrum is delineated by the observed position.
The upper panels ({\it a} and {\it b}) represent the model
 before core formation (isothermal phase)
 and the lower panels ({\it c} and {\it d}) represent the model
 after the core formation (first core phase).
The horizontal axis represents the recession velocity 
 ($-1.5{\rm km\,s^{-1}}\le V \le 1.5{\rm km\,s^{-1}}$).
The left panels ({\it a} and {\it c}) are for CS $J=2$--$1$
 and the right ones ({\it b} and {\it d}) are for CS $J=7$--$6$.
The maximum of the vertical axis is taken to be 40 K.
The positions of the spectra are labeled using the $x$ and $y$ distance
 from the center.
That is, the spectrum in the upper-right corner is $(4,4)$ and
 the upper-left one in the central four spectra is labelled $(-1,1)$.}
\end{figure}

In Figure \ref{fig:spectrum},
 we show the CS position-to-position spectra of $J=2$--$1$ and $J=7$--
$6$
 for the prestellar isothermal collapse phase [({\it a}) and ({\it b})]
 and for the protostellar first core phase [({\it c}) and ({\it d})].
The area covered by the spectra is the same as the grid of $L=9$
 shown in Figure \ref{fig:physical}.
These spectra are for the edge-on view of the disk ($\theta=90^\circ$).
The difference between
 the stages before and after the core formation
 is clearly seen in this 140-AU scale spectra.
In the isothermal collapse phase, since the gas is essentially isothermal
 with a temperature of $T_{\rm K}\simeq 10$ K, the brightness temperature
 is equal to $T_{\rm B}\simeq 7$--$8 {\rm K}$.
In the first core phase, the peak brightness temperature
 reaches $T_{\rm B}\sim 30 {\rm K}$ for $J=2$--$1$ and 
$T_{\rm B}\sim 25 {\rm K}$ for $J=7$--$6$.
In this scale, the first core with a high temperature contributes
 to the spectra. 
 (In Figure \ref{fig:physical} ({\it d}),
 gas with $T_{\rm K} \gtrsim 20{\rm K}$ extends to
 $r\lesssim 30{\rm AU}$ from the center.)

The line width also indicates a clear difference.
The line width of the first core phase
 (full-width half-maximum $\sim 2\,{\rm km\,s^{-1}}$) is 
 apparently larger than that of the isothermal collapse phase
 ($\sim 1\,{\rm km\,s^{-1}}$).
Figures  \ref{fig:spectrum} ({\it a}) and ({\it b}) indicate
 a two-peak signature.
The blue- and red-shifted components represent, respectively, 
 the approaching and departing sides of the contracting gas.
Since, in the contracting gas, the near side gas with a positive  recession
 velocity experiences self-absorption due to the foreground contracting
 envelope,
 the signature in which the blue peak (far side) is stronger
 than the red peak (near side)
 indicates that the gas is contracting \citep{zhou1993}. 
This shows that
 the brightness of the CS $J=2$--$1$ line includes contributions from the infalling gas
 not only in the isothermal collapse phase but also in the first core phase.
The difference between ({\it a}) and ({\it c}) shows that the infall
 speed is accelerated from the prestellar to the protostellar phase.
Outflow, which begins to appear in the first core phase,
 is seen as a ``low-velocity'' component in the positions
 $(\pm 1,+4)$, $(\pm 1,-4)$ in panels ({\it c}) and ({\it d}).
This component can be traced near the center between $(+1,-2)$ and $(+1,+2)$.
  
\subsection{Effect of the Viewing Angle}

\begin{figure}
\centering
(a)\hspace*{37mm}(b)\hspace*{37mm}(c)\hspace*{37mm}(d)\\
\includegraphics[width=40mm]{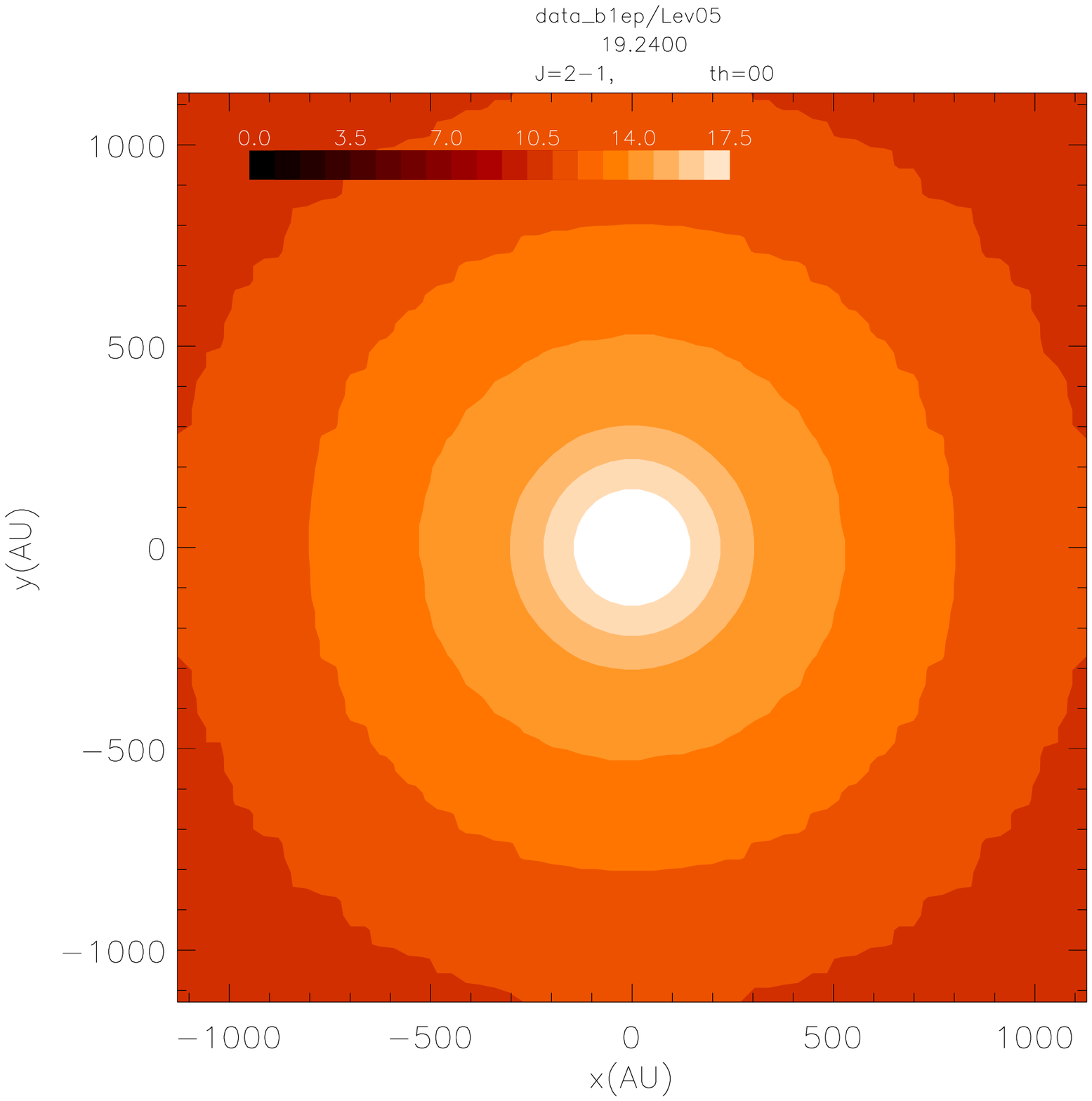}
\includegraphics[width=40mm]{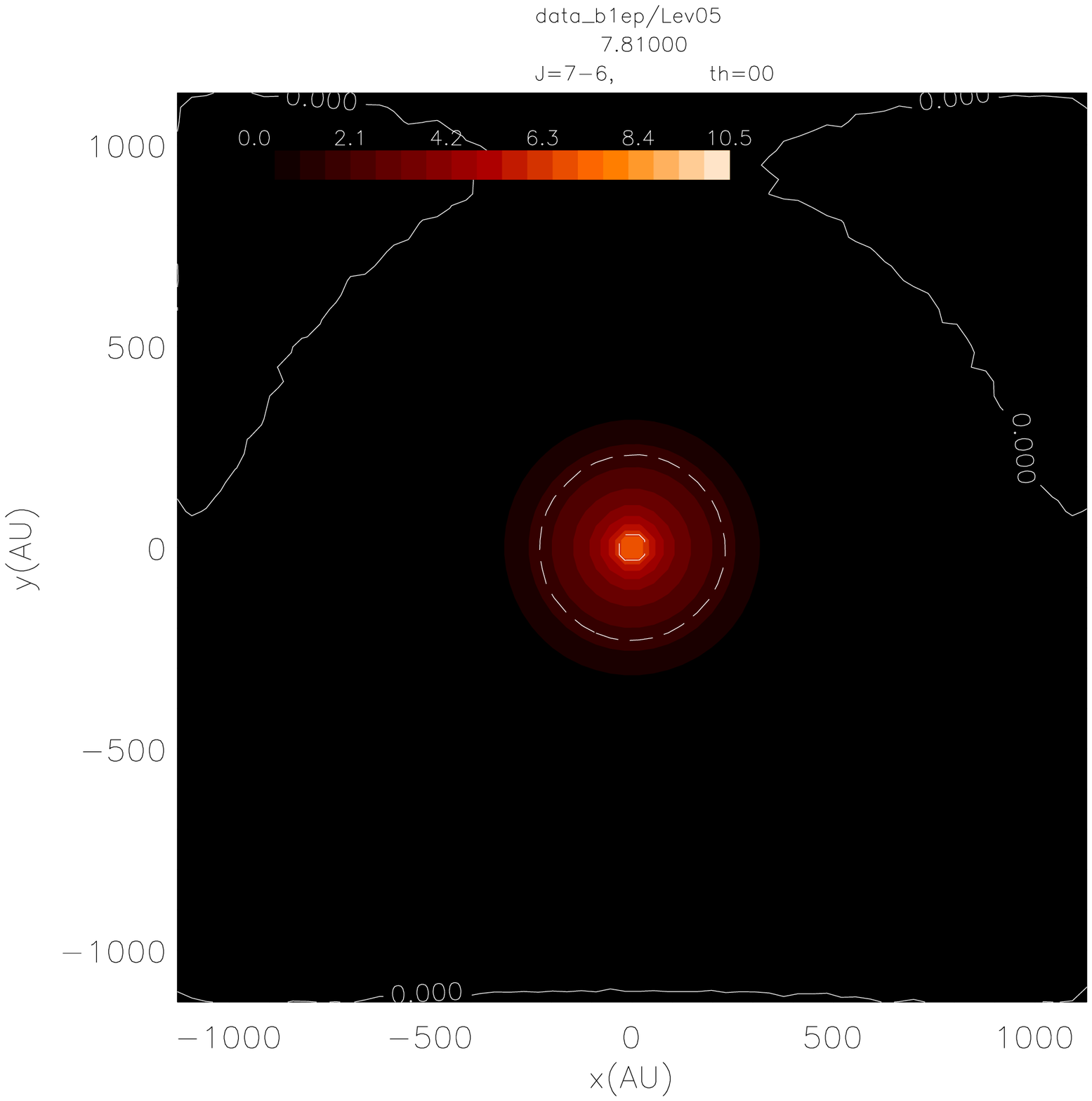}
\includegraphics[width=40mm]{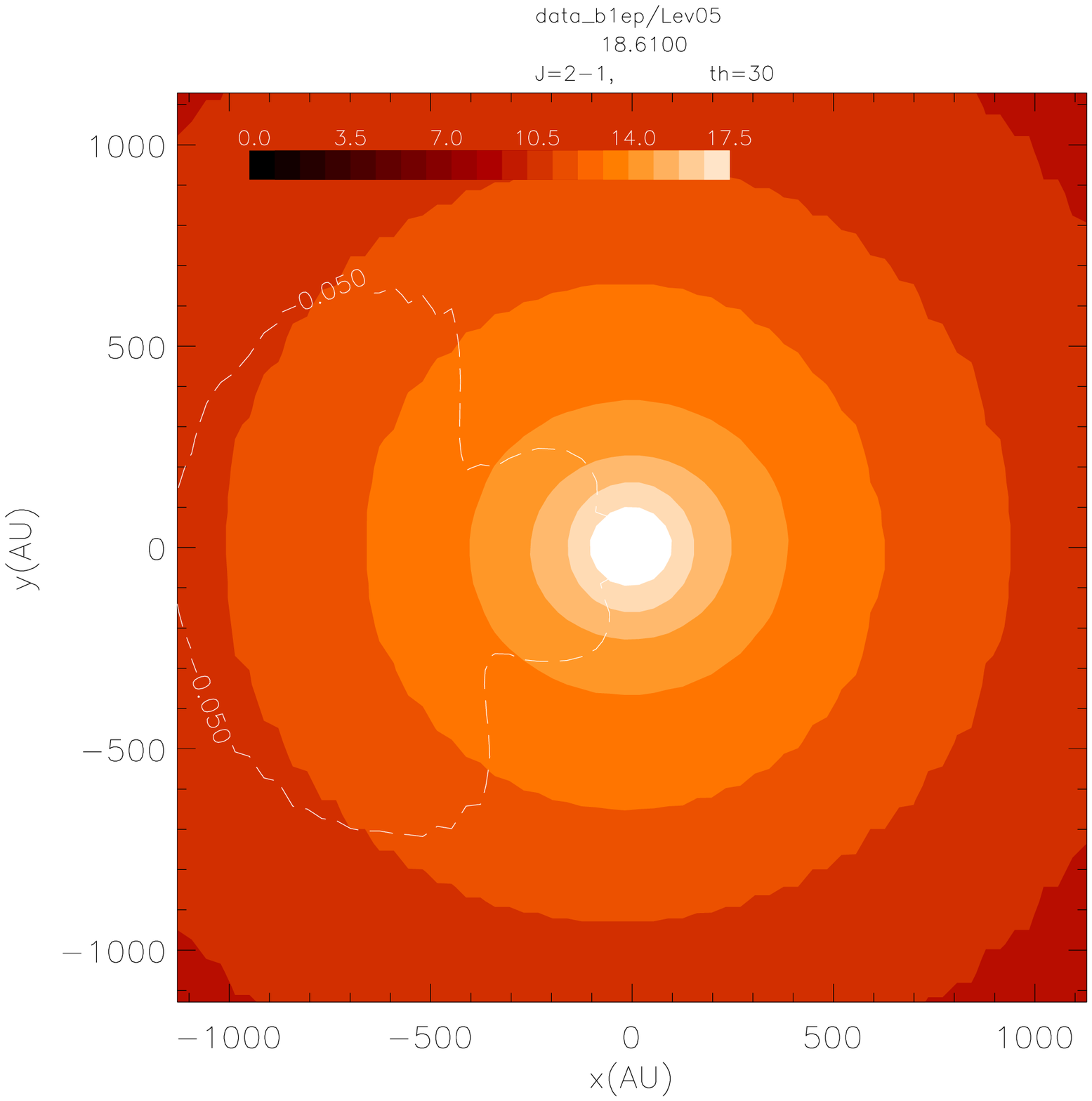}
\includegraphics[width=40mm]{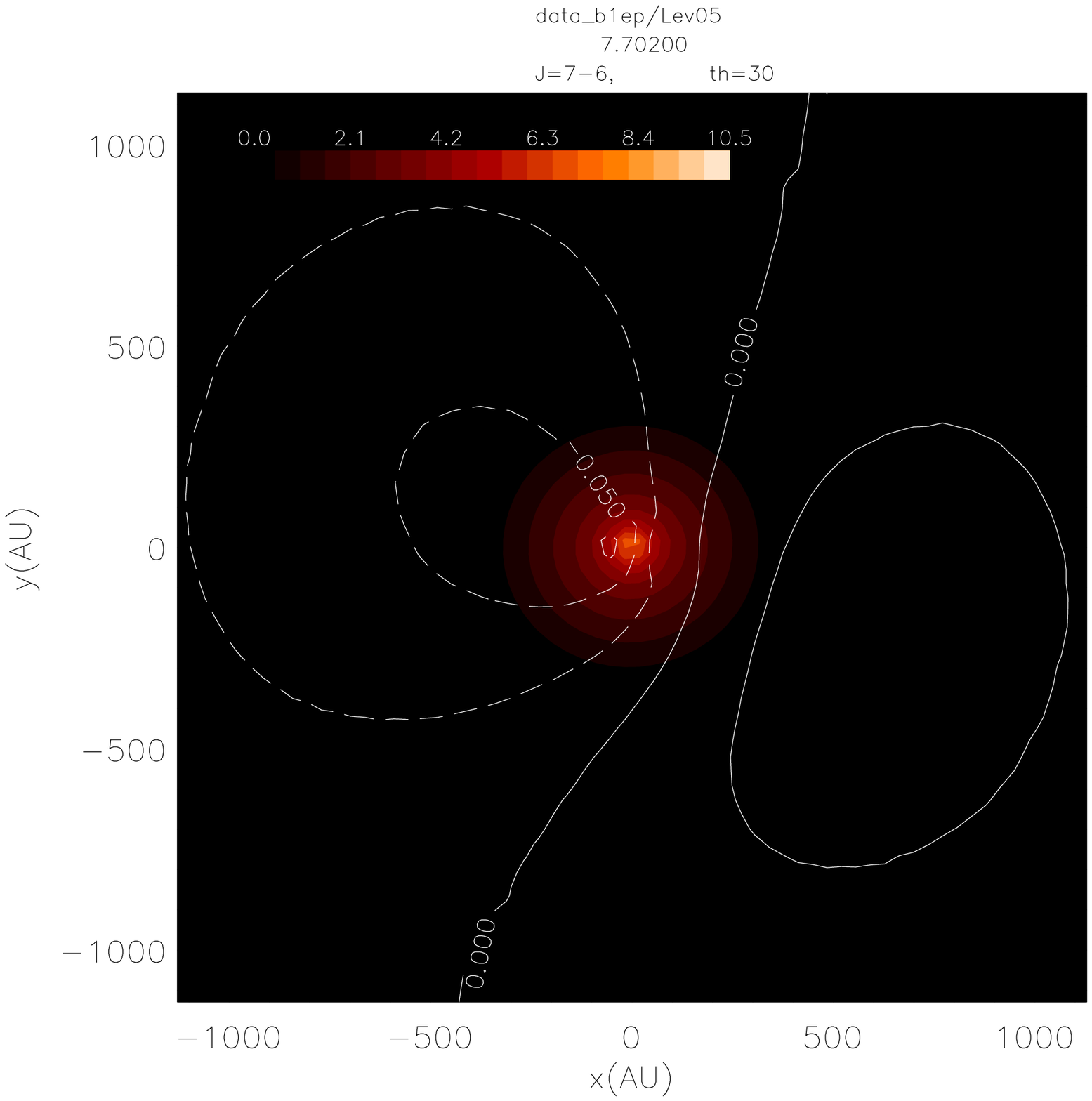}\\
(e)\hspace*{37mm}(f)\hspace*{37mm}(g)\hspace*{37mm}(h)\\
\includegraphics[width=40mm]{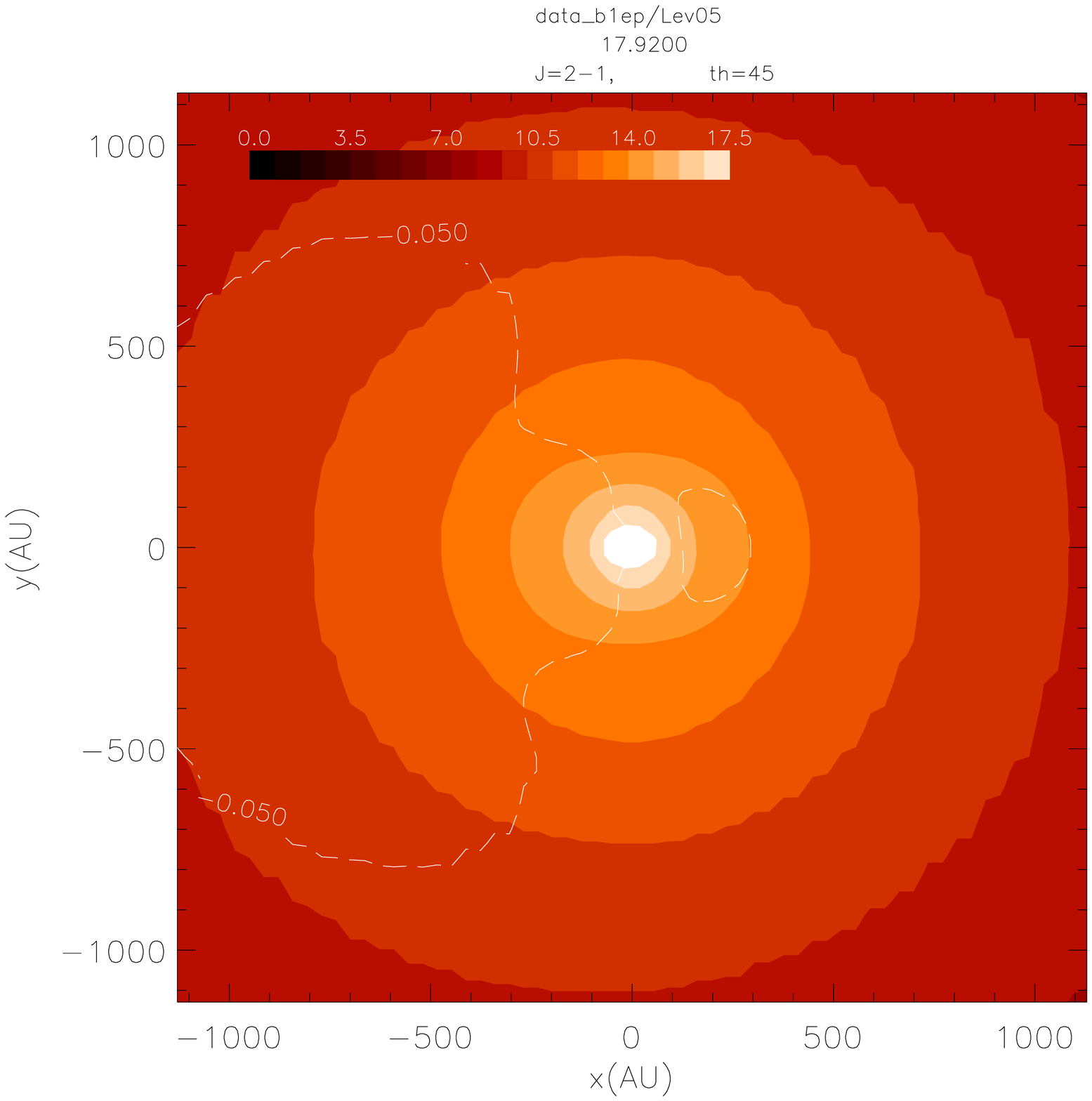}
\includegraphics[width=40mm]{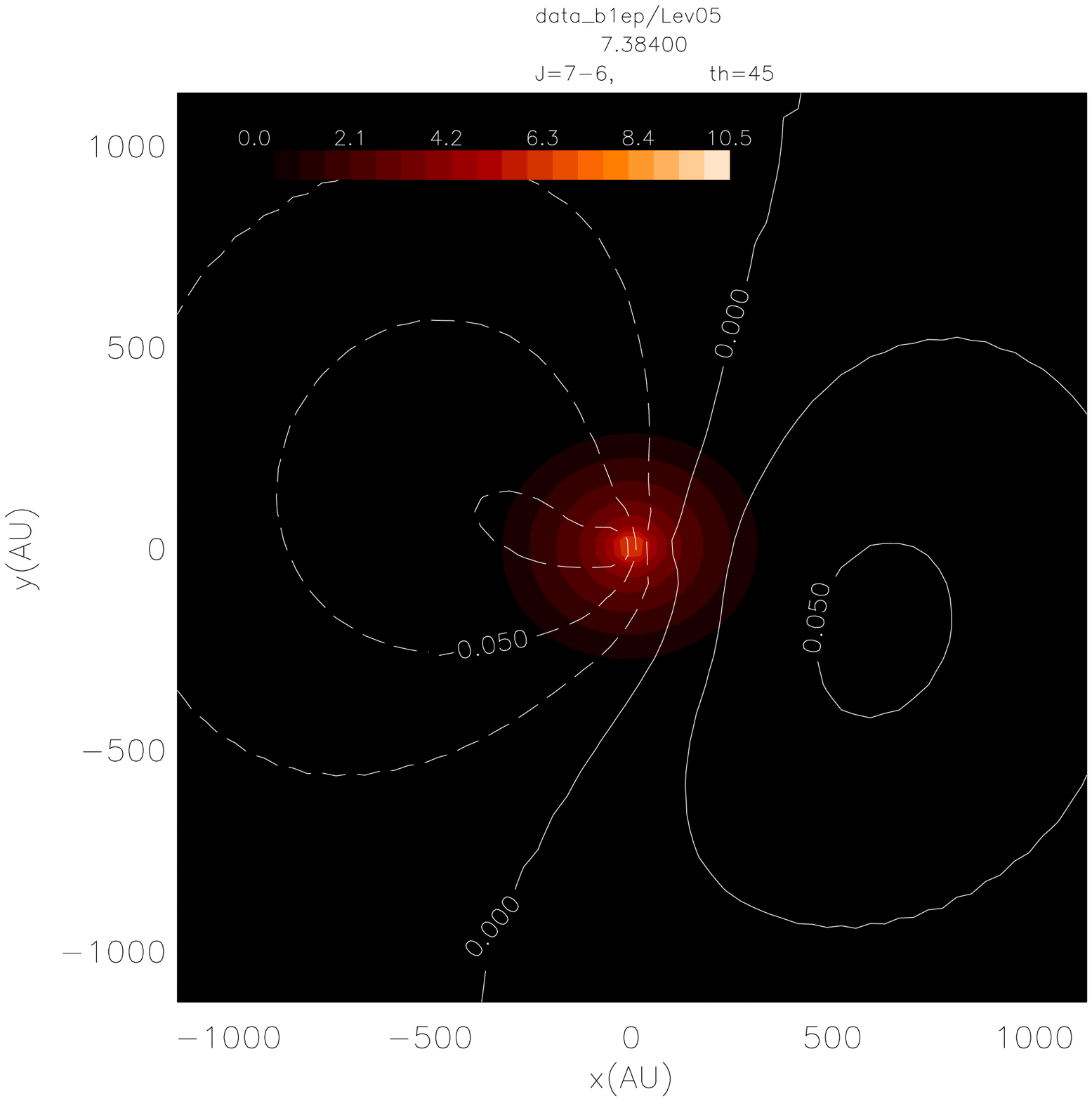}
\includegraphics[width=40mm]{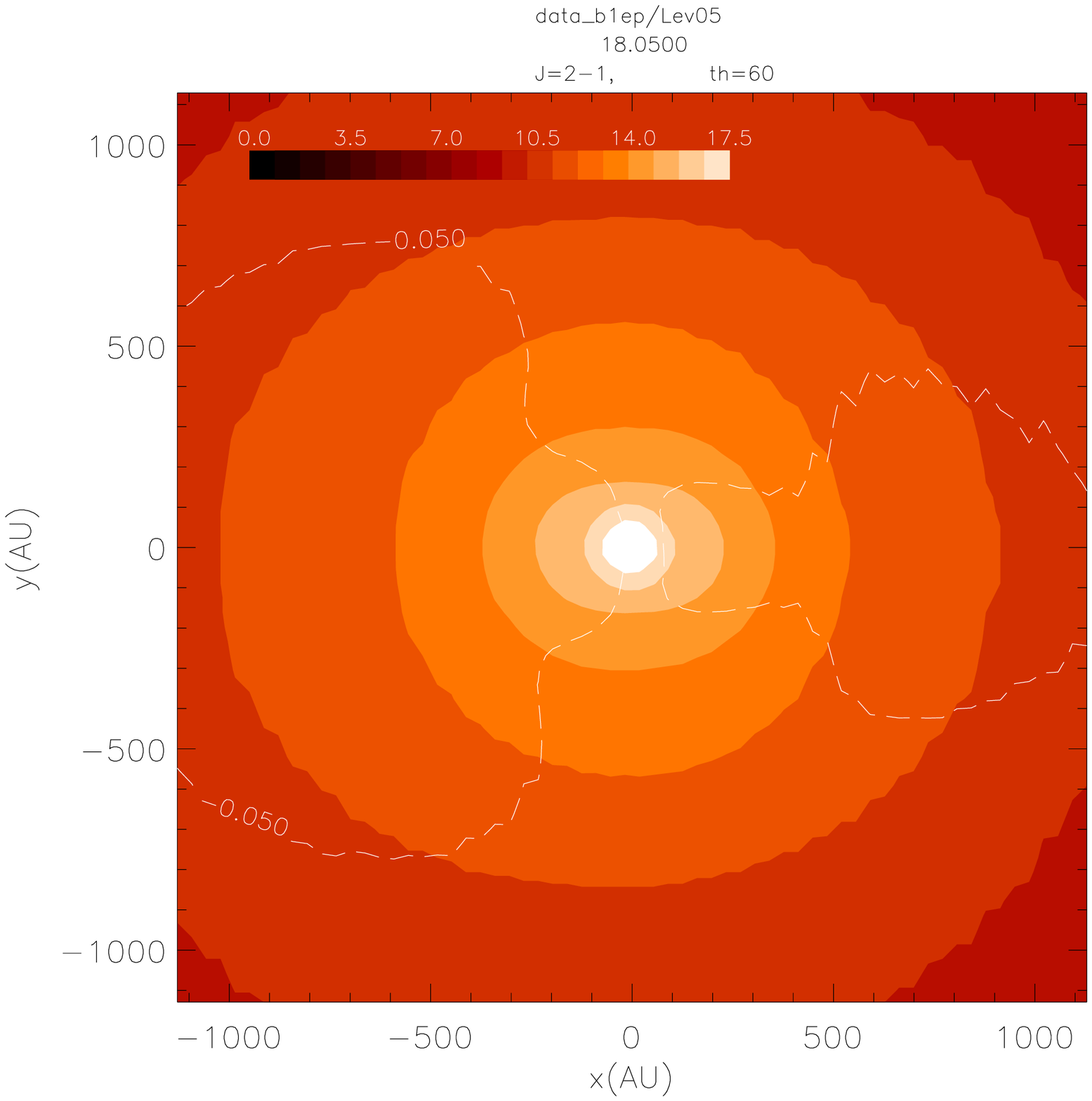}
\includegraphics[width=40mm]{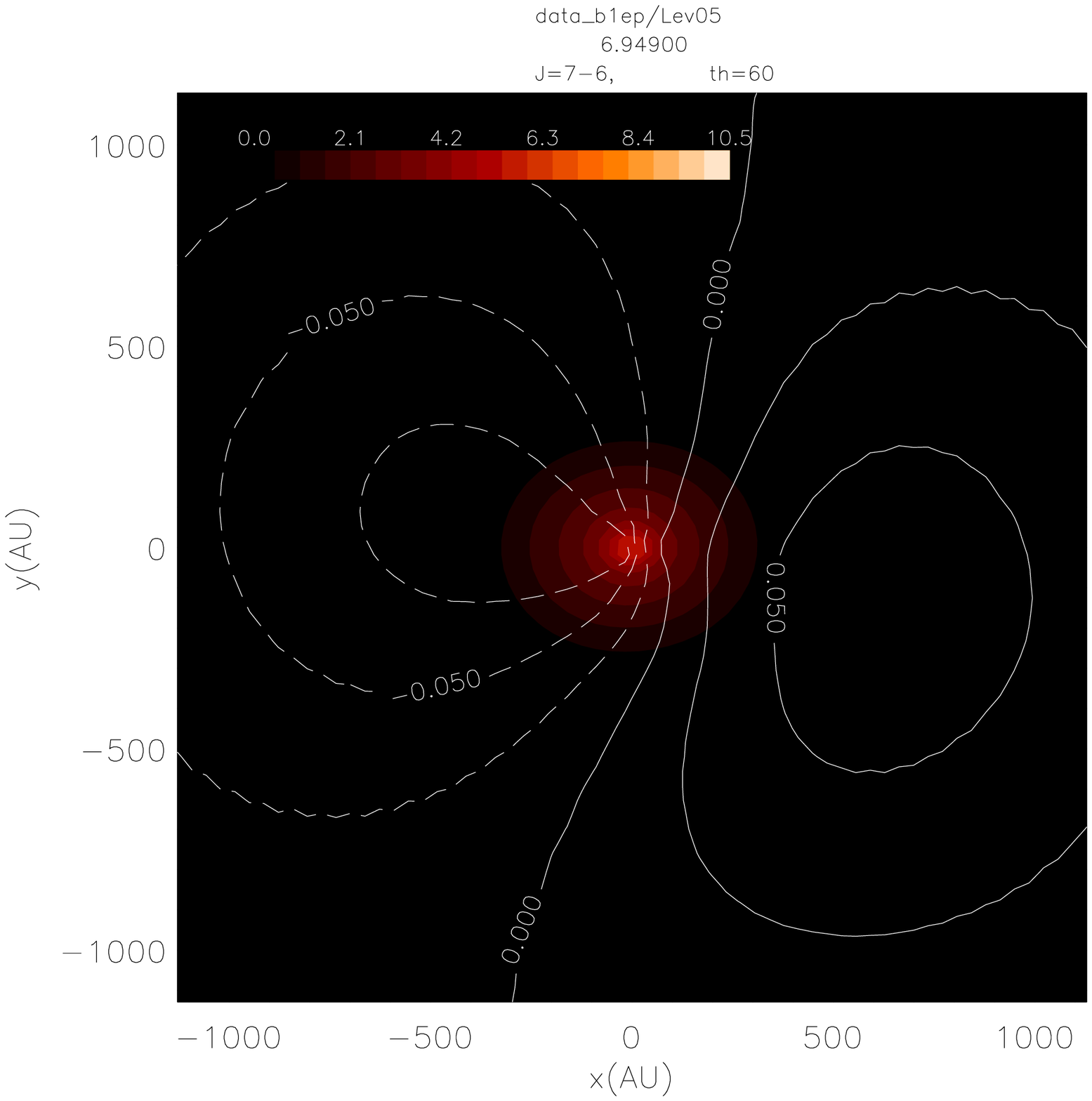}\\
(i)\hspace*{37mm}(j)\hspace*{37mm}(k)\hspace*{37mm}(l)\\
\includegraphics[width=40mm]{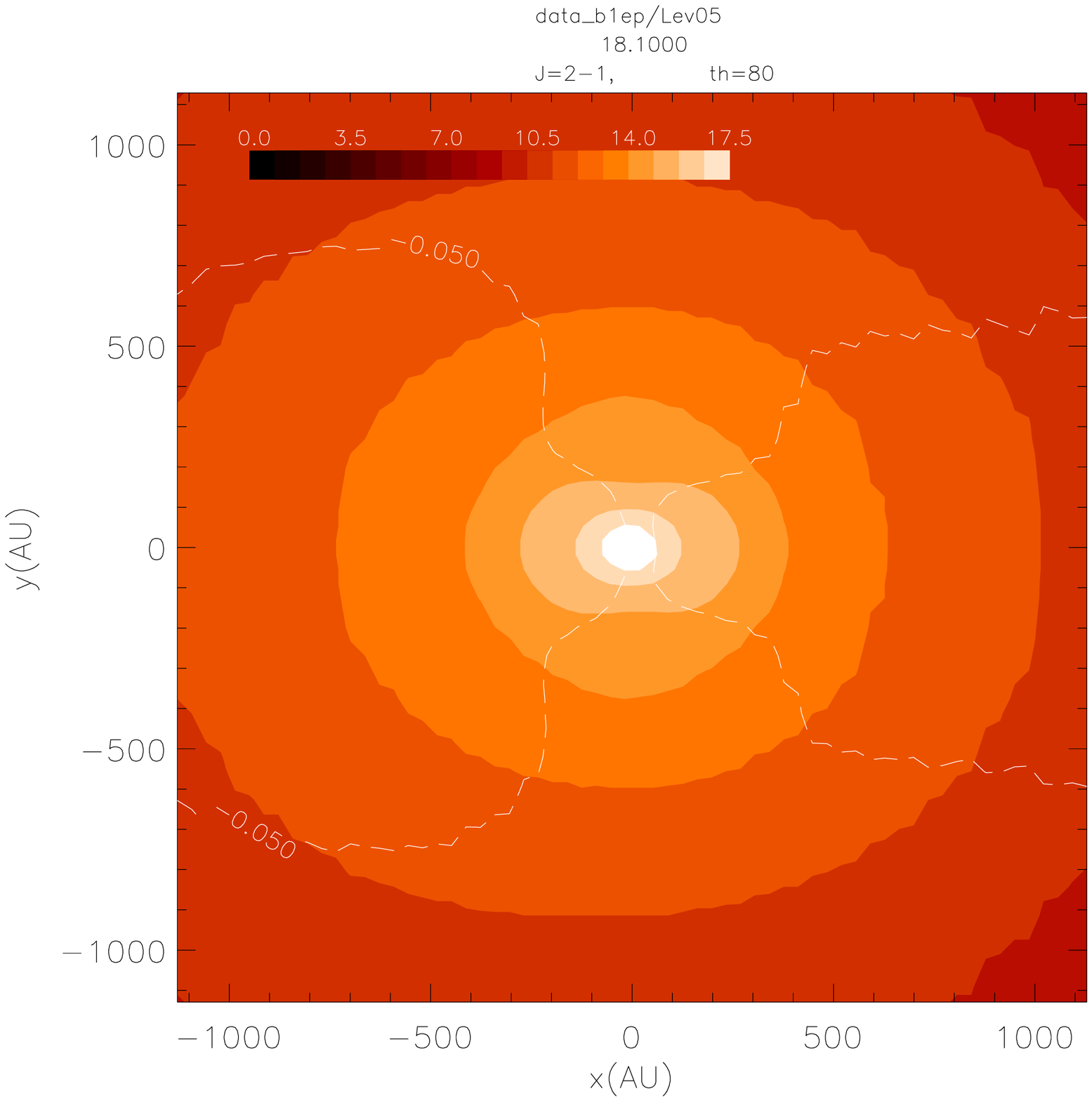}
\includegraphics[width=40mm]{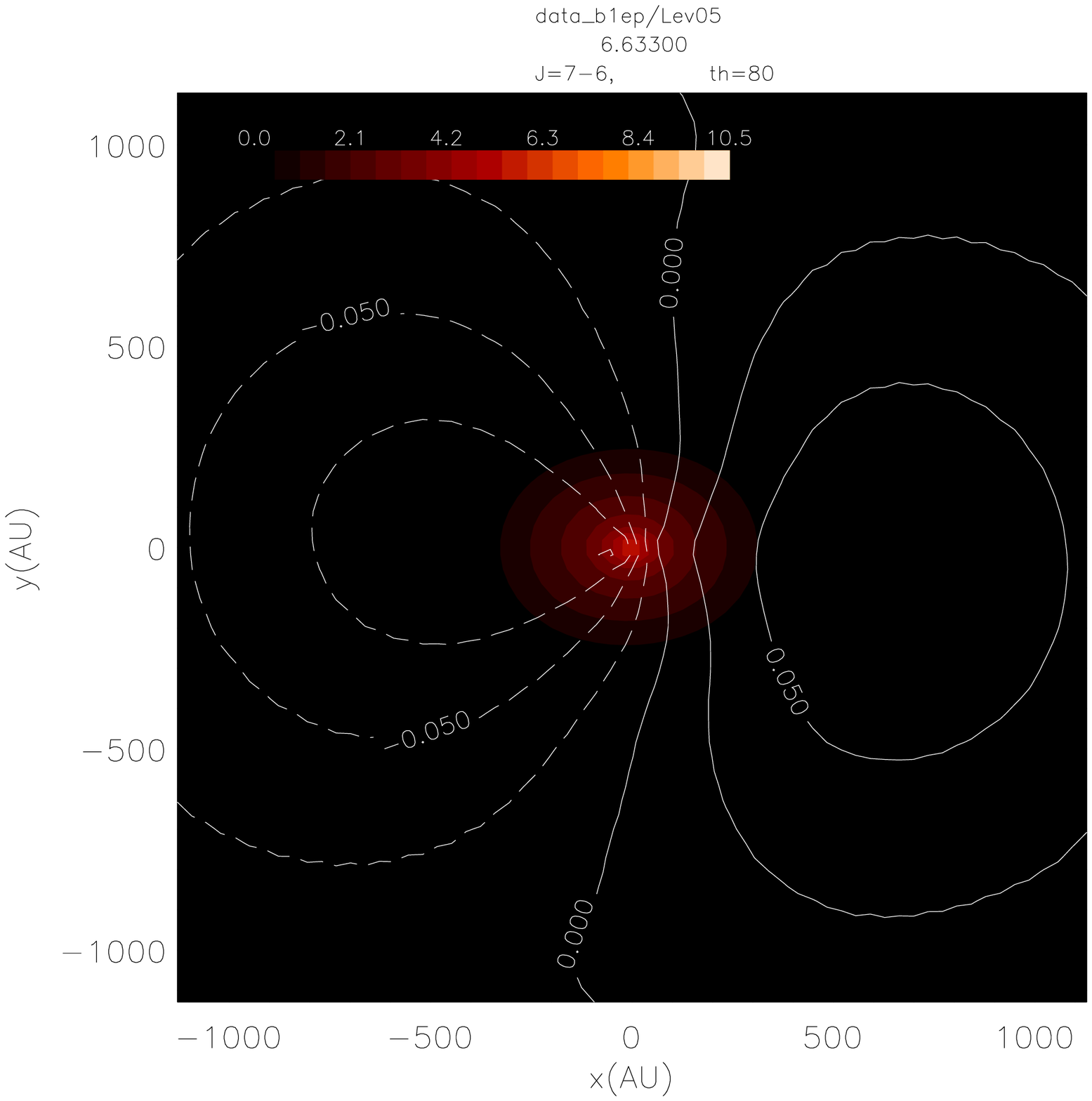}
\includegraphics[width=40mm]{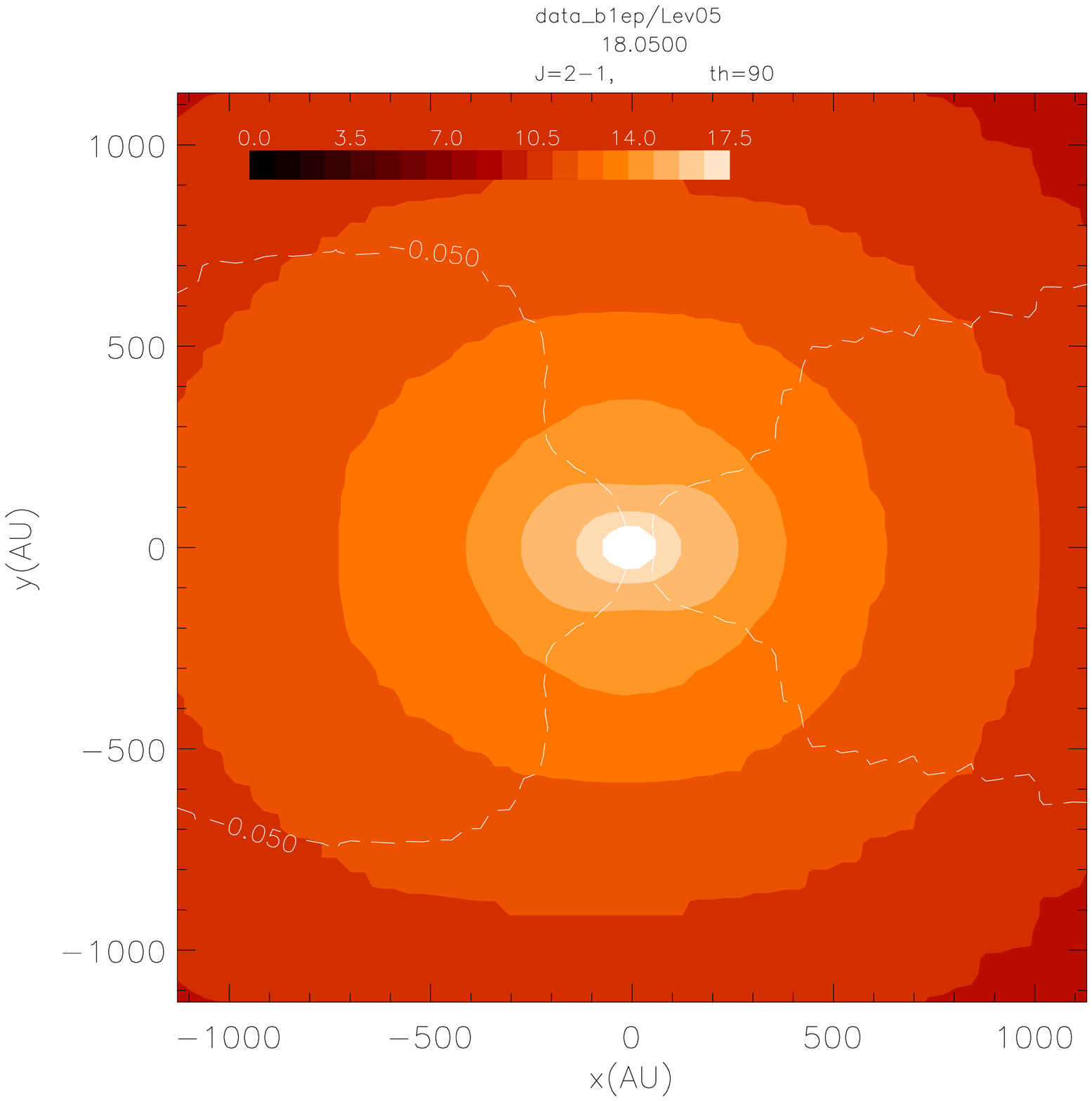}
\includegraphics[width=40mm]{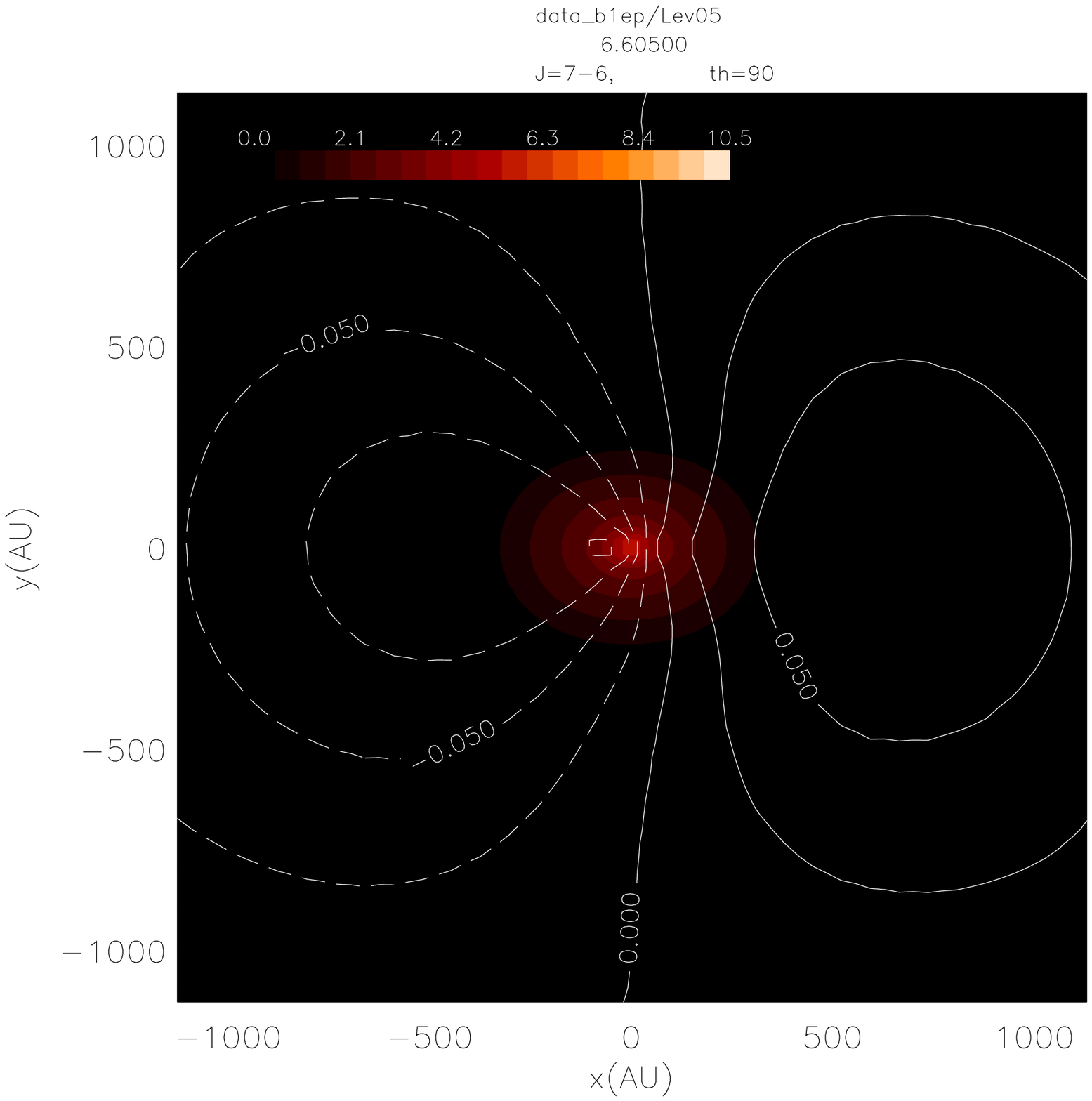}
\caption{\label{fig:theta-dep-pre}
CS $J=2$--$1$ and $J=7$--$6$ integrated intensity (false color)
 and intensity-weighted mean velocity $\langle V\rangle$ for different LOSs (contours) based on the data of Fig.\ref{fig:physical}({\it a}).
The panels ({\it a}), ({\it c}), ({\it e}), ({\it g}), ({\it i}), and ({\it k})
 are results for the CS $J=2$--$1$ lines
and panels ({\it b}), ({\it d}), ({\it f}), ({\it h}), ({\it j}), and ({\it l})
 are results for the CS $J=7$--$6$ lines.
The viewing angles $\theta$ for the respective panels are
  ({\it a}) and  ({\it b}): $\theta=0^\circ$ (pole-on),
  ({\it c}) and  ({\it d}): $\theta=30^\circ$,
  ({\it e}) and  ({\it f}): $\theta=45^\circ$,
  ({\it g}) and  ({\it h}): $\theta=60^\circ$,
  ({\it i}) and  ({\it j}): $\theta=80^\circ$, 
  and ({\it k}) and  ({\it l}): $\theta=90^\circ$ (edge-on).
The levels of the integrated intensity are shown in the color bar
 in the upper-left corner and the unit is ${\rm K\,km\,s^{-1}}$.
The solid and dashed contour lines of $\langle V\rangle$
 represent positive ($0\,{\rm km\,s^{-1}}\le\langle V\rangle\le 0.5\,{\rm km\,s^{-1}}$)
 and negative ($-0.5\,{\rm km\,s^{-1}}\le\langle V\rangle\le 0\,{\rm km\,s^{-1}}$)
 velocities, respectively.
The step of the contour is chosen to be $0.025\,{\rm km\,s^{-1}}$.}
\end{figure}

In Figure \ref{fig:theta-dep-pre}, we show 
 the integrated intensity distributions and 
 intensity-weighted mean velocity $\langle V \rangle$
 for the CS $J=2$--$1$ and $J=7$--$6$ lines for the prestellar isothermal
 collapse phase of Figure \ref{fig:physical} ({\it a}).
In this subsection we consider the effect of 
 the viewing angle $\theta$ of LOS.  

From the models of CS $J=2$--$1$, we can see that the spatial variation of the
 integrated intensity in CS $J=2$--$1$ is as small as $\sim 20\%$.
This is due to the fact that the line is relatively optically thick
 and the velocity gradient is not so large in this phase.
Therefore, the integrated intensity does not follow the density distribution
 (as was already shown from the similarity in the CS $J=2$--$1$ spectra
 in Fig.\ref{fig:spectrum}).

Panel ({\it a}) shows the pole-on view with $\theta=0^\circ$
 and panel ({\it k}) shows the edge-on view
 with $\theta=90^\circ$.
The direction of the minor axis of the integrated intensity
 coincides with the rotation axis.
The CS $J=7$--$6$ line has a low integrated intensity from $6.6$
 to $7.8\, {\rm K\,km\,s^{-1}}$, compared with $J=2$--$1$
 ($18$--$19\,{\rm K\,km\,s^{-1}}$).
However, an antisymmetric velocity pattern,
 which features global rotation motion,
 is well traced in CS $J=7$--$6$ but not in $J=2$--$1$.
The velocity pattern specific to the rotation motion is erased
 in the CS $J=2$--$1$ emission owing to its optical thickness.

\begin{figure}
\centering
(a)\hspace*{37mm}(b)\hspace*{37mm}(c)\hspace*{37mm}(d)\\
\includegraphics[width=40mm]{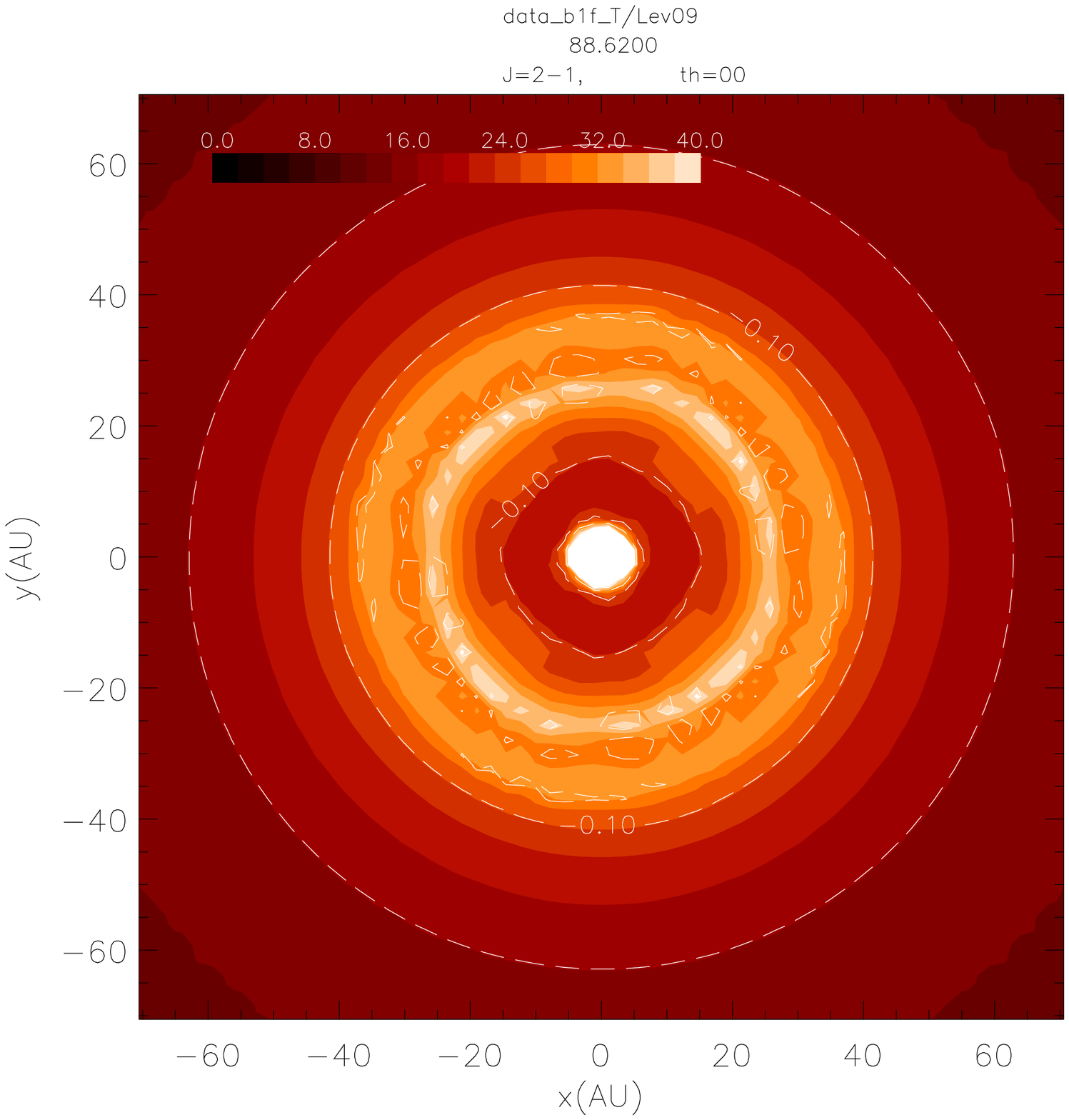}
\includegraphics[width=40mm]{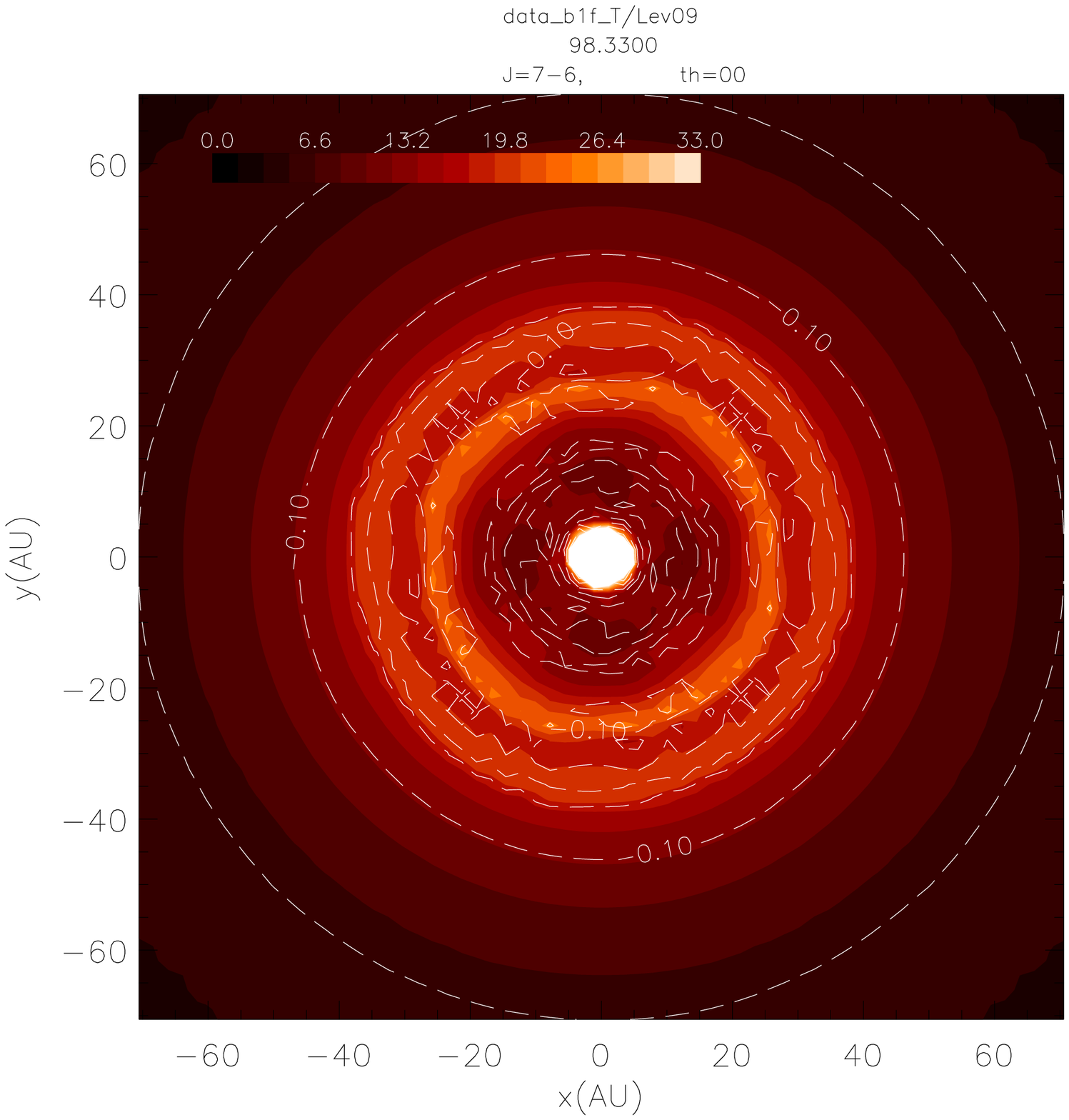}
\includegraphics[width=40mm]{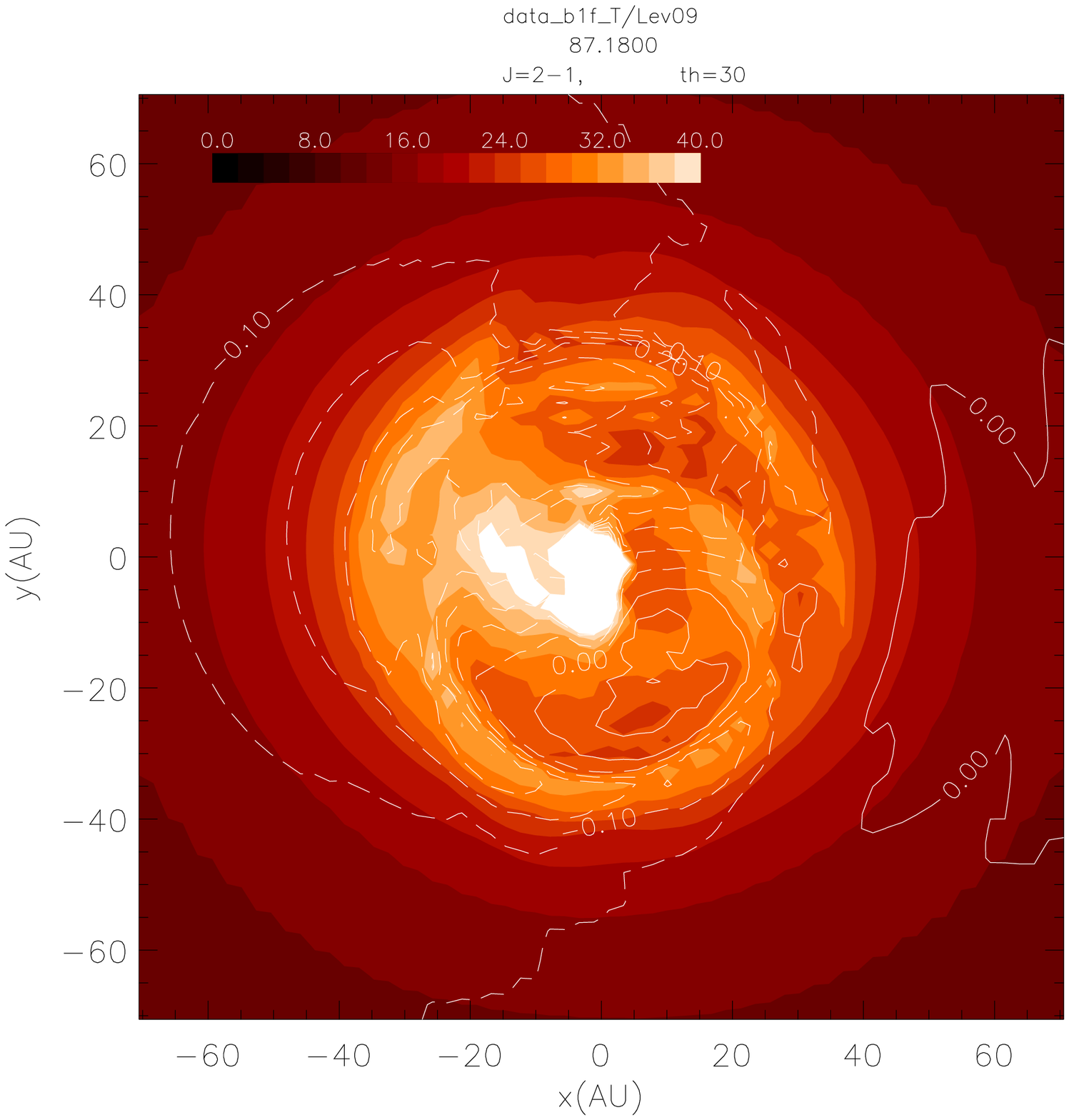}
\includegraphics[width=40mm]{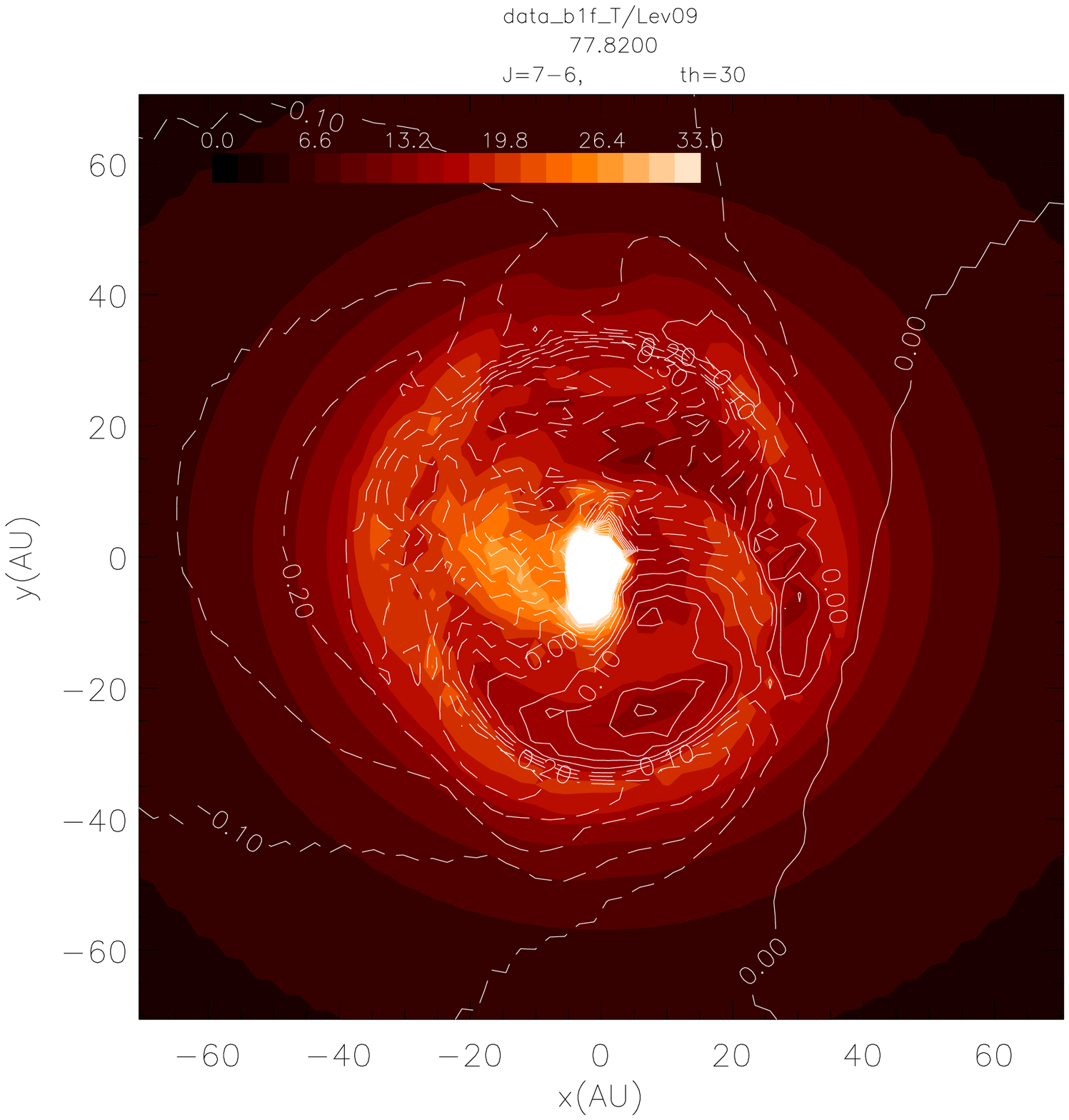}\\
(e)\hspace*{37mm}(f)\hspace*{37mm}(g)\hspace*{37mm}(h)\\
\includegraphics[width=40mm]{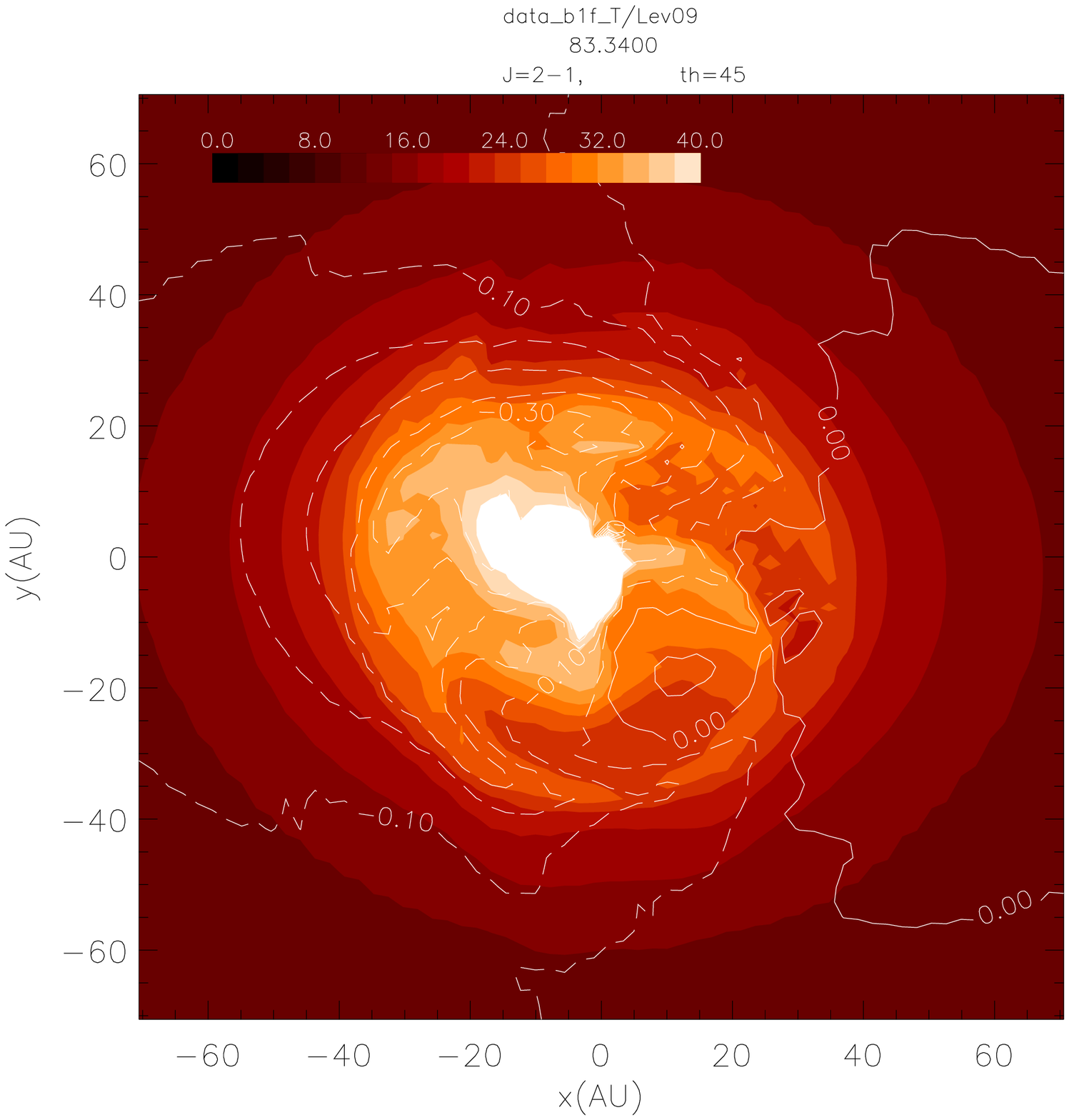}
\includegraphics[width=40mm]{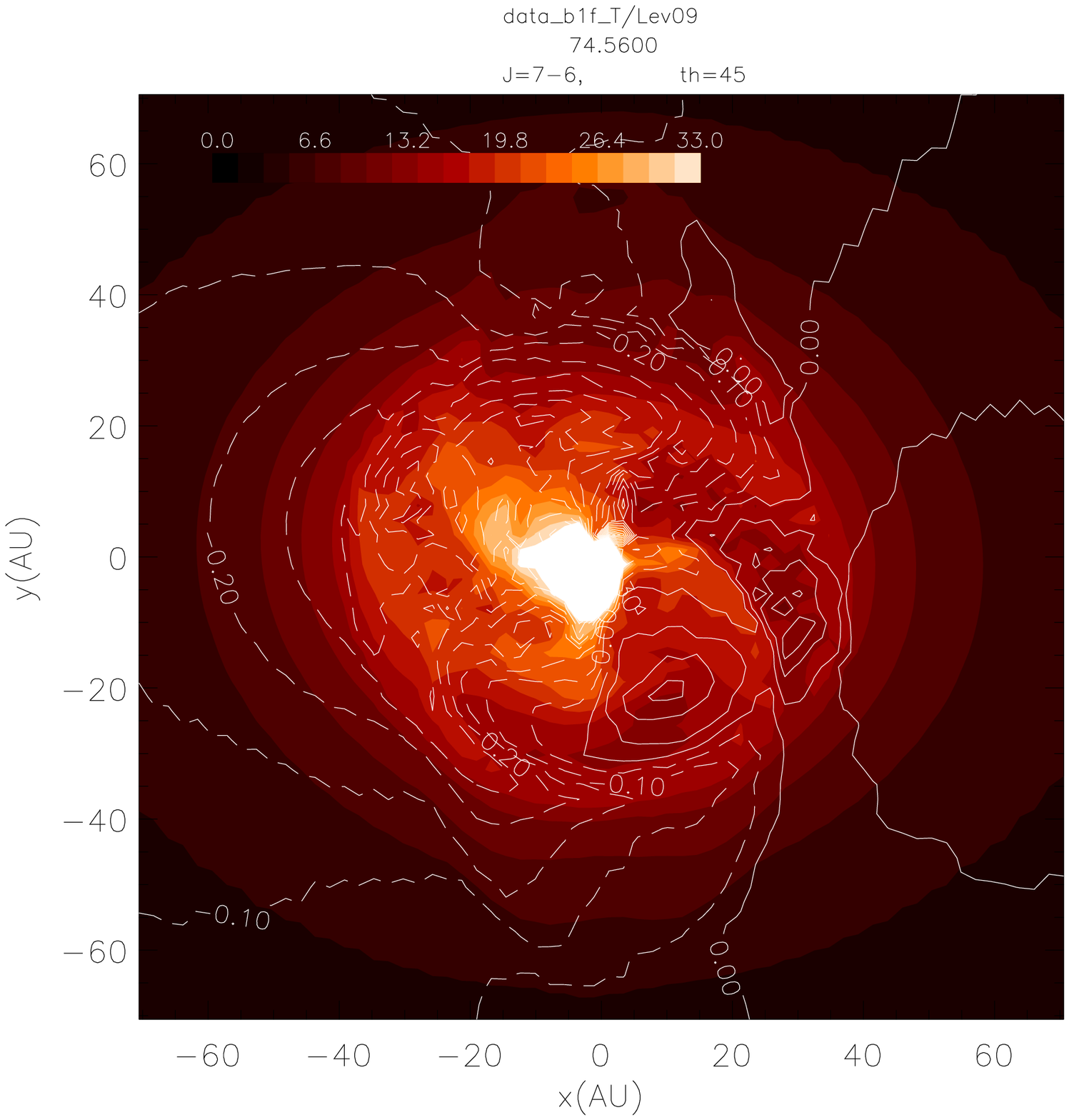}
\includegraphics[width=40mm]{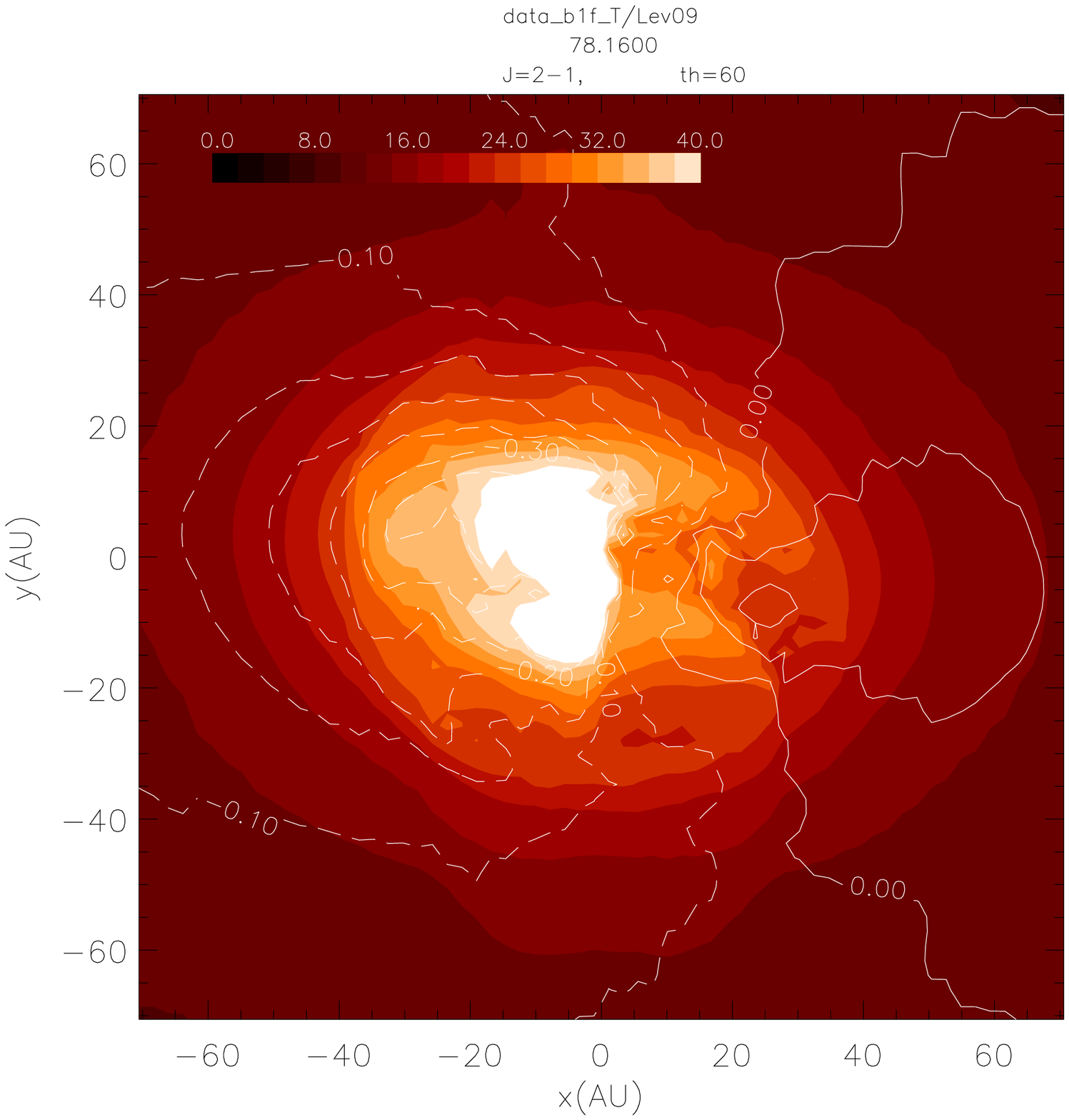}
\includegraphics[width=40mm]{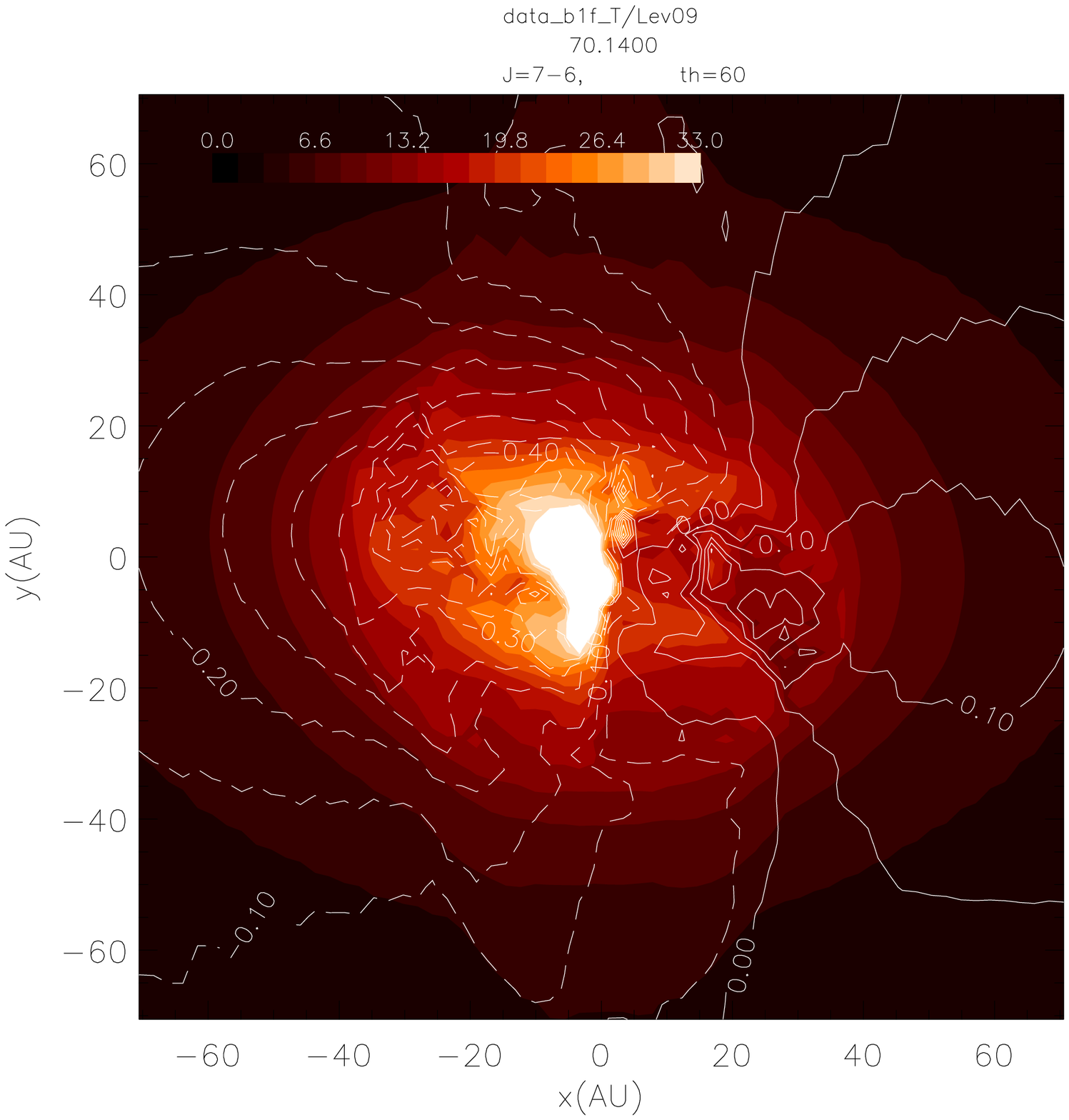}\\
(i)\hspace*{37mm}(j)\hspace*{37mm}(k)\hspace*{37mm}(l)\\
\includegraphics[width=40mm]{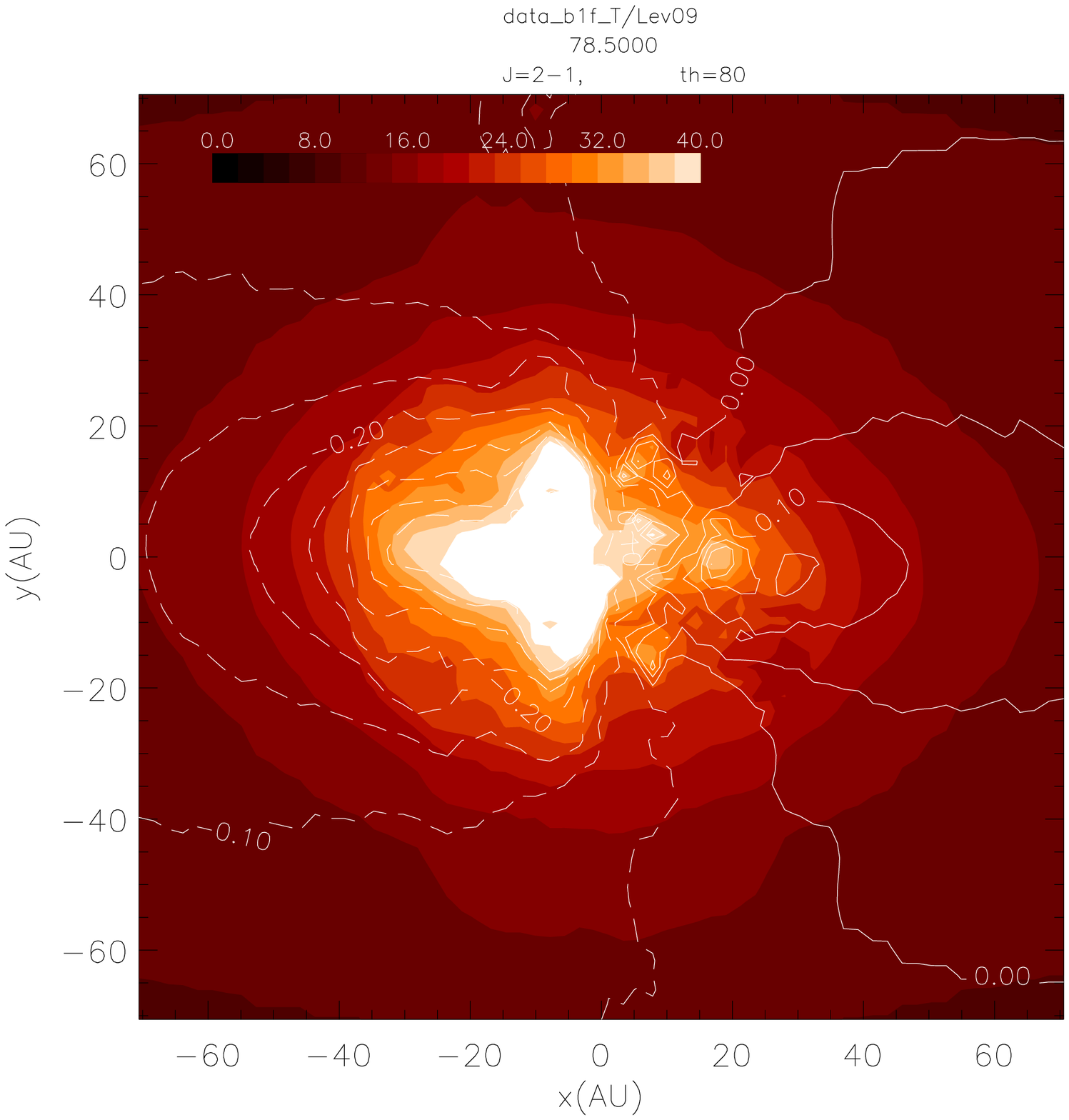}
\includegraphics[width=40mm]{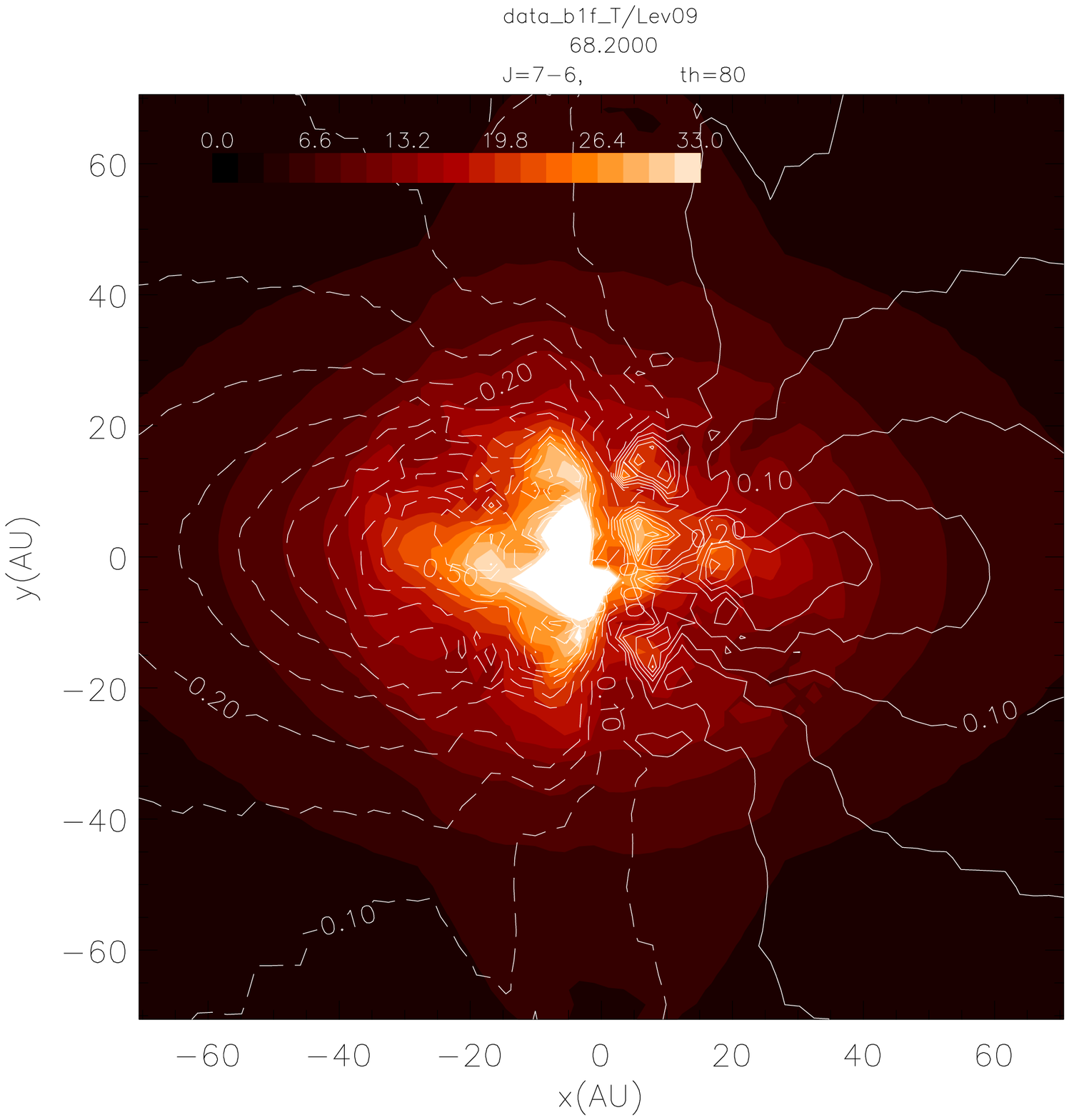}
\includegraphics[width=40mm]{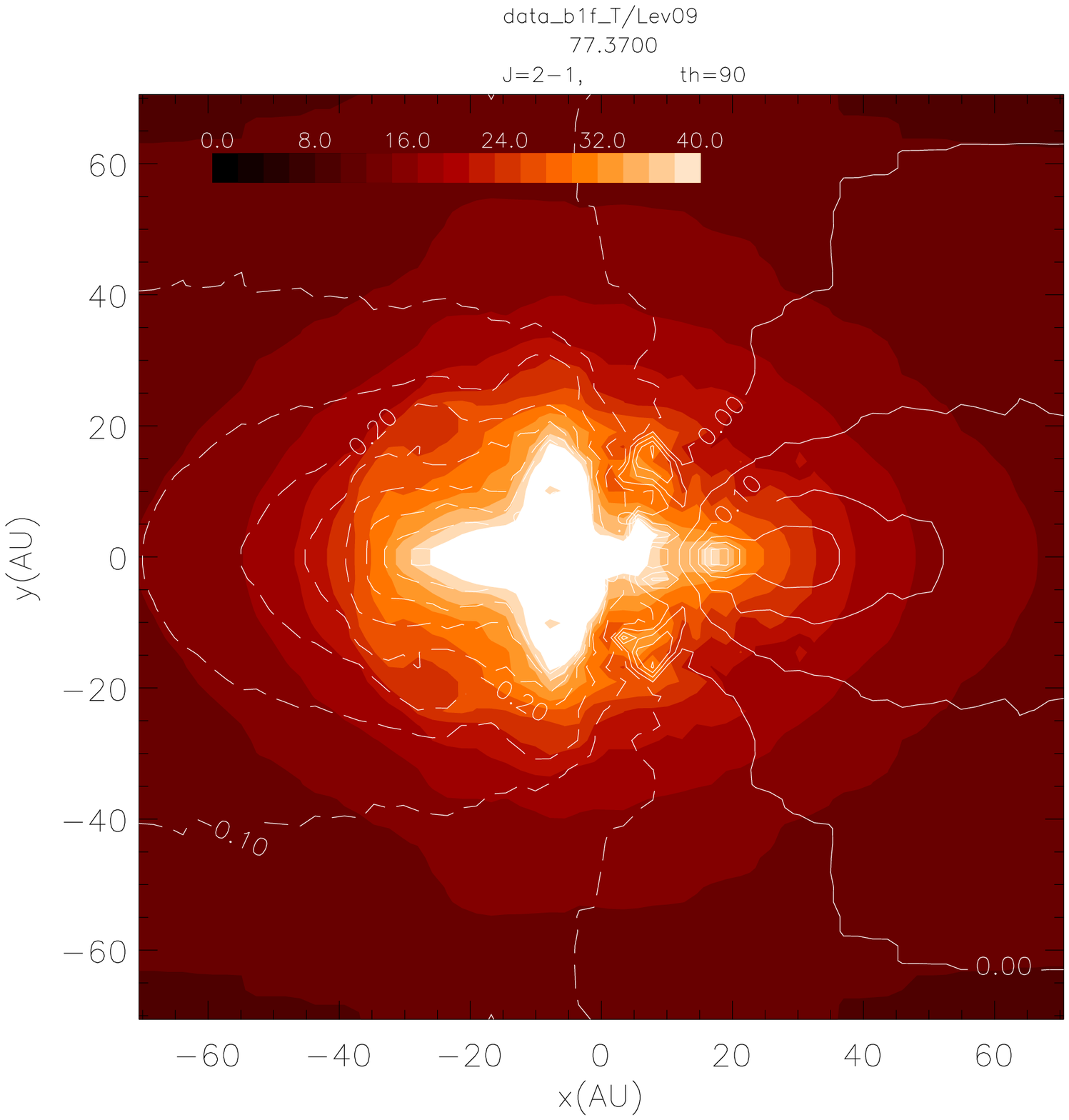}
\includegraphics[width=40mm]{f7k.eps}
\caption{\label{fig:theta-dep-post}
The same as Fig. \ref{fig:theta-dep-pre} but for the protostellar first core
 phase (Fig.\ref{fig:physical} ({\it c}) and ({\it d})).
The velocity range is 
 $0\,{\rm km\,s^{-1}}\le\langle V\rangle\le 1\,{\rm km\,s^{-1}}$ for positive (solid-line contour)
 velocity 
 and $-1\,{\rm km\,s^{-1}}\le\langle V\rangle\le 0\,{\rm km\,s^{-1}}$ for negative 
 (dashed-line contour) velocity.
The step of the contour is chosen to be $0.05\,{\rm km\,s^{-1}}$.}
\end{figure}

Figure \ref{fig:theta-dep-post} is the same as Figure \ref{fig:theta-dep-pre}
 but for the protostellar first core phase.
The model with $\theta=0^\circ$ (pole-on: panels ({\it a}) and ({\it b}))
 exhibits a ring-like enhancement of intensity.
This enhancement corresponds to the outflow lobe seen in
 Figure \ref{fig:physical} ({\it c}) and  ({\it d}).
It should be noted that overall $\langle V \rangle$ is negative
 when the system is viewed pole-on. 
However, since the ring has a more negative recession velocity
 than the rest of the field, this negative velocity originates
 from the outflowing gas.  

Viewing from $\theta=30^\circ$ [({\it c}) and ({\it d})],
 the feature becomes more complicated.
Namely, in addition to the ring-like feature
 a bar-like structure appears near the center
 in the integrated intensity distribution.
In the negative velocity region, an island of positive 
 velocity region appears at $-25{\rm AU}\lesssim x\lesssim 25{\rm AU}$ and
 $-30{\rm AU}\lesssim y\lesssim -5{\rm AU}$.
There are two possible explanations for this positive velocity region:
(1)infalling gas entering the disk with positive recession velocity absorbs
the emission coming from the hotter interior.  
If such gas is evacuated by the outflow, the
 blue-shift emission is strengthened.
However, if this is the case, a similar signature must appear
 in the pole-on model ({\it a} and {\it b}). 
Thus we consider the second explanation,
(2) a positive recession velocity from the 
 accelerated outflow in the far side.
This positive velocity region moves to the right with $\theta$
 and is merged with the signature of the rotating disk (the left-hand side has
 a negative velocity and the right-hand side has a positive velocity)
 for $\theta\gtrsim 60^\circ$  [({\it g})--({\it j})].     
Finally, the edge-on view [({\it k}) and ({\it l})]
 indicates a fat disk-like appearance in the integrated intensity distribution
 which is approximately symmetric against the rotation axis.
It should be noted that
 the gradient in $\langle V \rangle$ is larger than that
 of the isothermal collapse phase\footnote{The step of the isovelocity contour
 was chosen to be $0.025\, {\rm km\,s^{-1}}$ in Fig. \ref{fig:theta-dep-pre} but
 is $0.05\, {\rm km\,s^{-1}}$ in this figure.}.  

However, this distribution has a slight asymmetry in the brightness.
Globally, the $\langle V \rangle$ distribution is antisymmetric
 against the rotation axis, which apparently indicates that the disk is
 rotating. 
However, this also exhibits a deviation from the antisymmetric distribution.
A larger portion of the gas seems to have negative velocity.
This asymmetry against the rotation axis is due to the fact that
 the gas has both rotation and infall motions,
 which is seen more clearly
 in $\S\S$ \ref{sec:rot_and_infall} and \ref{sec:asymmetry}.

\subsection{Rotation and Infall Motion} 
\label{sec:rot_and_infall}

\begin{figure}
\centering
(a)\hspace*{70mm}(b)\\
\includegraphics[width=80mm]{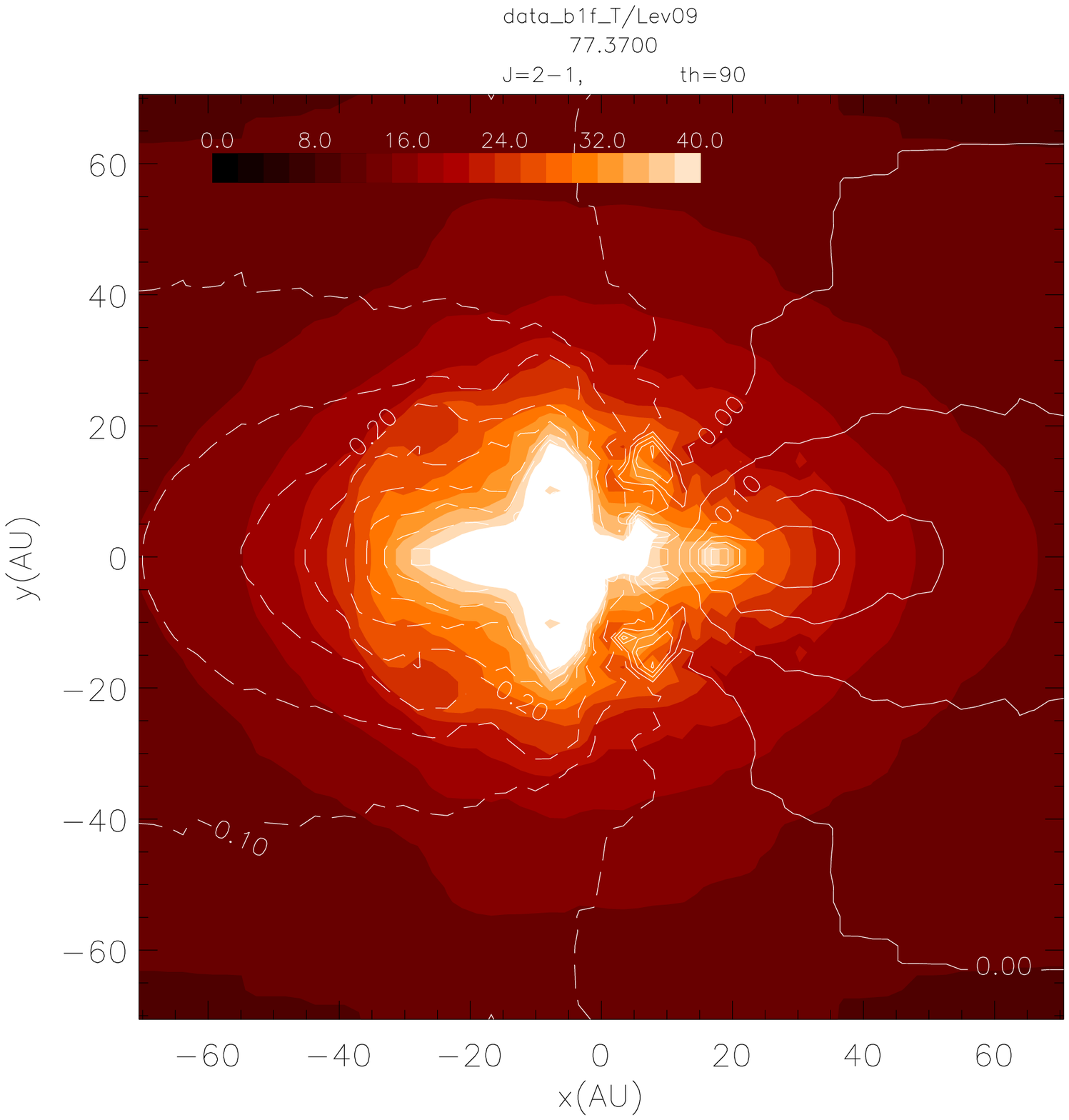}
\includegraphics[width=80mm]{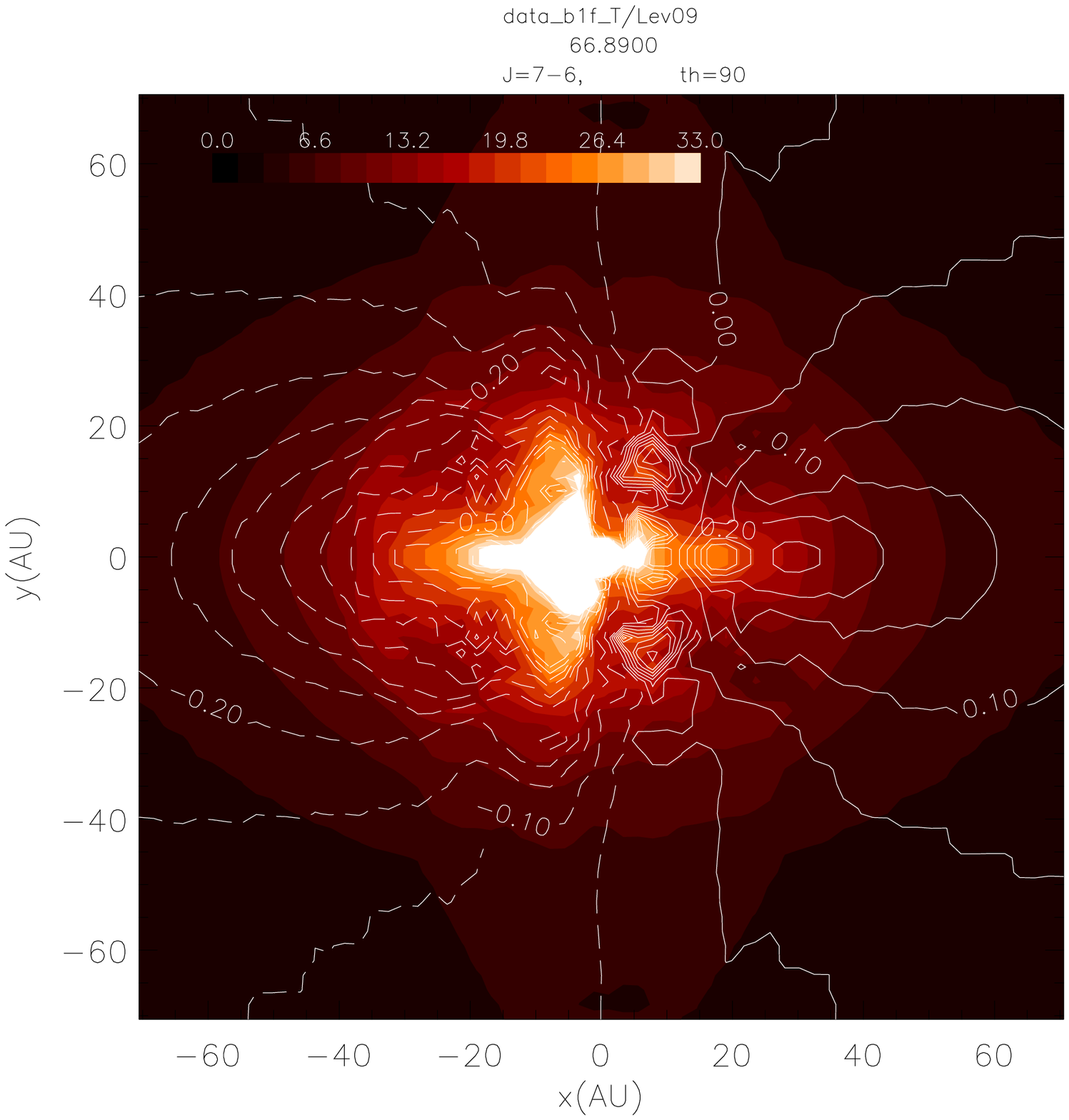}\\
(c)\hspace*{70mm}(d)\\
\includegraphics[width=80mm]{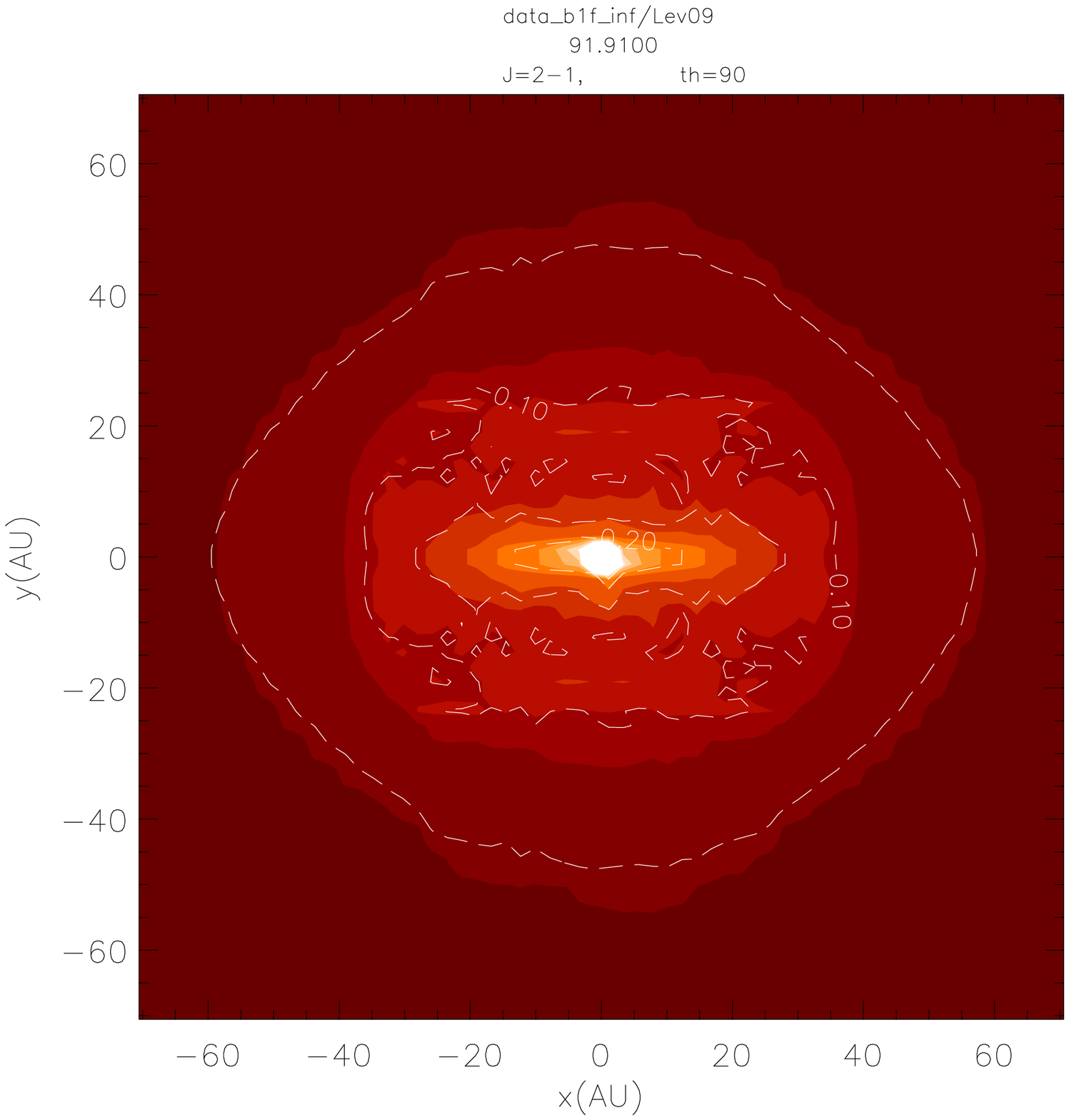}
\includegraphics[width=80mm]{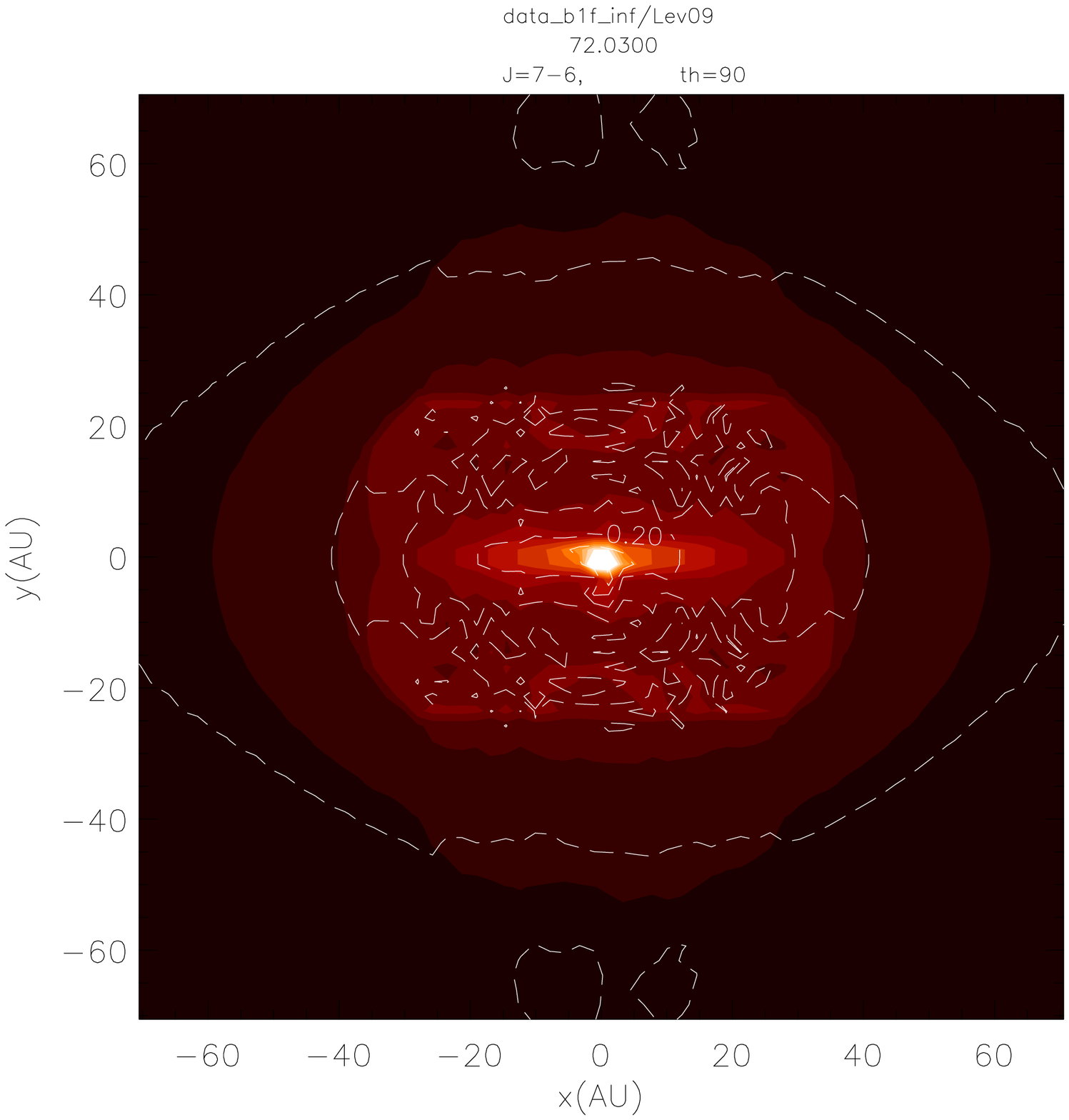}\\
(e)\hspace*{70mm}(f)\\
\includegraphics[width=80mm]{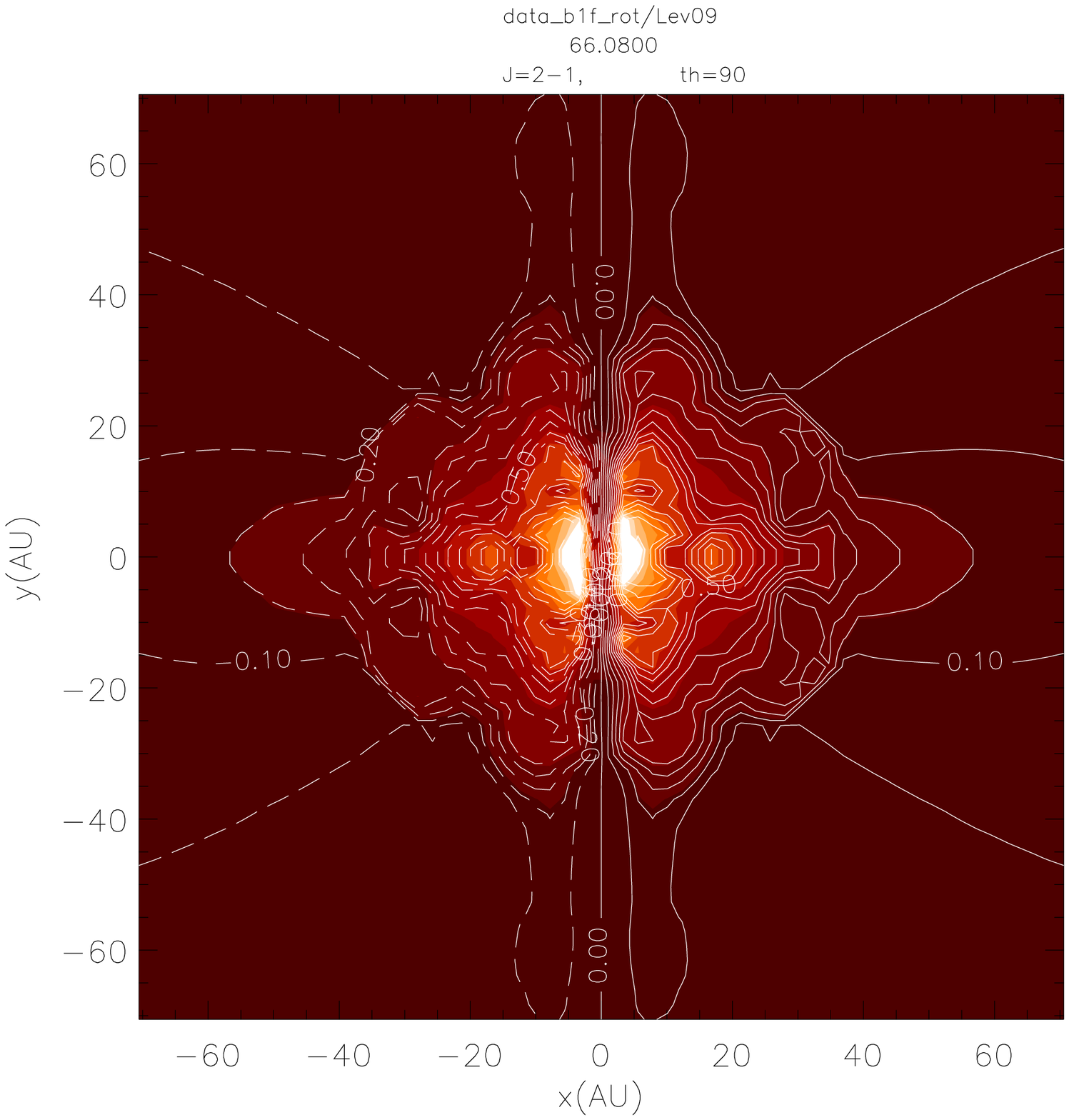}
\includegraphics[width=80mm]{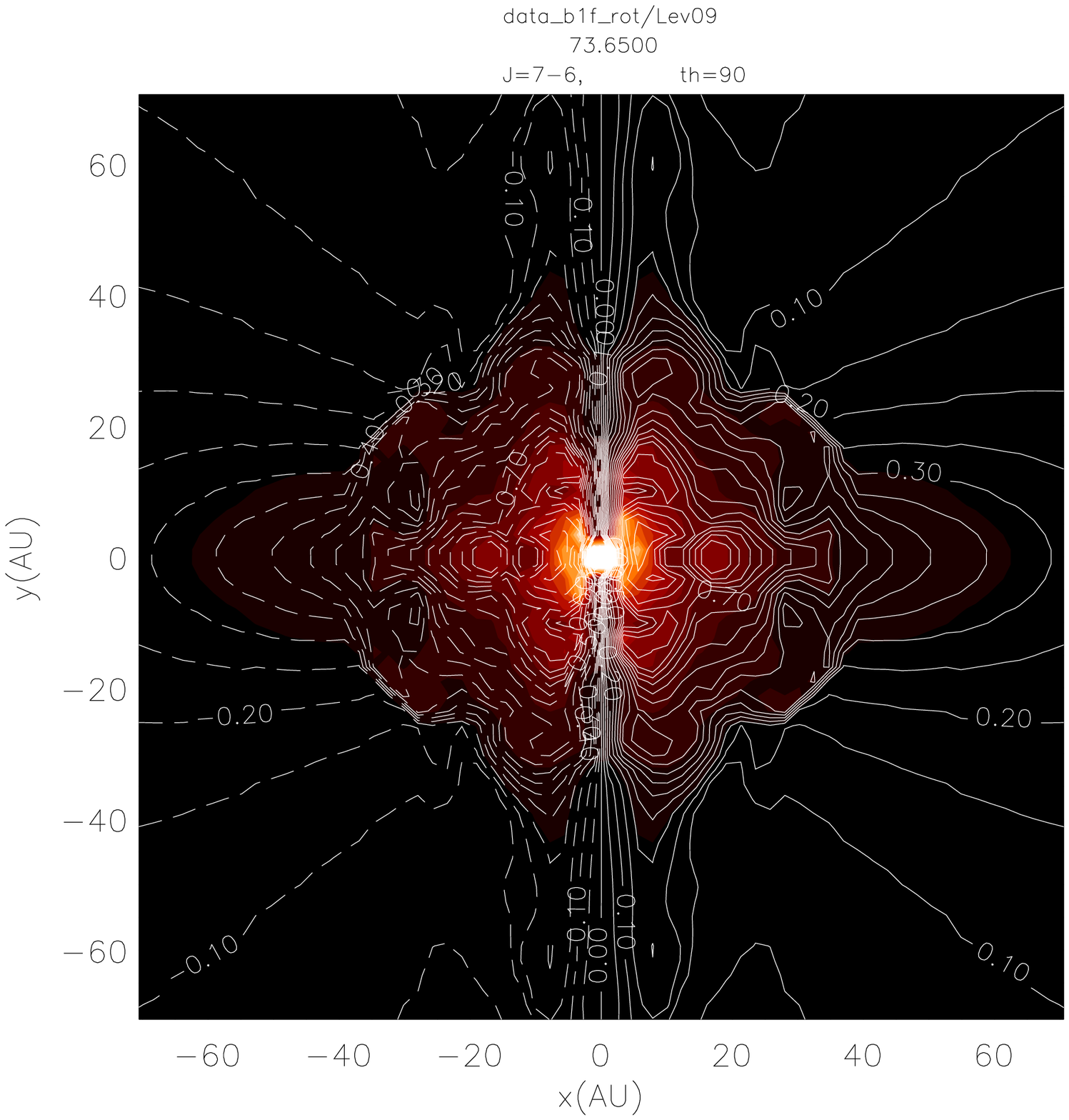}
\caption{\label{fig:comp_rot_inf}
Comparison between models in which only the inflow motion is assumed
[({\it c}) and ({\it d})]
and that for which only the rotation motion is assumed
 [({\it e}) and ({\it f})].
Panels ({\it a}), ({\it c}), and ({\it e}) are for CS $J=2$--$1$
 and panels ({\it b}), ({\it d}), and ({\it f}) are for CS $J=7$--$6$.
The calculations are made for the edge-on view with $\theta=90^\circ$.
}
\end{figure}

To investigate the origin of the asymmetric distributions,
 we show the integrated intensity and
 intensity-weighted average velocity for the CS $J=2$--$1$ ({\it a})
 and $J=7$--$6$ ({\it b}) lines in Figure \ref{fig:comp_rot_inf}.
To compare these, we made the same plots for a rotation model ({\it c} and
 {\it d}) in which 
 we preserved the rotation motion and artificially removed the infall velocity
 and for an infall model ({\it e} and {\it f}) in which 
 we preserved the infall motion and artificially removed the rotation velocity.
The rotation and inflow velocities are calculated
  using the  tangential unit vector $\boldsymbol{e}_\phi=\boldsymbol{e}_r\times \boldsymbol{e}_z$ as
\begin{equation}
\boldsymbol{v}_{\rm rotation}=\left(\boldsymbol{v}\cdot\boldsymbol{e}_\phi\right)\boldsymbol{e}_\phi,
\end{equation}
\begin{equation}
\boldsymbol{v}_{\rm inflow}=\boldsymbol{v}-\boldsymbol{v}_{\rm rotation},
\end{equation}
respectively. 

As expected, the rotation and infall models have
 completely different integrated intensity
 and the intensity-weighted velocity distributions.
In the infall model ({\it c} and {\it d}),
 the integrated intensity distribution exhibits a compact core in
 a fat spheroid. 
The observed first moment velocity is symmetric against the $z$-axis.
In contrast, in the rotation model ({\it e} and {\it f}),
 a flat disk is seen in the integrated intensity distribution
 and an outflow extends to $z\lesssim 70 {\rm AU}$ in
 the first moment velocity, which indicates an antisymmetric pattern.
Thus, an accurate model in which gas rotates and simultaneously inflows 
 has no symmetry except against the $x$-axis ($y=0$).
This asymmetry can be qualitatively understood from the fact that
 the superposition of symmetric and antisymmetric velocities
 has no symmetry.

We should also note from Figure \ref{fig:comp_rot_inf}
 ({\it a} and {\it b})
 that the gas to the left of the plot is relatively brighter than the gas on the right
 ({\it a} and {\it b}). 
However, the infall and rotation models have symmetric integrated intensity
  distributions against the rotation axis.
This can be understood as being due to the effect of rotating inflow gas. 
We discuss the reason in more detail in section \ref{sec:asymmetry}.
  
\section{Discussion}

\subsection{Comparison with Isothermal Model}

\begin{figure}
\centering
(a)\hspace*{70mm}(b)\\
\includegraphics[width=80mm]{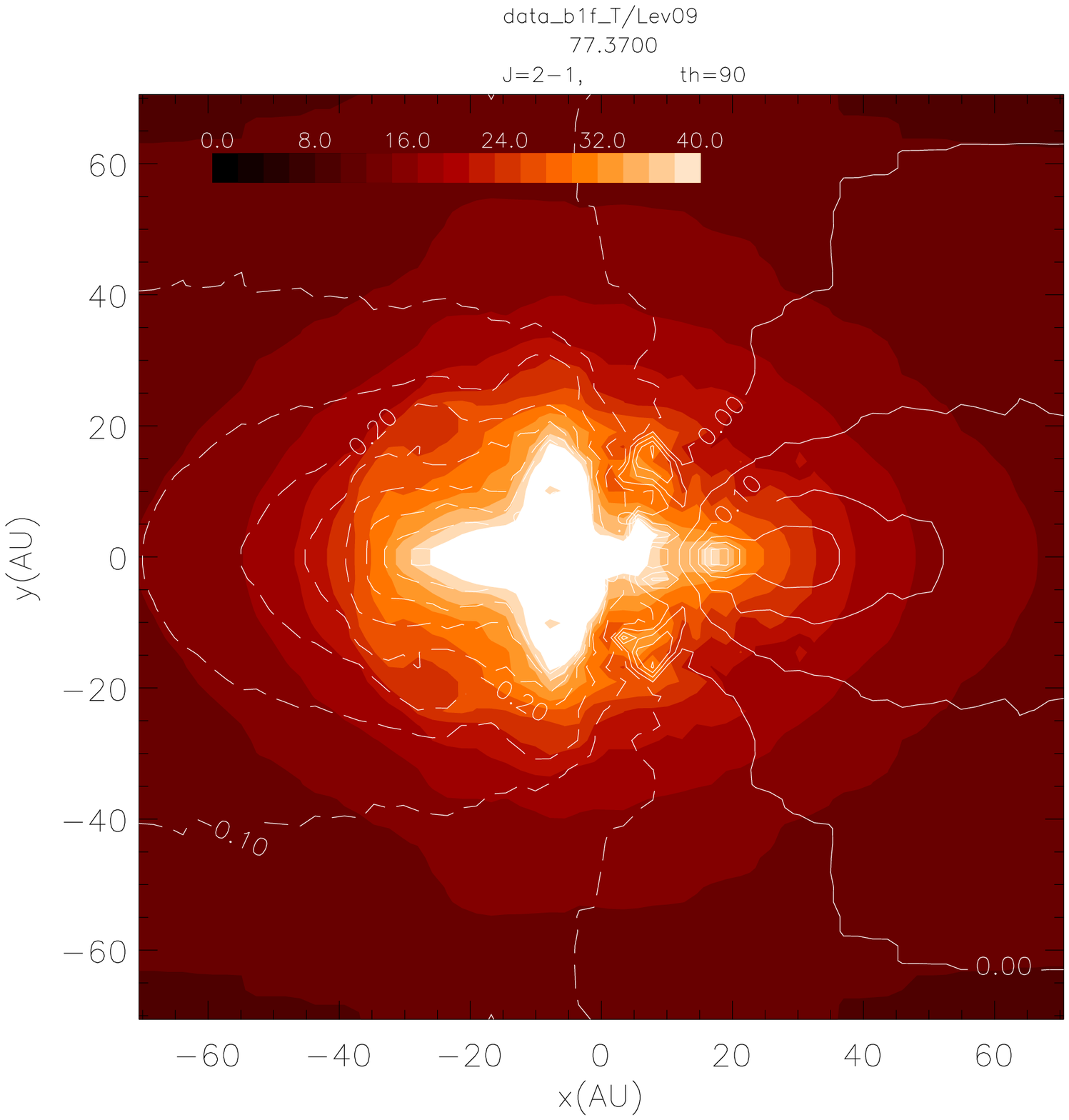}
\includegraphics[width=80mm]{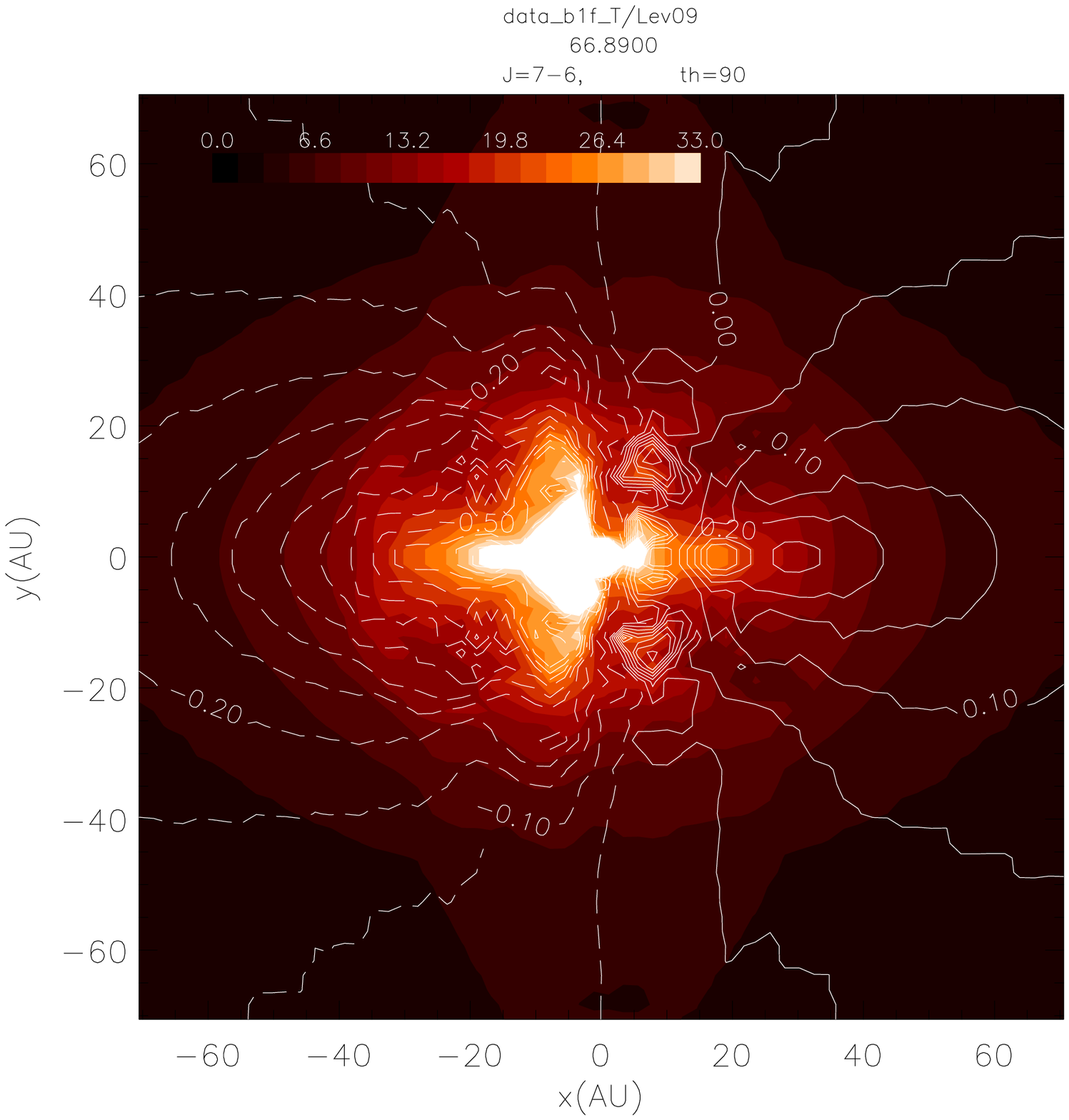}\\
(c)\hspace*{70mm}(d)\\
\includegraphics[width=80mm]{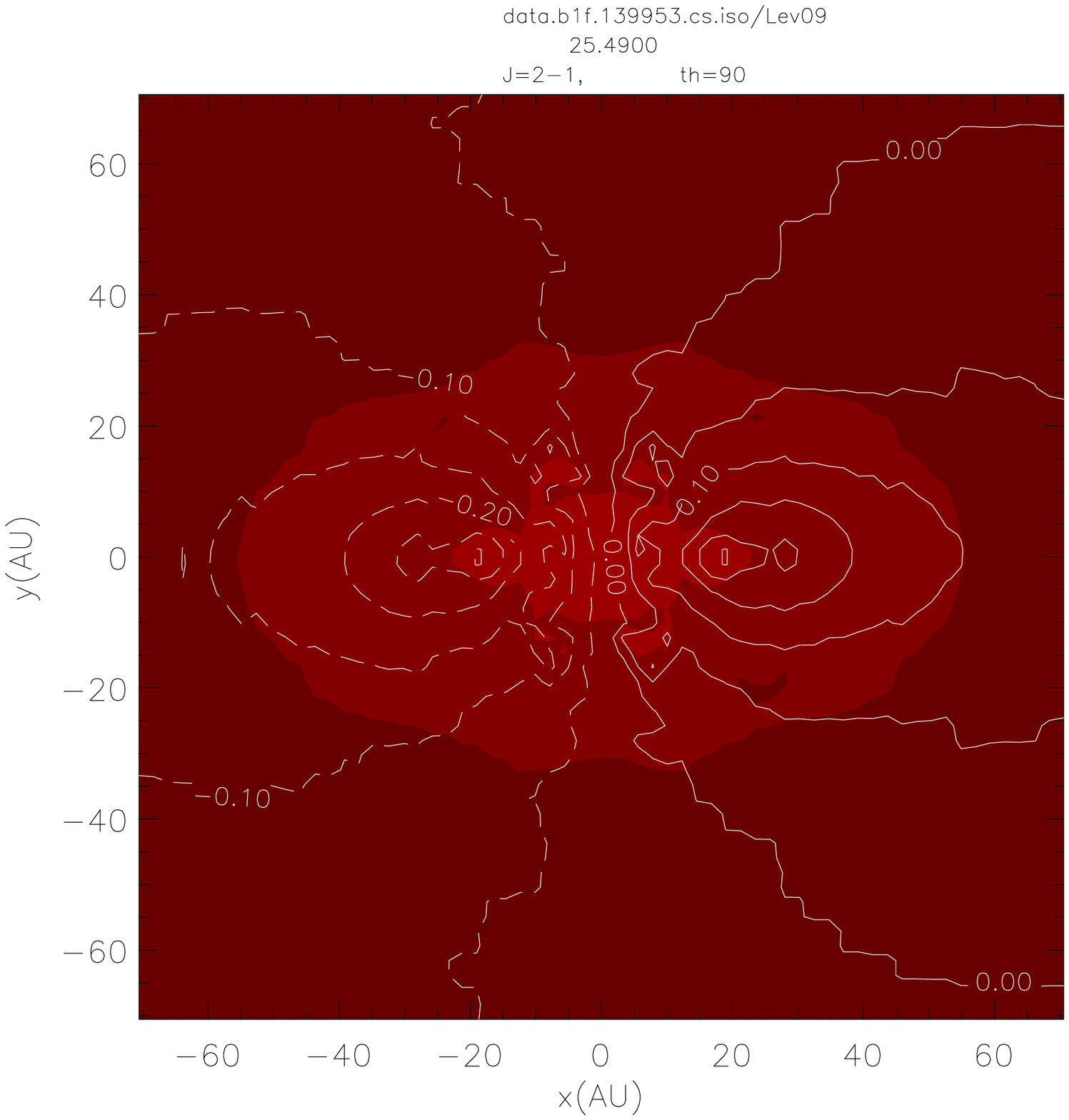}
\includegraphics[width=80mm]{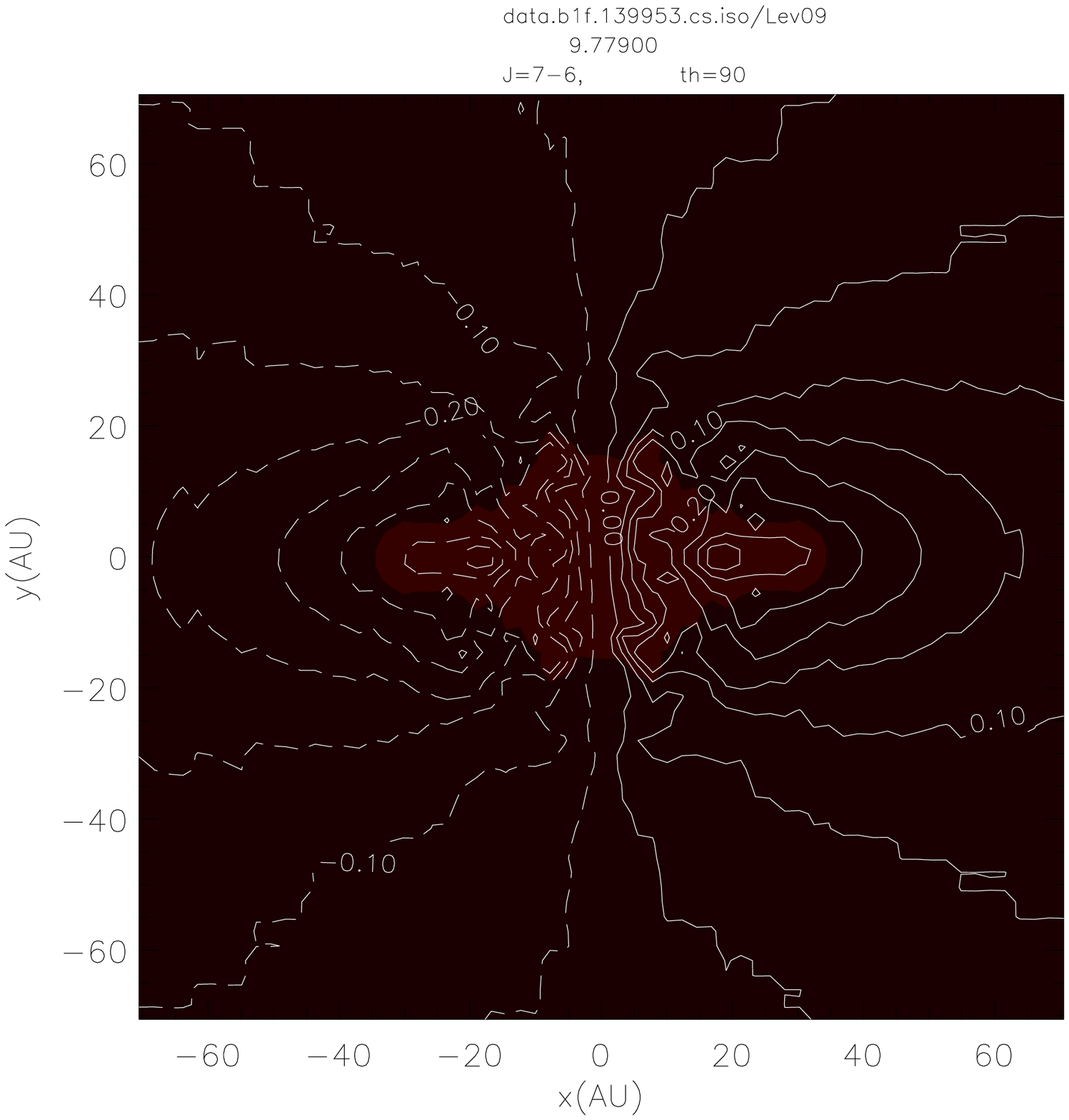}\\
\caption{\label{fig:comp_iso_poly}
Comparison with an isothermal model [({\it c}) and ({\it d})].
Panels ({\it a}) and ({\it c}) are for CS $J=2$--$1$
 and panels ({\it b}) and ({\it d}) are for CS $J=7$--$6$.
The calculations are made for the edge-on view with $\theta=90^\circ$.
}
\end{figure}

In Figure \ref{fig:comp_iso_poly},
 we compare the radiation model [({\it a}) and ({\it b})]
 with the isothermal model  [({\it c}) and ({\it d})],
 in which the temperature distribution is artificially 
 replaced with the isothermal distribution of kinetic temperature
 $T_{\rm K}=10\,{\rm K}$. 
Panels ({\it a}) and ({\it c}) show the CS $J=2$--$1$ line
 and ({\it b}) and ({\it d}) show the  CS $J=7$--$6$ line.
The figure indicates that the isothermal model has extremely
 low brightness even in the protostellar first core phase
 (the CS $J=2$--$1$ line has a peak of $25.5\,{\rm K\,km\,s^{-1}}$ while
 the radiative model predicts $89.6\,{\rm K\,km\,s^{-1}}$.
The CS $J=7$--$6$ line has a peak of $\simeq 10\,{\rm K\,km\,s^{-1}}$
 in the isothermal model while the radiative model predicts
 $\simeq 130\,{\rm K\,km\,s^{-1}}$). 
Since the peak of the integrated intensity is at
 $\simeq 17\,{\rm K\,km\,s^{-1}}$ for $J=2$--$1$ and
 $\simeq 6\,{\rm K\,km\,s^{-1}}$ for $J=7$--$6$ in the isothermal collapse phase,
 this isothermal model predicts a similar brightness
 as for the isothermal collapse phase.
This indicates that if we ignore the radiation transfer in determining
 the kinetic temperature 
 the emission is significantly underestimated in the protostellar phase,
 at least at this scale,
which is not unexpected.

Another difference between the two models is
 that the velocity gradient in the disk
 is significantly underestimated in the isothermal model.
That is,
 the contour of
 $\langle V\rangle\simeq 0.2\, {\rm km\,s^{-1}}$  runs at
 $|x| \simeq 55\, {\rm AU}$ in ({\it a})
 but $|x| \simeq 40\, {\rm AU}$ in ({\it c}).
This is a natural outcome of low temperature for the isothermal model
 $T_{\rm K}=10\,{\rm K}$. 
At this temperature the molecules are essentially in the ground state.
The effect of the optical thickness does not play an important role
 in the isothermal model,
 in which the integrated intensity follows the
 column density distribution
 and the isovelocity seems to follow the mass average velocity.
This clearly shows that 
 both the gas distribution and the gas dynamics
 are likely to be misinterpreted if we ignore the radiation transfer.
 
\subsection{Asymmetry against the Rotation Axis}
\label{sec:asymmetry}
\begin{figure}
\centering
\includegraphics[width=80mm]{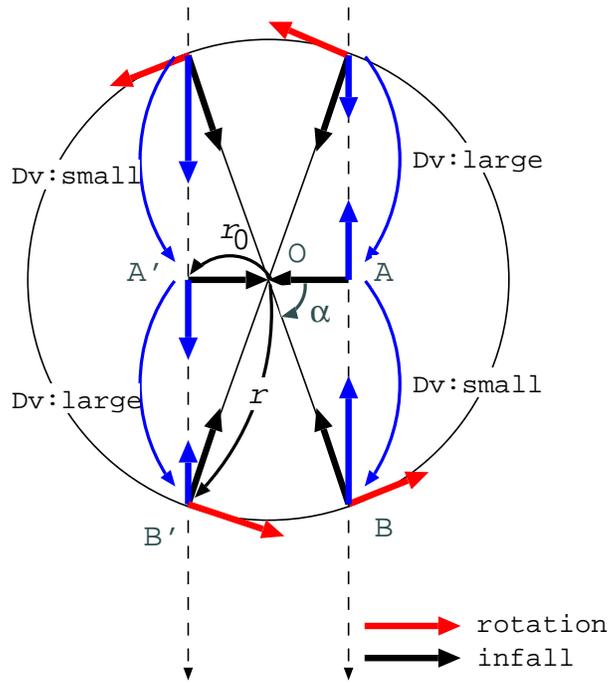}
\caption{\label{fig:asymmetry}Explanation why asymmetry arises.}
\end{figure}

In this subsection we consider why the left side of
 the disk and outflow is observed to be brighter than the right side.
We assume a gas with both contraction and rotation motions,
 in which the infall $v_{\rm in}$ and rotation $v_\phi$ 
 speeds depend only on the radius $r$ as 
 $v_{\rm in}(r)$ and $v_{\phi}(r)$.
Consider two LOSs A--B and A$'$--B$'$,
 symmetric with respect to the center O,
 as shown in Figure \ref{fig:asymmetry}.
Points A and A$'$ are tangential points of the LOSs. 
Asymmetry arises from the configuration that
 emission from the gas with higher $T_{\rm ex}$ near the tangential point
 is absorbed by the foreground cooler gas.
The velocity difference between the emitter and the absorber
 has an essential role in the self-absorption.  
The LOS velocities at points A and A$'$ are
 respectively $v_{\phi}(r_0)\equiv v_{\phi\,0}$ and $-v_{\phi}(r_0)\equiv -v_{\phi\,0}$,
 where $r_0$ represents the distance between A (and A$'$) and O.
Those at points B and B$'$ are equal to
 $v_{\phi}(r)\cos\alpha+v_{\rm in}(r)\sin\alpha$ and
 $-v_{\phi}(r)\cos\alpha+v_{\rm in}(r)\sin\alpha$, respectively.
Here $\angle$ AOB is denoted by $\alpha$ and $r$ represents the distance
 between B (and B$'$) and O.
The relative velocity of point B observed from point A becomes
\begin{equation}
\Delta v_{AB}=v_\phi(r)\cos\alpha-v_{\phi\,0}+v_{\rm in}(r)\sin\alpha,
\end{equation}  
while that of A$'$B$'$ is equal to
\begin{equation}
\Delta v_{A'B'}=-v_\phi(r)\cos\alpha+v_{\phi\,0}+v_{\rm in}(r)\sin\alpha.
\end{equation}  
Since $v_{\rm in}(r)\sin\alpha > 0$, 
 $\Delta v_{AB} > \Delta v_{A'B'}$ and $|\Delta v_{AB}| > |\Delta v_{A'B'}|$,
 when $v_\phi(r)\cos\alpha-v_{\phi\,0} > 0$.
When $v_\phi(r)\cos\alpha-v_{\phi\,0} < 0$, $\Delta v_{AB} < \Delta v_{A'B'}$
 and $|\Delta v_{AB}| < |\Delta v_{A'B'}|$.
This is valid irrespective of $\alpha$.
For rigid body rotation, the rotation speed satisfies
 $v_\phi(r)\cos\alpha-v_{\phi\,0} = 0$.
Since the rotation law of this kind of object is between
 the Kepler rotation $v_\phi\propto r^{-1/2}$
 and the rigid body rotation $v_\phi\propto r^{+1}$,
 the inequality $v_\phi(r)\cos\alpha-v_{\phi\,0} < 0$ is satisfied
 for $r > r_0$.
In this case, the magnitude of the relative velocity
 for A$'$B$'$, $|\Delta v_{A'B'}|$, is
 larger than that for AB, $|\Delta v_{AB}|$, 
 that is, $|\Delta v_{A'B'}|>|\Delta v_{AB}|$.
This means that the approaching side (A$'$B$'$) has a larger
 velocity gradient than the receding side (AB),
 when inflow and rotation motions coexist.

Since the excitation temperature decreases with radius,
 the emissions from the inner radii (A and A$'$) are more or less 
 absorbed by the intervening foreground gas (B and B$'$).
Absorption depends not only on the excitation temperature 
 but also the velocity difference between the emitter and absorber
 if we consider a line with a relatively large optical depth that
 shows a ``self-absorption" feature.
A larger velocity gradient leads to a less efficient absorption.
In this case, since $|\Delta v_{A'B'}|>|\Delta v_{AB}|$,
 self-absorption due to the foreground gas is less efficient
 for the LOS A$'$B$'$.
As a result, when the gas has both infalling motion 
 and rotation motion, the approaching side (A$'$B$'$) is brighter
 than the receding side (AB).
This explains the asymmetry around the rotation axis.
That is, the rotation and simultaneous contraction motions
 of a gas leads to an approaching-side-enhanced asymmetry,
 if the excitation temperature decreases with radius.   

\subsection{Observability of the First Core Stage}

To find the first core, we have to know how it can be observed.
In Figures \ref{fig:theta-dep-pre} and \ref{fig:theta-dep-post},
 we have shown how the observational features change after the first
 core is formed.   
In the face-on view, we have found that a bright spot and a ring are typical
 for the first core phase.
The former indicates a first core and the latter indicates molecular
 outflow driven magnetically.
Since the second core is believed to have a much brighter intensity 
 owing to its deeper gravitational potential,
 objects found with the intensity calculated here
 $\sim 50\,{\rm K\,km\,s^{-1}}$ are regarded as being
 candidates for the first core. 
In the edge-on view, the first core phase has a typical appearance of
 a disk and magnetically driven molecular outflow, both of which have
 strong blue-intense asymmetry.  This indicates strong evidence that the rotation and infall motions coexist.
In Figure \ref{fig:comp_iso_poly}, 
 the lowest level of the integrated intensity represents
 the outflow at the 70-AU scale.
The height of the molecular outflow
 is approximately proportional to the age after the first core is formed,
 since the molecular outflow began to be accelerated just after the
 first core formation.
If we measure the length of the molecular outflow as
 $\lesssim 100 \,{\rm AU}$,
 the central core must be a first core.

\section{Summary}

We have calculated the non-LTE radiation transfer for CS rotation transitions,
 which gives an expectation of the observation features of the first core.
A  Monte-Carlo code is developed to solve the non-LTE radiation transfer
 based on the nested grid hierarchy.
Incorporating the nested grid hierarchy enables us to calculate
 the case in which a large optical depth is expected for the envelope
 in the foreground.
Viewing from the rotation axis,
 the line width allows identification of the transition between before and after the first 
 core formation.
The spectra of the disk show blue-intense asymmetry,
 which indicates the inflow.
The molecular outflow launched from the vicinity of the first core
 appears as a 50-AU-scale ring in the pole-on view
and is observed as a rotating funnel in the edge-on view.
The rotating and inflowing disk has a characteristic feature
 which is seen
 in neither a purely rotating disk nor a purely inflowing disk:
The approaching side is brighter than the receding side in the integrated intensity    and the velocity gradient is larger in the approaching side than in the receding side.
This enables us to identify the rotating inflowing disk.
This asymmetry is also seen in the molecular outflow gas.
     
\bigskip
This work was supported in part by JSPS 
 Grant-in-Aid for Scientific Research (A) 21244021 in the fiscal year 2009--2010. 
K. Tomida was supported by the Research Fellowship from JSPS
 for Young Scientists.
Numerical computations were carried out in part on NEC SX-9 and Cray XT4 at 
 the Center for Computational Astrophysics, CfCA, of National Astronomical 
 Observatory of Japan.

\clearpage

\appendix
\section*{Non-LTE Radiation Transfer Calculation on Nested Grid Hierarchy}
\begin{figure}[h]
\centering
\includegraphics[width=120mm]{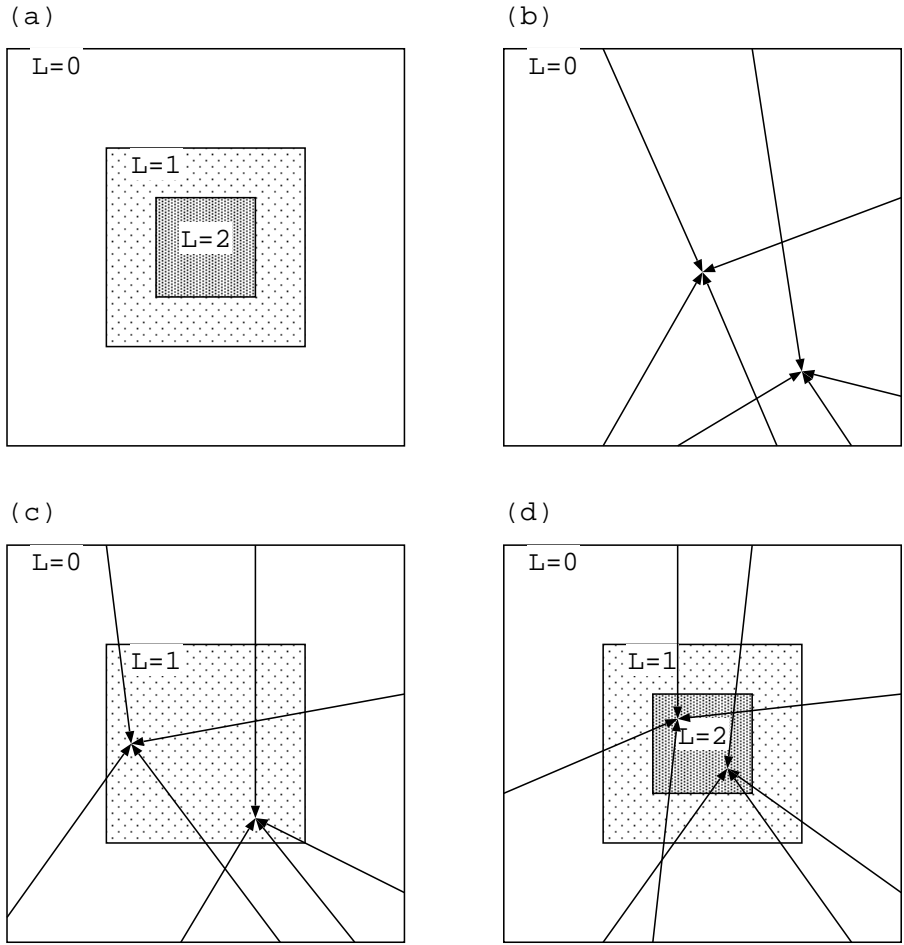}
\caption{\small\label{fig:nest-nonLTE}
Method of solving non-LTE radiative transfer problem on the nested grid.
(a) Geometry of the nested grid, where the $L=1$ grid covers the central
 1/4 of the $L=0$ grid, the $L=2$ grid covers the central
 1/4 of the $L=1$ grid, and so on.
Although 2-dimensional cross-cut is shown in this figure, actual
 radiative transfer problem is solved in 3-dimensional geometry. 
(b) In this method, we first solve the non-LTE problem of $L=0$ with
a CMB outer boundary condition.
This gives the distributions of the absorption coefficient and emissivity
 in the $L=0$ grid, which enables us to calculate the intensity at the outer 
 boundary of $L=1$. 
(c) We solve the non-LTE problem of $L=1$ with
 the intensity at the boundary obtained in (b).
This gives the distributions of absorption coefficient and emissivity
 in $L=1$ grid, which enables us to calculate the intensity at the outer 
 boundary of $L=2$. 
(d) We solve the non-LTE problem of $L=2$ with
 the intensity at the boundary obtained in (c).
This procedure continues until we have reached the target level.}
\end{figure}

In this Appendix, we describe the numerical method to solve the non-LTE
 radiation transfer calculation on the nested grid hierarchy.
The nested grid uses a number of grids whose spatial resolutions
 are different. 
The finer grid covers a central region
 and the coarser one covers the whole cloud ({Fig.\ref{fig:nest-nonLTE}\it a}). 
Each level of grid contains the same number of grid points.
The boundary condition for the coarsest (level $L=0$) grid is given
 as the cloud is immersed in a vacuum which is filled  with the CMB.
On the contrary, the boundary condition for the grids contained in the 
 $L=0$ grid $(L > 0)$ is determined by the strength of intensity $I_\nu$ at
 the boundary which is calculated using the emissivity and absorption
 coefficient of the outer grids.     
\begin{enumerate}
\item 
Solve $L=0$ grid with the CMB condition at the outer boundary.
Non-LTE radiative transfer calculation
 gives distributions of emissivity and absorption coefficient obtained in $L=0$ level
 ({Fig.\ref{fig:nest-nonLTE}\it b}).
\item   
Solve $L=1$ grid.
\begin{enumerate}
\item 
Generate rays passing $L=1$ grid. 
\item 
Integrate radiative transfer equation along the rays
 for the part covered by $L=0$ grid but outside $L=1$ grid.
Record the intensity at the outer boundary of $L=1$.
\item 
By iteration, obtain the equilibrium solution for $L=1$ grid
 with the boundary condition obtained in the previous step (2-b),
giving the distributions of emissivity and absorption coefficient 
in $L=1$ level (Fig.\ref{fig:nest-nonLTE}{\it c}).
\end{enumerate}
\item 
Solve $L=2$ grid.
\begin{enumerate}
\item 
Generate rays passing $L=2$ grid. 
\item 
Integrate radiative transfer equation along the rays
 inside $L=0$ grid but outside $L=2$ grid using emissivity
 and absorption coefficients obtained in both $L=0$ and $L=1$ grids.
When a point is covered both by $L=0$ and $L=1$ level grids,
 the emissivity and absorption coefficients for the finer level ($L=1$) are taken
 according to the basic idea of the nested grid.
Record the intensity at the outer boundary of $L=2$.
\item 
With iteration, obtain the equilibrium solution for $L=2$ grid
 with the boundary condition obtained in the previous step (3-b).
This gives the distributions of emissivity and absorption coefficient
 in $L=2$ level (Fig.\ref{fig:nest-nonLTE}{\it d}).
\end{enumerate}
\item
When the target grid level is reached, we stop this procedure.
\end{enumerate}

\clearpage
\begin{table}
\begin{tabular}{lll}   
Transition CS & Frequency (GHz) & ALMA Band \\
\hline
\hline
$J=1-0$ & 48.9909549 & ---\\
$J=2-1$ & 97.9809533 & Band 3 \\
$J=7-6$ & 342.8828503 & Band 7 \\
$J=8-7$ & 391.8468898 &  Band 8  \\
\hline
\end{tabular}
\centering\caption{Frequencies in laboratory frame of the transitions
appeared in this paper.\label{tbl:freq}}
\end{table}

\clearpage
%
%

%
%

%
%

%
%

%
%

%
%

%
%

%
%

%
%

\end{document}